\documentclass[aps,prd,twocolumn,groupedaddress,nofootinbib]{revtex4-2}

\usepackage{mathtools}
\usepackage{amsmath}   
\usepackage{amssymb}
\usepackage{graphicx}
\usepackage{xcolor}
\usepackage{multirow}
\usepackage{lipsum}
\usepackage{soul}
\usepackage{aas_macros} 
\usepackage{rotating} 
\usepackage{svg}

\usepackage{soul}
\newcommand{\addtext}[1]{{\color{brown} #1}}

\newcommand{\oforder}[1]{\mathcal{O}(#1)}

\newcommand{\E}[1]{\times 10^{#1}}

\newcommand{\Msun}{M_\odot}

\newcommand{\orcid}[1]{\href{https://orcid.org/#1}{\includegraphics[width=9pt]{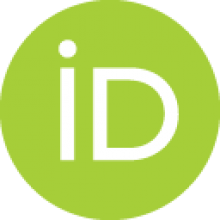}}}

\begin{document}

\title[Collapsar Disk Outflow Nucleosynthesis]{Collapsar disk outflows II: Heavy element production}

\author{Coleman Dean\orcid{0000-0001-9364-4785}}
\email[]{cddean@ualberta.ca}
\affiliation{Department of Physics, University of Alberta, Edmonton, AB T6G 2E1, Canada}

\author{Rodrigo Fern\'andez\orcid{0000-0003-4619-339X}}
\affiliation{Department of Physics, University of Alberta, Edmonton, AB T6G 2E1, Canada}

\date{\today}

\begin{abstract}
We investigate nucleosynthesis in the sub-relativistic outflows from black hole (BH) accretion disks formed in failed supernovae
from rapidly-rotating Wolf-Rayet stars. These disks reach the neutrino-cooled regime during a portion of their evolution, undergoing significant
neutronization and thus having the potential to support the $r$-process. Here, we analyze the formation of heavy elements
in the ejecta from global, axisymmetric, long-term, viscous hydrodynamic simulations of these systems that include neutrino emission
and absorption, Newtonian self-gravity, a pseudo-Newtonian potential for the BH gravity, and a 19-isotope nuclear
reaction network. Tracer particles are used for post-processing with a larger reaction network. 
In addition to analyzing models introduced in a previous paper, we present new models in which we modify the rotation
profile of the progenitor star, to maximize neutrino reprocessing of circularized mass shells. 
All of our models produce several $M_\odot$ of oxygen, followed by about a solar mass of carbon, neon, and nickel, with
other alpha elements produced in smaller quantities. 
Only one of our models, with the lowest strength of viscous angular 
momentum transport, yields significant amounts of first $r$-process peak elements, with negligible yields at higher nuclear masses.
The rest of the set, including models with a modified rotation profile, produces very small or negligible quantities of elements beyond the iron group. Models that produce the heaviest
elements (up to $A\sim 200$) do so along the proton-rich side of the valley of stability at high entropy ($s/k_B \sim 80$), 
pointing to the $rp$-process as a mechanism that operates in collapsars. 
The absence of neutron-rich ejecta proves to be insensitive to 
changes in the rotation profile of the star, suggesting that heavy $r$-process elements are difficult to produce in collapsars if no 
large-scale poloidal magnetic field is present in the disk to drive outflows during neutronization. 
\end{abstract}

\maketitle


\section{Introduction \label{sec:intro}} 

The majority of chemical elements other than hydrogen and helium are made in stellar interiors, or in stellar explosions, 
through a variety of nuclear processes (e.g., \cite{burbidge_1957,johnson_2019}). Massive stars are expected to produce 
mostly alpha chain elements during stellar evolution, and iron-group elements through explosive
nucleosynthesis during the supernova (SN) (e.g., \cite{janka_2023}). 
Whether and how the rapid-neutron capture process ($r$-process) 
occurs in these SNe is an area of active theoretical research (e.g., \cite{wanajo_2018,witt_2021,wang_2023}). Observationally, 
there may be a need for an $r$-process source that operates with a shorter time-delay after star formation than neuron 
star (NS) mergers, at least to explain the europium enrichment of some extremely metal-poor stars 
in the halo of our galaxy \citep{mathews_1990}. A source that operates on the timescale of massive star evolution ($\sim 10^6$\,yr) would bridge this gap.

Collapsars are a subset of massive stellar explosions, corresponding to a failed supernova of a rapidly-rotating progenitor
that forms a black hole (BH) accretion disk \cite{woosley_1993}. If a relativistic jet is launched and successfully traverses the 
stellar envelope, a long-duration gamma-ray burst (lGRB) is produced (e.g., \cite{macfadyen_1999}). The accretion disk can, by itself,
launch a sub-relativistic outflow that can eject the stellar envelope, possibly accounting for the broad-line type Ic (Ic-BL) SNe
associated with lGRBs \cite{macfadyen_2003}.

If the accretion disk in a collapsar reaches the neutrino-cooled (``NDAF") regime, then significant neutronization
can occur, providing the conditions for the $r$-process to operate \cite{macfadyen_1999,kohri_2005}. However, 
one of the main remaining uncertainties  
is how much of the neutron-rich material from the disk makes 
its way into the outflow and is ejected.

In addition, not all solar heavy element abundances can be explained by the $s$- and $r$-processes. Particularly, 
proton-rich isotopes such as $^{96}$Ru and $^{92}$Mo, as well as as $^{92}$Nb, which are shielded on the $(N,Z)$-plane 
by other stable isotopes and cannot be produced by $\beta$-decays from the neutron rich side of the valley of stability. 
While some of these $p$-nuclei have been proposed to be produced by the photodisintegration of $s$- and $r$- process 
nuclei by gamma rays (the $\gamma$-process) in core-collapse (CC) SNe environments \cite{woosley_1978,howard_1991}, the 
$\gamma$-process alone cannot explain all $p$-nuclei in the solar abundance. Other $p$-nuclei production processes often 
require some explosive astrophysical environments with proton rich $Y_e>0.5$ ejecta \cite{wallace_1981,frohlich_2006}. 
Recent viscous hydrodynamic simulations of collapsar disk outflows \cite{just_2022,fujibayashi_2022} 
produce at least some fraction of the ejecta with proton-rich conditions ($Y_e>0.5$), suggesting nucleosynthesis 
may occur in these explosions on the proton-rich side of the ``valley of stability" via the rapid proton capture process ($rp$-process). 

Heavy element production has been investigated both in collapsar jets and in disk winds, on the assumption that neutronized 
material will make its way out of the star. One-dimensional, steady-state nucleosynthesis calculations including the production 
of $^{56}$Ni and $r$-process elements have been performed for stellar mass BH-accretion disk outflows by \cite{pruet_2003}, \cite{fujimoto_2004}, and \cite{surman_2006}. 
Nucleosythesis calculations on magnetically-driven jets have found that a small amount $r$-process 
element production is possible when neutron-rich material from the disk is ejected via the jet \cite{fujimoto_2007,ono_2012,nakamura_2013}. 
More recently, general-relativistic magnetohydrodynamic (GRMHD) simulations of collapsar disks that start from equilibrium tori and 
neglect the infalling mantle of the star produce $r$-process elements in a mass accretion rate-dependent way \cite{siegel_2019,miller_2020}. 
Additionally, the electron fraction of the ejecta in these simulations, and the extent of $r$-process element production, differs based 
on the neutrino transport schemes used.

From the observational perspective, the presence of $r$-process elements in collapsar ejecta have been predicted semi-analytically 
to produce a detectable near-infrared excess during the photospheric phase, relative to a non-$r$ process enriched CCSN \cite{barnes_2022}.
A study of the late time light-curves of four SNe-GRB  (attributed to collapsar sources) was inconclusive in determining the presence 
of $r$-process enrichment, with light-curves consistent with both no $r$-process enrichment or little ($0.01 - 0.15\Msun$) $r$-process 
contribution \cite{rastinejad_2023}.
 
In order to generate a complete description of the elements produced in collapsars, as well as to answer the question of whether 
$r$-process elements can be produced in these explosions, there is a need for global, long-term simulations of collapsar disk 
outflows that include the physics relevant for nucleosynthesis that is not present in existing work. Axisymmetric, viscous 
hydrodynamic models allow for longer term simulations than 3D GRMHD calculations, following disk formation from the collapsing 
progenitor, and  evolving the disk wind driven shock through the stellar mantle of the star. Recent works (\cite{just_2022}, 
\cite{fujibayashi_2022}, \cite{fujibayashi_2023}, and \cite{paper_1} [hereafter Paper I]) use this method, and find that there 
is insufficient neutronization in ejected material to produce significant quantities of $r$-process elements. 

In Paper I, we introduced a numerical approach to conduct global simulations of collapsars, starting from pre-collapse, rotating 
Wolf-Rayet stars. Progenitors are evolved from core collapse to BH formation with a general relativistic, spherically-symmetric, 
neutrino radiation-hydrodynamics code that accounts for approximate rotation effects. Results are then mapped into a axisymmetric 
hydrodynamics code for longer-term evolution ($\gtrsim 100$\,s), including neutrino emission and absorption, viscous angular 
momentum transport, self-gravity, and a 19-isotope nuclear network, capturing BH accretion disk formation self-consistently, and 
outflow production until the shocked wind reaches the stellar surface. Results show that the disk wind is consistently capable of 
driving a successful stellar explosion. However, we found insufficient neutronization of the disk outflow to produce significant 
heavy $r$-process elements. The ejecta nevertheless contains sufficient $^{56}$Ni to power a type Ic-BL light curve.  

Here we perform detailed nucleosynthesis analysis of the models from Paper I, and present additional models that employ progenitors 
with a modified rotation profile, aimed at maximizing the chance of ejecting neutron-rich matter in the disk outflow. 

The paper is structured as follows: Section \ref{sec:methods} provides an overview of the numerical setup presented in Paper I, 
placement and post-processing of tracer particles, modification of stellar rotation profiles, and a list of models evolved. 
Section \ref{sec:results} compares the results of the simulations using progenitors with modified rotation relative to those from Paper I, as
well as our nucleosynthesis results, and comparison with previous work. Section \ref{sec:summary} contains a summary and discussion. 


\section{Methods \label{sec:methods}}

The numerical setup used for our simulations is discussed in detail in Paper I, here we provide a brief summary of the 
numerical hydrodynamic methods employed. We also describe the use of tracer particles and a nuclear reaction network for 
nucleosynthesis in post-processing, as well as modifications to the stellar rotation profile. 

\subsection{Summary of Computational Approach}\label{sec:computational_approach}

We employ two rapidly rotating Wolf-Rayet progenitor stars (\texttt{16TI} and \texttt{35OC}) from \cite{woosley_2006b}. 
The stars are evolved from the onset of core collapse until BH formation with the spherically-symmetric, general relativistic 
neutrino radiation-hydrodynamic code \texttt{GR1D} version 1 \cite{oconnor_2010}. Rotational effects are included in approximate 
form during this evolution stage. At the point of BH formation, the stellar progenitor is mapped into a two-dimensional (2D) 
axisymmetric viscous hydrodynamic setup based on the astrophysical hydrodynamics code \texttt{FLASH} version 3.2 \cite{fryxell00,dubey2009}. 

The public version of \texttt{FLASH} has been modified to include viscous angular momentum transport, a customized implementation 
of multipole (Newtonian) self-gravity, and a neutrino leakage scheme that accounts for emission and absorption. The pseudo-Newtonian 
potential of Ref. \cite{artemova1996} is included to account for the gravity of a spinning BH, with mass and angular momentum
updated at every timestep by accreting matter. We employ the (Helmholtz) equation of 
state of \cite{timmes2000}, extended to higher and lower densities relative to the table in \texttt{FLASH}, as outlined in Paper I. 

\begin{figure*}
    \centering
    \includegraphics[width=\textwidth]{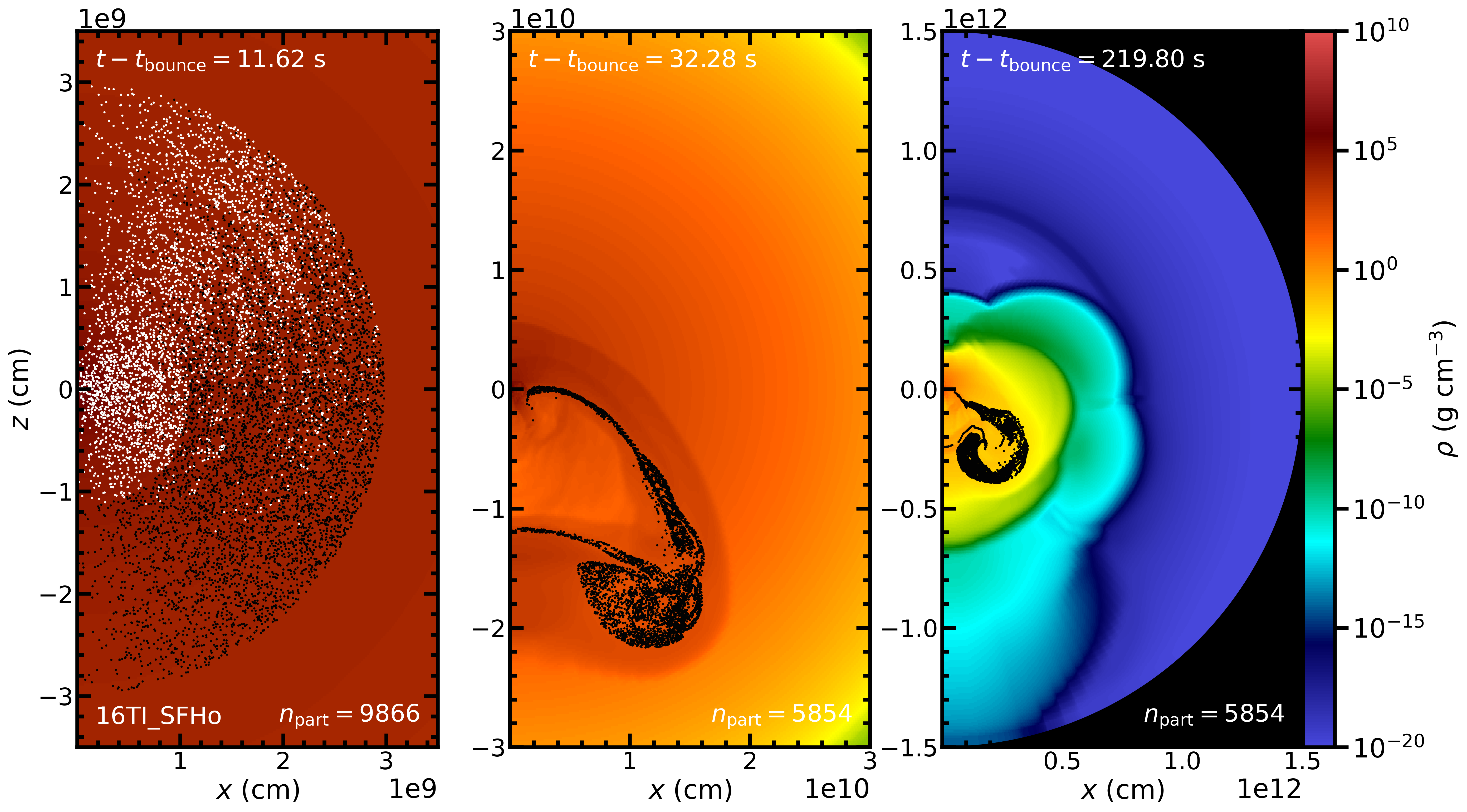}
    \caption{Snapshots of the density in model \texttt{16TI\_SFHo} at three points in the simulation, as labelled, corresponding 
    to initial particle placement (left), shortly after the start of rapid shock expansion (middle), and end of the simulation (right). 
    Tracer particle positions are overlayed, with white dots corresponding to particles that end up accreted onto the BH, and ejected 
    particles marked in black. The number of particles remaining in the computational domain is listed in each panel, 
    including the small number that never exceed $1\,$GK and are discarded from the analysis. 
    The black region visible in the third panel marks the outermost edge of the spherical grid. 
    }
    \label{fig:particle_progression}
\end{figure*}

Finally, all simulations include the 19-isotope nuclear reaction network of \cite{weaver1978}, as implemented in \texttt{FLASH}, 
to track nuclear energy changes during the simulation. This also allows us to obtain detailed composition information for the set 
of 19 isotopes: n, p$^+$, $^{3}$He, $^{4}$He, $^{12}$C, $^{14}$N, $^{16}$O, $^{20}$Ne, $^{24}$Mg, $^{28}$Si, $^{32}$S, $^{36}$Ar, 
$^{40}$Ca, $^{44}$Ti, $^{48}$Cr, $^{52}$Fe, $^{54}$Fe, $^{56}$Fe, and $^{56}$Ni. Above a temperature $T_{\rm NSE} = 5\,{\rm GK}$,
we supplement this network with a nuclear statistical equilibrium (NSE) solver, based on that reported in 
\cite{seitenzahl_2008}\footnote{Available at \url{https://cococubed.com/}.}.

\subsection{Nucleosynthesis in Post-Processing \label{sec:skynet}}

To explore heavy element nucleosynthesis over a much wider range of isotopes than those in the embedded 19-isotope network, 
we use a standard post-processing approach with passive tracer particles and an external nuclear reaction network. Particles 
record thermodynamic and kinematic quantities, as well as other source terms such as energy or lepton number rates of change, 
which act on individual fluid parcels for the duration of the simulation. 

For each model, $\sim 10^4$ particles are initialized at the time of disk formation. Particles are initialized at random locations, 
with tracers having equal mass and following the density distribution, constrained to the radial range $10^5\,$cm - $3\E{9}\,$cm 
in all models except \texttt{35OC\_SFHo\_$k_{\rm rad}$10}, for which the radial range is $10^6\,$cm - $10^{10}\,$cm. This radial 
range was chosen as a first attempt to obtain a reasonable sampling of the neutrino-reprocessed matter that is used consistently in 
all simulations. Particles do not sample all ejected material, however. In Sec.~\ref{sec:optimal_particle_placement} we quantify 
sampling of the ejecta by the particles, and develop suggestions for initial particle placement in future collapsar simulations.

Figure \ref{fig:particle_progression} shows a density colour map of model \texttt{16TI\_SFHo} with overlaid particle positions, 
shortly after particle placement (left panel), midway through the simulation after the onset of rapid expansion (middle panel), 
and at the end of the simulation (right panel). Black dots denote particles that eventually make their way into the outflow, and 
white dots show particles accreted onto the BH. The number of particles out of the initial $10^4$ that remain within the 
computational domain are labelled in each snapshot. The middle and right panels illustrate that the shock wave in this model 
expands towards the $-\hat{z}$ pole due to the instability to axisymmetric perturbations described in Paper I. It is 
apparent that as a result of this asymmetric expansion, there is additional accretion from the $+\hat{z}$ pole direction, with 
the majority of particles initialized in that  hemisphere not making their way into the outflow. Nucleosynthesis calculations are 
performed on particles that are unbound (positive Bernoulli parameter) at the end of the simulation. While the particles do not 
completely sample the ejecta, ejected particles mostly represent the peak of the mass ejection distribution 
(Section~\ref{sec:optimal_particle_placement}).   

\begin{figure*}
    \centering
    \includegraphics[width=0.7\textwidth]{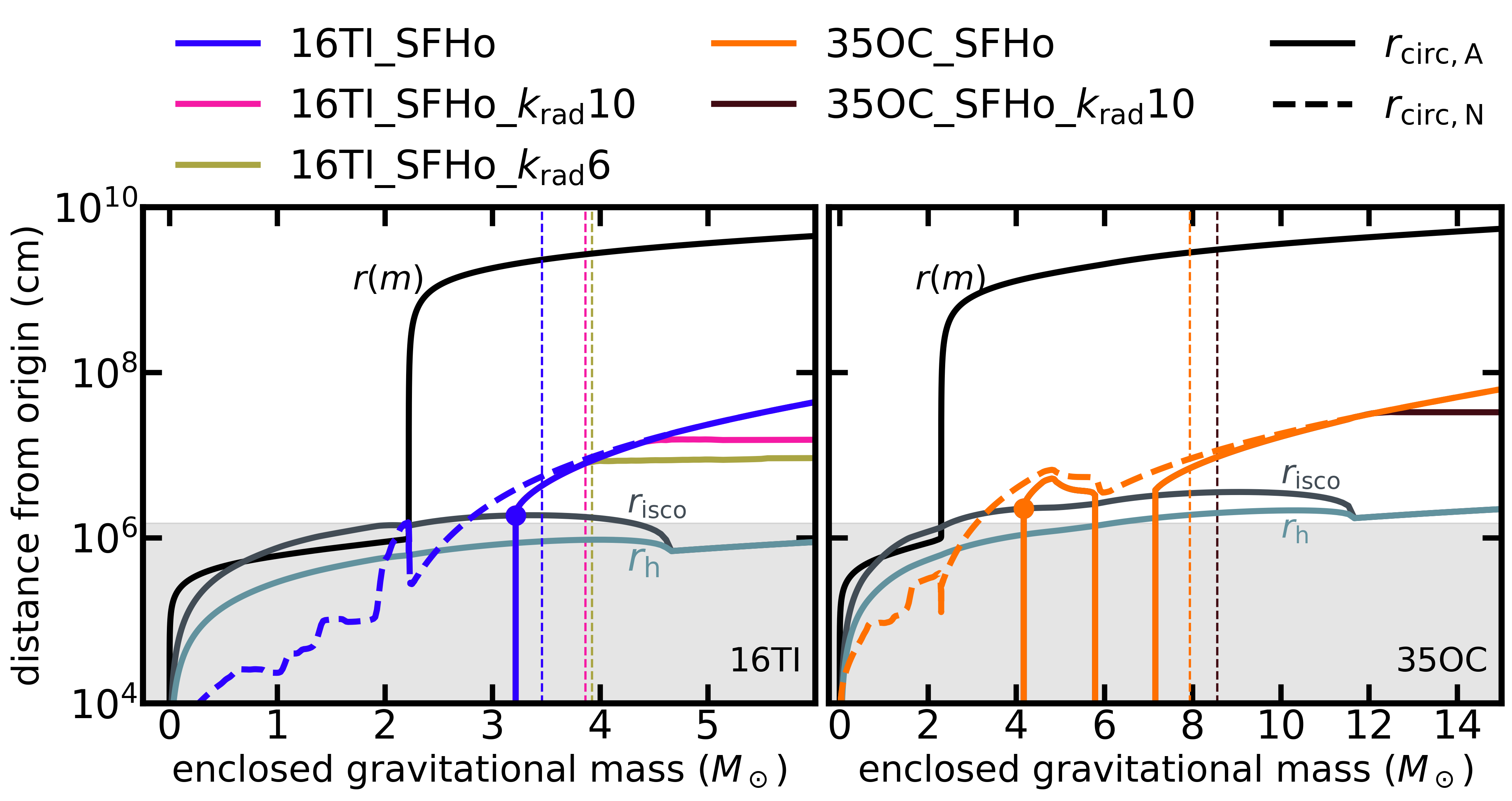}
    \caption{Characteristic radial distances as a function of enclosed gravitational mass for stellar progenitors \texttt{16TI} 
    and \texttt{35OC} from \cite{woosley_2006b} at the last snapshot in \texttt{GR1D} prior to BH formation. The radial coordinate 
    is $r(m)$ (black solid), with the the PNS surface corresponding to the vertical segment at $M\simeq 2M_\odot$. Colored lines 
    without label show circularization radii obtained with the pseudo-Newtonian potential employed by the code (solid) and a 
    Newtonian potential with the same enclosed mass (dashed, c.f. Eq.~\ref{eq:rcirc_N}), with colors corresponding to different 
    models, as labeled above. Models \texttt{16TI\_SFHo} and \texttt{35OC\_SFHo} were presented in Paper I, while remaining 
    models have the modified rotation profile discussed in Section \ref{sec:modified_angz} (Equation~\ref{eq:j_modified}), with the 
    value of $k_{\rm rad}=\{6,10\}$ indicated in the model name. Finally, each panel shows the event horizon radius 
    $r_{\rm h}$ and radius of the innermost stable circular orbit $r_{\rm isco}$ for a BH with the same enclosed mass and angular 
    momentum as in the progenitor with unmodified $j(M)$. The gray shaded region is excised from the computational domain when mapping 
    to \texttt{FLASH}. The BH mass when disk formation is predicted to occur is marked with a circle, while the BH mass at actual disk 
    formation in each simulation is marked with a thin vertical dashed line.}
    \label{fig:radr_mgrav}
\end{figure*}

Tracer particles are post processed with the nuclear reaction network {\tt SkyNet} \cite{lippuner_2017_skynet}, using the same settings 
as in Refs. \cite{lippuner_2017,Fernandez2020BHNS,fernandez_2022}. The network uses $\sim 7800$ isotopes and $>10^5$ reactions, including 
strong forward reaction rates from the REACLIB database \cite{cyburt_2010}, with inverse rates computed from detailed balance; spontaneous 
and neutron-induced fission rates from \cite{frankel_1947}, \cite{mamdouh_2001}, \cite{wahl_2002}, and \cite{panov_2010}; weak rates from
\cite{fuller_1982}, \cite{oda_1994}, \cite{langanke_2000}, and the REACLIB database; and nuclear masses from the REACLIB database, including
experimental values where available, or otherwise theoretical masses from the Finite-Range Droplet Macroscopic model (FRDM) of \cite{moeller_2016}. 
Post-processing of tracer particles begins the last time the temperature exceeds $6\E{9}\,$K. When the temperature exceeds $5\E{9}\,$K the 
abundances are evolved in NSE. Trajectories that never exceed $5\E{9}\,$K are evolved from the time they reach maximal temperature, assuming 
NSE. Particles that never exceed $10^9\,$K are discarded. Trajectories are evolved in \texttt{SkyNet} for $30\,$years, extrapolating from the 
end of the \texttt{FLASH} simulation ($\sim 100-400\,$s) assuming homologous evolution of density with time ($\rho\propto t^{-3}$).

\subsection{Modified Progenitor Rotation Profile \label{sec:modified_angz}}

The progenitor stars \texttt{16TI} and \texttt{35OC} have specific angular momenta $j$ such that the circularization 
radius\footnote{We neglect the difference between circularization radii obtained with Newtonian or pseudo-Newtonian potentials, as the 
difference matters only prior to disk formation (Figure~\ref{fig:radr_mgrav}).}

\begin{equation}
\label{eq:rcirc_newtonian}
r_{\rm circ} = \frac{j^2}{GM}
\end{equation}
increases monotonically with increasing distance from the center of the star outside the mass coordinate of disk formation. 
In Eq.~(\ref{eq:rcirc_newtonian}), $M$ is the enclosed
gravitational mass. As a result, mass shells located further out circularize at larger distances from the BH
upon collapse, thus reaching lower maximal temperatures. Given the sensitivity of neutrino emissivities to temperature,
this configuration limits the range of stellar shells that are subject to significant neutronization when contributing to 
the shocked accretion disk, and hence limits the occurrence of neutron-rich nucleosynthesis.

\begin{table*}
\caption{List of models studied in this paper, and key quantities. The first five models were introduced in Paper I, and the 
last three (\texttt{16TI\_SFHo\_$k_{\rm rad}$6}, \texttt{16TI\_SFHo\_$k_{\rm rad}$10}, and \texttt{35OC\_SFHo\_$k_{\rm rad}$10}) 
are new. Columns from left to right show the model name, progenitor star from \cite{woosley_2006b}, EOS used in \texttt{GR1D} 
evolution to BH formation, viscosity parameter used in the 2D post-BH evolution in \texttt{FLASH}, and angular momentum profile 
factor $k_{\rm rad}$ (Equation~\ref{eq:j_modified}). Subsequent columns show times relative to core bounce time in \texttt{GR1D}
(BH formation time in \texttt{GR1D} $t_{\rm bh}$, shocked disk formation time in \texttt{FLASH} $t_{\rm df}$, 
shock breakout time $t_{\rm sb}$, and the maximum simulation time $t_{\rm max}$), and BH masses at various points in time (BH mass at disk 
formation, BH mass at the transition to the ADAF phase [Sec.~\ref{sec:paper1_summary}], and final BH mass). 
\label{tab:progenitors}}
\begin{ruledtabular}
\begin{tabular}{lccccccccccc}
Model & Progenitor & EOS & $\alpha$ & $k_{\rm rad}$ & $t_{\rm bh}$ & $t_{\rm df}$ & $t_{\rm sb}$ & $t_{\rm max}$ & $M_{\rm bh}(t_{\rm df})$ & $M_{\rm bh}(t_{\rm ADAF})$ & $M_{\rm bh}(t_{\rm max})$  \\ 
 & & & & & (s) & (s) & (s) & (s) & (M$_\odot$) & (M$_\odot$) & (M$_\odot$) \\
\noalign{\smallskip}
\hline
\noalign{\smallskip}
\texttt{16TI\_SFHo}                  & 16TI & SFHo & 0.03 & --- & 2.72 & 11.1 & 116 & 219.8 & 3.5 & 3.7 & 4.4 \\ 
\noalign{\smallskip}
\texttt{16TI\_SFHo\_$\alpha$01}      &      &      & 0.1  & --- & 2.72 & 11.0 & 236 & 427.1 & 3.5 & --\footnote{The \texttt{16TI\_SFHo\_$\alpha$01} model does not exhibit an NDAF phase, starting in the ADAF phase at the time of disk formation.} & 4.1  \\
\texttt{16TI\_SFHo\_$\alpha$001}     &      &      & 0.01 & --- & 2.72 & 10.6 & 153 & 295.4 & 3.4 & 4.4  & 4.6  \\
\texttt{16TI\_DD2}                   &      & DD2  & 0.03 & --- & 5.24 & 9.9  & 168 & 302.0 & 2.9 & 3.2  & 3.7  \\
\texttt{35OC\_SFHo}                  & 35OC & SFHo &      & --- & 0.99 & 10.8 & 68  & 102.8 & 7.9 & 10.2 & 11.7 \\
\noalign{\smallskip}
\texttt{16TI\_SFHo\_$k_{\rm rad}$6}  & 16TI &      &      & 6   & 2.72 & 12.2 & 147 & 440.0 & 3.9 & 4.4  & 6.6  \\
\texttt{16TI\_SFHo\_$k_{\rm rad}$10} &      &      &      & 10  & 2.72 & 11.9 & 111 & 217.8 & 3.9 & 4.2  & 5.4  \\ 
\texttt{35OC\_SFHo\_$k_{\rm rad}$10} & 35OC &      &      &     & 0.99 & 11.3 & 72  & 111.1 & 8.6 & 10.9 & 12.9 \\
\end{tabular}
\end{ruledtabular}
\end{table*}

In an attempt to assess the sensitivity of the nucleosynthesis output to the angular momentum profile of the
star, we evolve models with a modified $j(M)$ in the star, starting from core-collapse. In both progenitor models, 
the original rotation profile is sub-Keplerian over a significant fraction 
of the enclosed stellar mass, increasing outward to exceed Keplerian rotation in the outermost 
$\sim 0.3\%$ and $\sim 0.1\%$ of enclosed mass in models \texttt{16TI\_SFHo} and \texttt{35OC\_SFHo}, respectively.
The dynamical time at these locations is $\sim670\,$s and $\sim665\,$s, respectively. Effectively, the super-Keplerian 
layers are frozen over the timescale of the simulation.

The most optimistic case would be obtained if most of the stellar shells circularize close to the BH, so that
matter is hot enough for strong neutrino reprocessing, while avoiding too short an accretion timescale, and keeping the
disk formation process unaltered. We thus modify the angular momentum profile of the star so that a 
\emph{constant circularization radius} is obtained outside a fixed mass shell, with the modified circularization radius satisfying
\begin{equation}
\label{eq:rcirc_N}
r^*_{\rm circ} = \operatorname{min}(r_{\rm circ},k_{\rm rad}\cdot r_{\rm in}),
\end{equation}
where $r_{\rm in}$ is the inner radial boundary of the computational domain and $k_{\rm rad}=6-10$ is a constant found by
trial-and-error. The resulting functional form is shown in Figure~\ref{fig:radr_mgrav}.

The modified angular momentum profile $j^*$ in the progenitor is then
\begin{equation}
\label{eq:j_modified}
j^*(M) = 
\begin{cases}
j(M) & r_{\rm circ}<k_{\rm rad}r_{\rm in} \\
\left(G M\cdot k_{\rm rad} r_{\rm in}\right)^{1/2} & r_{\rm circ}\ge k_{\rm rad}r_{\rm in}.
\end{cases}
\end{equation}
Note that the modified specific angular momentum still increases with enclosed mass $(\propto M^{1/2})$.

\subsection{Models Evolved}

Table~\ref{tab:progenitors} lists all the models studied in this paper. The first five were presented in Paper I. The baseline model is 
\texttt{16TI\_SFHo}, with \texttt{16TI\_SFHo\_$\alpha$01} and \texttt{16TI\_SFHo\_$\alpha$001} changing the magnitude of the viscosity 
parameter $\alpha$, and \texttt{16TI\_DD2} using a different EOS during the \texttt{GR1D} evolution. Model \texttt{35OC\_SFHo} changes 
the progenitor star, keeping everything else constant.  

Three new models are presented here: \texttt{16TI\_SFHo\_$k_{\rm rad}$6}, \texttt{16TI\_SFHo\_$k_{\rm rad}$10}, and 
\texttt{35OC\_SFHo\_$k_{\rm rad}$10}, which use the modified angular momentum profile of Equation~(\ref{eq:j_modified}),
otherwise with the same settings as in models \texttt{16TI\_SFHo} and \texttt{35OC\_SFHo}, respectively. The parameter 
$k_{\rm rad}$ is set such that the circularization radius plateaus with increasing enclosed mass at 
$\sim 2\, r_{\rm circ}(M_{\rm df})$ (Figure \ref{fig:radr_mgrav}) in the corresponding un-modified model from Paper I 
(\texttt{16TI\_SFHo\_$k_{\rm rad}$6} and \texttt{35OC\_SFHo\_$k_{\rm rad}$10}, c.f. Figure~\ref{fig:radr_mgrav}). 
Additionally, we explore a model with $k_{\rm rad}$ consistent with that of the other progenitor (\texttt{16TI\_SFHo\_$k_{\rm rad}$10}). 

The BH mass at several times in the simulation is shown in Table \ref{tab:progenitors}. 
BH masses can increase by $\sim\,5-50\%$ during the NDAF phase, depending on the model.

The maximum simulation time $t_{\rm max}$ for each model is also listed in Table~\ref{tab:progenitors}, with the same 
criterion for stopping the simulation as in Paper I (the shock reaching the outermost radius of the computational domain, 
where $P\simeq 150$\,dyn\,cm$^{-2}$).


\begin{figure*}
    \centering
    \includegraphics[width=0.8\textwidth]{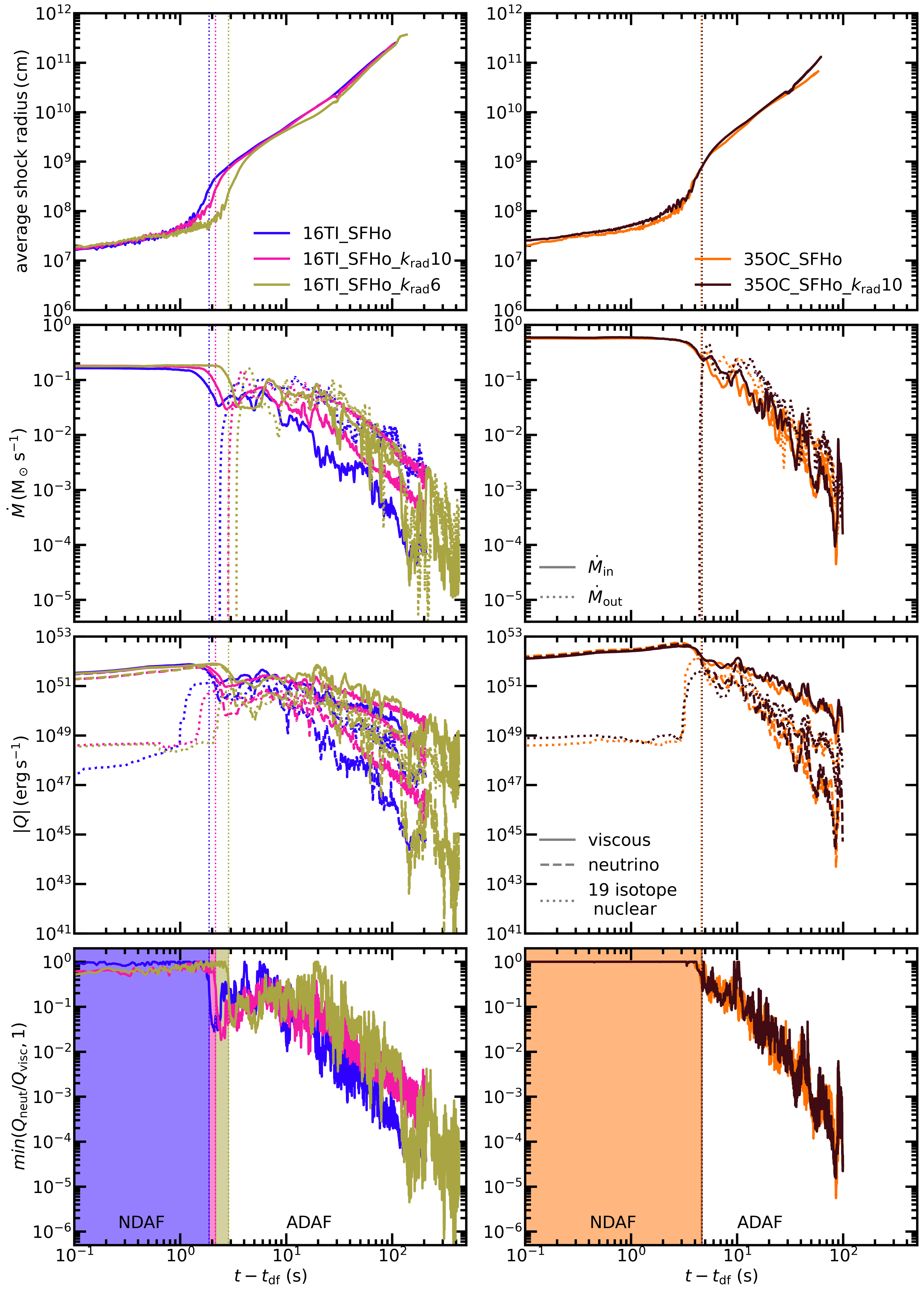}
    \caption{Evolution of models with modified angular momentum profile and the unmodified control set, as labelled 
    (c.f. Table~\ref{tab:progenitors}). \emph{Top}: Evolution of the average shock radius as a function of post-disk formation time. 
    The vertical dotted lines in each panel represent the transition from the NDAF to ADAF phase, as defined by Eq.~(\ref{eq:ndaf-adaf}). 
    \emph{Second from top}: Mass accretion rate $\dot{M_{\rm in}}$ across the inner radial boundary, and mass outflow rate $\dot{M}_{\rm out}$ 
    across a spherical surface at $R_{\rm ej} = 10^9\,$cm as a function of time after disk formation. \emph{Third from top}: Absolute value of 
    viscous heating, net neutrino cooling, and nuclear energy injection as function of post-disk formation time. \emph{Bottom}: Ratio of 
    postshock-integrated net neutrino cooling to viscous heating as a function of post-disk formation time, capped at unity. The early NDAF 
    phase (Eq.~\ref{eq:ndaf-adaf}) is shaded, while the later ADAF phase extends beyond the vertical dotted line. Curves in Rows 2 and 3 are 
    smoothed with a moving average of width $0.5\,$s, for visibility.}
    \label{fig:evolution}
\end{figure*}

\begin{figure*}
    \centering
    \includegraphics[width=\textwidth]{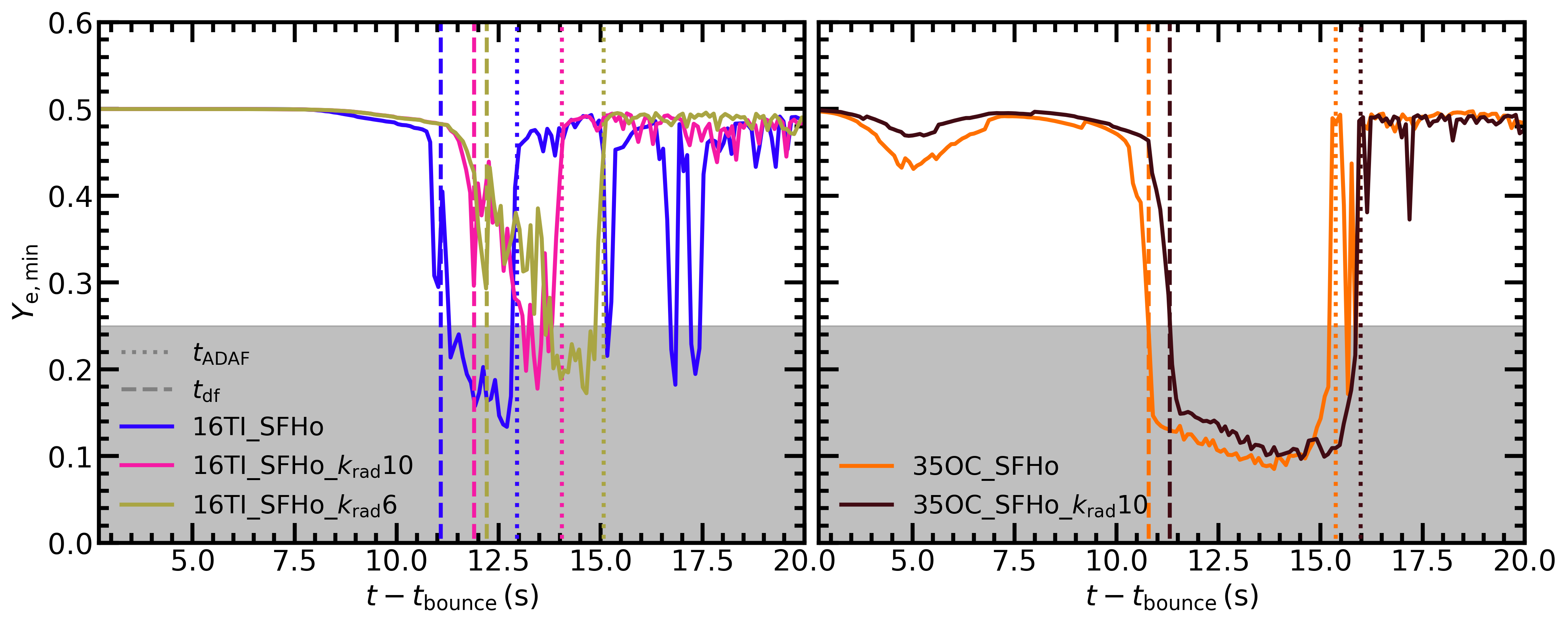}
    \caption{Minimum electron fraction in the simulation domain as a function of time, for different models, as labelled. Disk formation 
    times are marked with a dashed line, and the transition from NDAF to ADAF phase (Eq.~\ref{eq:ndaf-adaf}) is marked with a dotted line. 
    The shaded region shows $Y_e < 0.25$, which is an approximate measure of the value below which lanthanides can be produced (e.g., \cite{lippuner_2015}).}
    \label{fig:minYe}
\end{figure*}

\section{Results \label{sec:results}}

\subsection{Summary of Models from Paper I \label{sec:paper1_summary}}

After BH formation, the stellar mantle starts to accrete radially through the event horizon, 
adding to the mass (and modifying the spin) of the 
central BH. As the specific angular momentum $j$ of infalling material increases, the circularization radius of infalling matter eventually exceeds the innermost stable circular orbit 
$r_{\rm isco}$ of the BH, forming an accretion disk. Figure \ref{fig:radr_mgrav} illustrates the variation of characteristic 
radii as a function of enclosed gravitational mass for each progenitor star, using a snapshot at the end of the \texttt{GR1D} evolution. 

At this point, due to the angular momentum of accreted material, we see the buildup of matter at the equator (i.e. the formation of 
a ``dwarf disk"), before transition to a thermalized disk and the emergence of a shock wave that separates the disk from the 
supersonically infalling star. Most models show poloidal oscillation of this shock wave during an initial NDAF phase, which occurs 
when viscous heating of the disk is balanced by neutrino cooling. 

As the temperature and density of the disk drops, viscous heating becomes dominant over neutrino cooling, leading to a phase of rapid 
shockwave expansion during the Advection Dominated Accretion Flow (ADAF, \cite{narayan_1994}) 
phase. On a timescale of $\sim 100-400$\,s, the expanding 
shock wave in all cases reaches the stellar surface and the outer boundary of the computational domain. Paper I evolves five models 
varying the stellar progenitor, equation of state (EOS) used during \texttt{GR1D} evolution, and the strength of viscous angular 
momentum transport ($\alpha$ parameter).

\subsection{Models with Modified Angular Momentum Profile}

\begin{figure*}
    \centering
    \includegraphics[width=\textwidth]{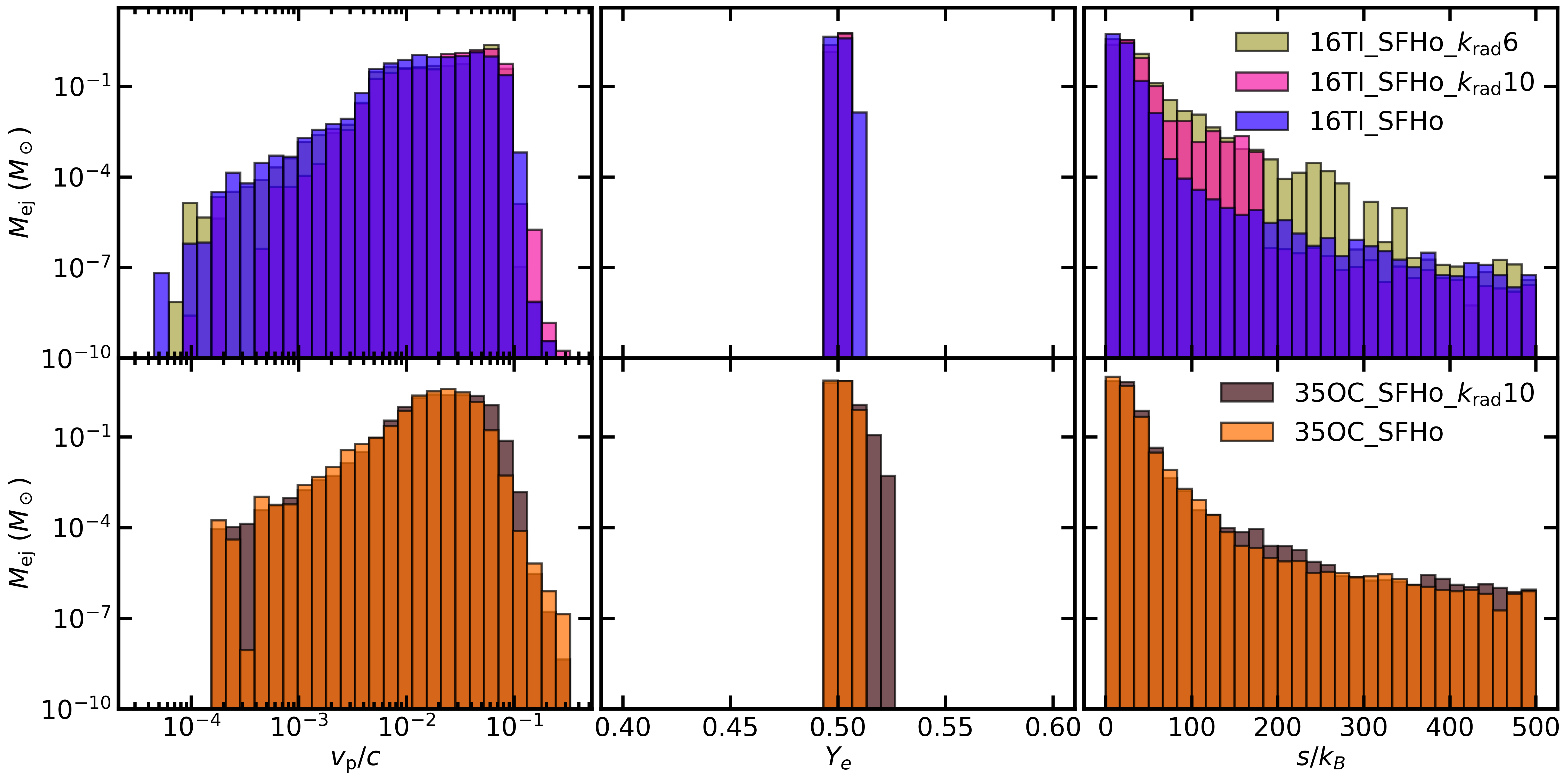}
    \caption{Unbound mass histograms at the end of the simulation, for models with modified rotation profile and their corresponding 
    unmodified counterparts, as labelled (c.f. Table~\ref{tab:progenitors}). Only matter with positive Bernoulli parameter and $v_r>0$ 
    is considered. Histograms are binned by poloidal velocity $v_{\rm p}=(v_r^2+v_\theta^2)^{1/2}$ (left), electron fraction $Y_e$ (center), and entropy per baryon 
    $s/k_{\rm B}$ (right).}
    \label{fig:model_histograms}
\end{figure*}

\begin{table*}
\caption{Bulk outflow properties, obtained by integrating unbound material at the end of the simulation, in models with modified angular 
momentum profile and their corresponding base models (c.f. Table~\ref{tab:progenitors}). Columns from left to right show model name, 
ejecta mass, ejecta kinetic energy at the end of the simulation, asymptotic ejecta kinetic energy, mass-weighted average expansion 
velocity at infinity (Eq.~\ref{eq:v_infinity}), and minimum electron fraction of outflowing material (c.f. Fig.~\ref{fig:minYe}). 
\label{tab:supernova}
} 
\begin{ruledtabular}
\begin{tabular}{lccccc}
Model & $M_{\rm ej}$ (M$_\odot$) & $K_{\rm ej}$ ($10^{51}\,$erg) & $K_\infty$ ($10^{51}\,$erg) & $\langle v_\infty \rangle$ ($10^3\,$km\,s$^{-1}$) & 
  $Y_{\rm e,min}(t_{\rm max})$ \\ 
\noalign{\smallskip} \hline
\texttt{16TI\_SFHo}                   & 8.19 & 9.07 & 9.2  & 8.7  & 0.498 \\ 
\texttt{16TI\_SFHo\_$k_{\rm rad}$10}  & 7.93 & 14.1 & 14.4 & 12.0 & 0.498 \\ 
\texttt{16TI\_SFHo\_$k_{\rm rad}$6}   & 7.01 & 13.8 & 14.0 & 12.3 & 0.495 \\ 
\noalign{\smallskip}
\texttt{35OC\_SFHo}                   & 15.1 & 9.45 & 10.6 & 7.7 & 0.497 \\ 
\texttt{35OC\_SFHo\_$k_{\rm rad}$10}  & 14.3 & 13.0 & 14.1 & 8.7 & 0.500 \\ 
\end{tabular}
\end{ruledtabular}
\end{table*}

The three new models with modified angular momentum profile (\texttt{16TI\_SFHo\_$k_{\rm rad}$6}, \texttt{16TI\_SFHo\_$k_{\rm rad}$10}, 
and \texttt{35OC\_SFHo\_$k_{\rm rad}$10}) evolve in a qualitatively similar way to their corresponding unmodified baseline model runs 
(\texttt{16TI\_SFHo} and \texttt{35OC\_SFHo}, respectively). Figure \ref{fig:evolution} illustrates differences in the evolution between 
these models.  

The BH formation time in the \texttt{GR1D} evolution is unchanged, as the angular momentum profile remains unchanged in the core of the 
star. Disk formation in each of the models with modified angular momentum occurs $\sim 0.5 - 2\,$s later than in their corresponding base 
model, owing to the increased accretion rate during this initial evolution phase (Figure \ref{fig:evolution}). The BH masses at the time 
of disk formation also exceed those in the base models by $\sim 0.5\Msun$ (Table~\ref{tab:progenitors}).

Like in most models from Paper I, and particularly in the baseline cases, the accretion disk begins in an NDAF stage and
transitions to an ADAF phase at a time $\sim1-5\,$s after disk formation (with the exception of 
model \texttt{16TI\_SFHo\_$\alpha$001}, for which the NDAF phase lasts $\sim12\,$s). 
For quantitative analysis, we define the transition
between NDAF and ADAF phases in terms of the ratio of the absolute value of the (post-shock integrated) net neutrino cooling $Q_\nu$ to 
viscous heating $Q_{\rm visc}$, according to: 
\begin{equation}
\label{eq:ndaf-adaf}
\begin{cases}
    |Q_{\nu}/Q_{\rm visc}| \ge 0.3 & \rm{NDAF~phase} \\
    |Q_{\nu}/Q_{\rm visc}| < 0.3   & \rm{ADAF~phase} \\
\end{cases}
\end{equation}
Figure~\ref{fig:evolution} shows that this criterion consistently captures the sudden drop \addtext{--} and onset of large amplitude
fluctuations -- in the accretion rate onto the BH, which is a characteristic of the ADAF phase.
Previous time-dependent studies of collapsar disks refer to the \emph{ignition} accretion rate \cite{chen_2007,de_2021} to mark the 
transition from NDAF to ADAF. This quantity is
defined as the minimum accretion rate for which the condition between local (specific) energy source terms satisfies 
$q_{\rm visc}-q_{\rm cool} = q_{\rm visc}/2$ somewhere in the disk, with $q_{\rm cool}$ including net neutrino cooling and nuclear dissociation. This condition is
equivalent to $q_{\rm cool}/q_{\rm visc} = 0.5$, which is compatible with (but not identical to) our criterion.

\begin{figure*}
    \centering
    \includegraphics[width=0.7\textwidth]{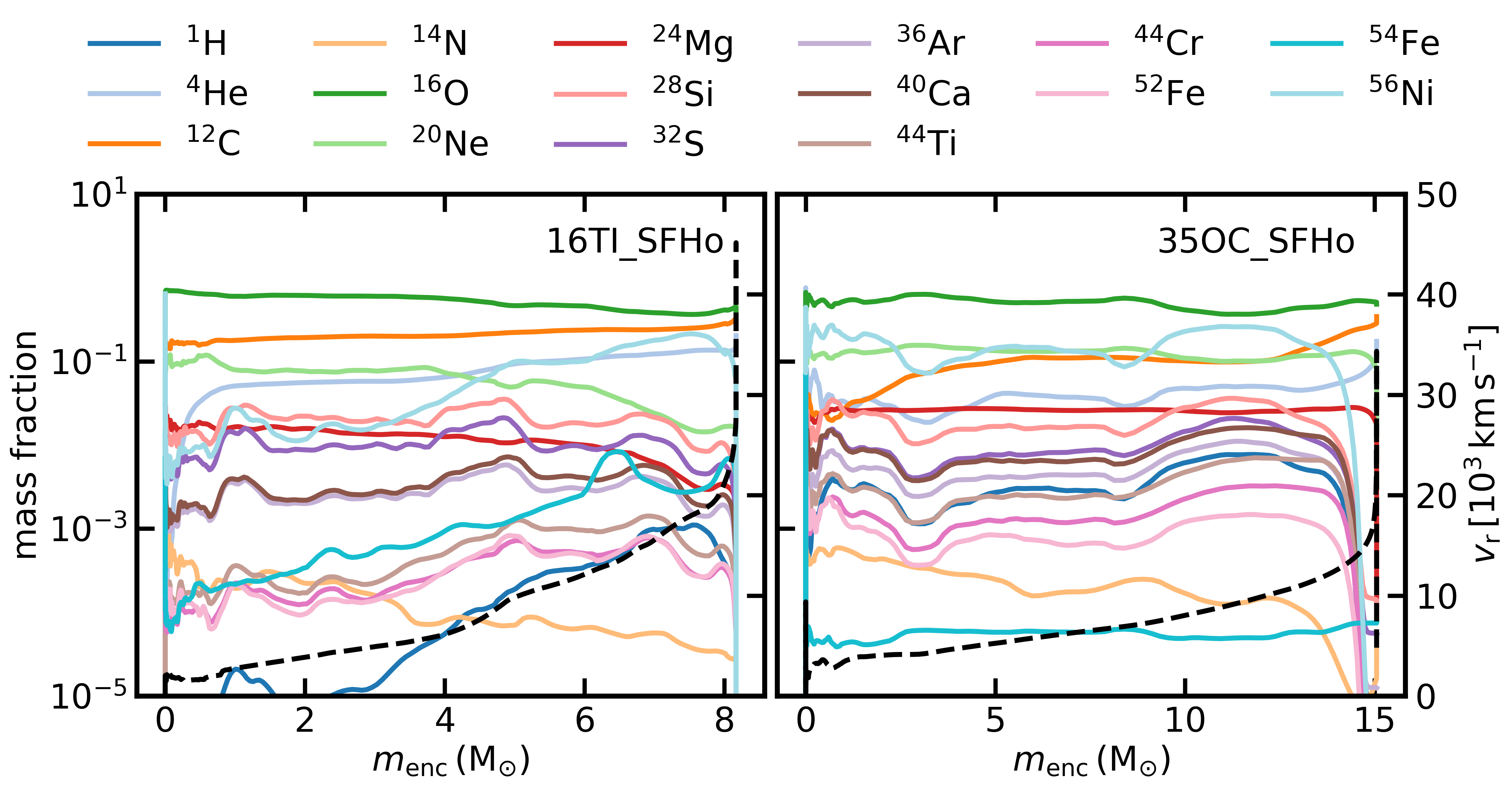}
    \caption{Angle-averaged mass fractions in unbound material of selected isotopes from the 19-isotope network, as a function of angle-averaged 
    enclosed mass, for models \texttt{16TI\_SFHo} (left) and \texttt{35OC\_SFHo} (right). The black dashed line shows the angle-averaged 
    radial velocity of the ejecta as a function of enclosed mass (right y-axis). The mass fraction of $^1$H represents a combination of 
    p$^+$ and $^1$H.}
    \label{fig:lagrangian_composition}
\end{figure*}

Like in the base models, we see that after the transition to the ADAF phase, when rapid expansion of the shock begins, the 
shock front geometry freezes following the oscillatory instability to axisymmetric perturbations in the NDAF phase. This leads 
to the shock expanding to the stellar surface, often towards one of the poles ($+\hat{z}$ or $-\hat{z}$).  We see very similar 
shock breakout times in models \texttt{16TI\_SFHo} and \texttt{16TI\_SFHo\_$k_{\rm rad}$10}, while model \texttt{16TI\_SFHo\_$k_{\rm rad}$6} 
follows $\sim 30\,$s later owing to the more significant change to the angular momentum profile. Similarly, models \texttt{35OC\_SFHo} and 
\texttt{35OC\_SFHo\_$k_{\rm rad}$10} evolve to shock breakout on very similar timescales ($\sim 5\,$s difference). 

The minimum electron fraction in the computational domain occurs within the disk. This quantity drops to its lowest value shortly 
after disk formation, remaining low until the transition from the NDAF phase to the ADAF phase and the onset of rapid expansion of 
the shock wave (Figure \ref{fig:minYe}). The duration of the phase with significant neutronization is $\sim 2\,$s for models 
\texttt{16TI\_SFHo} and \texttt{16TI\_SFHo\_$k_{\rm rad}$10}, but extends slightly longer in model \texttt{16TI\_SFHo\_$k_{\rm rad}$6} 
($\sim 3$\,s), which has a $j$ profile that results in a smaller circularization radius (Eq.~\ref{eq:j_modified}). The duration of 
neutronization in models \texttt{35OC\_SFHo} and \texttt{35OC\_SFHo\_$k_{\rm rad}$10} are also consistent with one another ($\sim 4.5\,$s), 
with the onset of neutronization delayed from the base \texttt{35OC\_SFHo} model by $\sim 0.5\,$s (Figure \ref{fig:evolution}). Overall, 
modification of the angular momentum profile did not produce a significant lengthening of the neutronization phase relative to the base 
models, except for \texttt{16TI\_SFHo\_$k_{\rm rad}$6}. Even in that case, however, this longer neutronization resulted in only a small 
effect on the electron fraction of the ejecta. 

Figure \ref{fig:model_histograms} shows the velocity, electron fraction, and entropy distributions of unbound material in the computational domain (positive Bernoulli parameter) at the end of the 
simulation, for models with modified angular momentum profile, as well as for their corresponding base models. Histograms from models with modified rotation show the same overall features as their fiducial 
counterparts: broad velocity distribution peaking at $\sim 6\E{-2}\,$c, narrow electron fraction distribution, with most ejecta 
having $Y_e\sim 0.5$, and an entropy distribution that peaks at $\sim 10$\,k$_{\rm B}$ per baryon and has a long tail reaching several hundred $k_B$ or more.

\begin{figure*}
    \centering
    \includegraphics[width=0.7\textwidth]{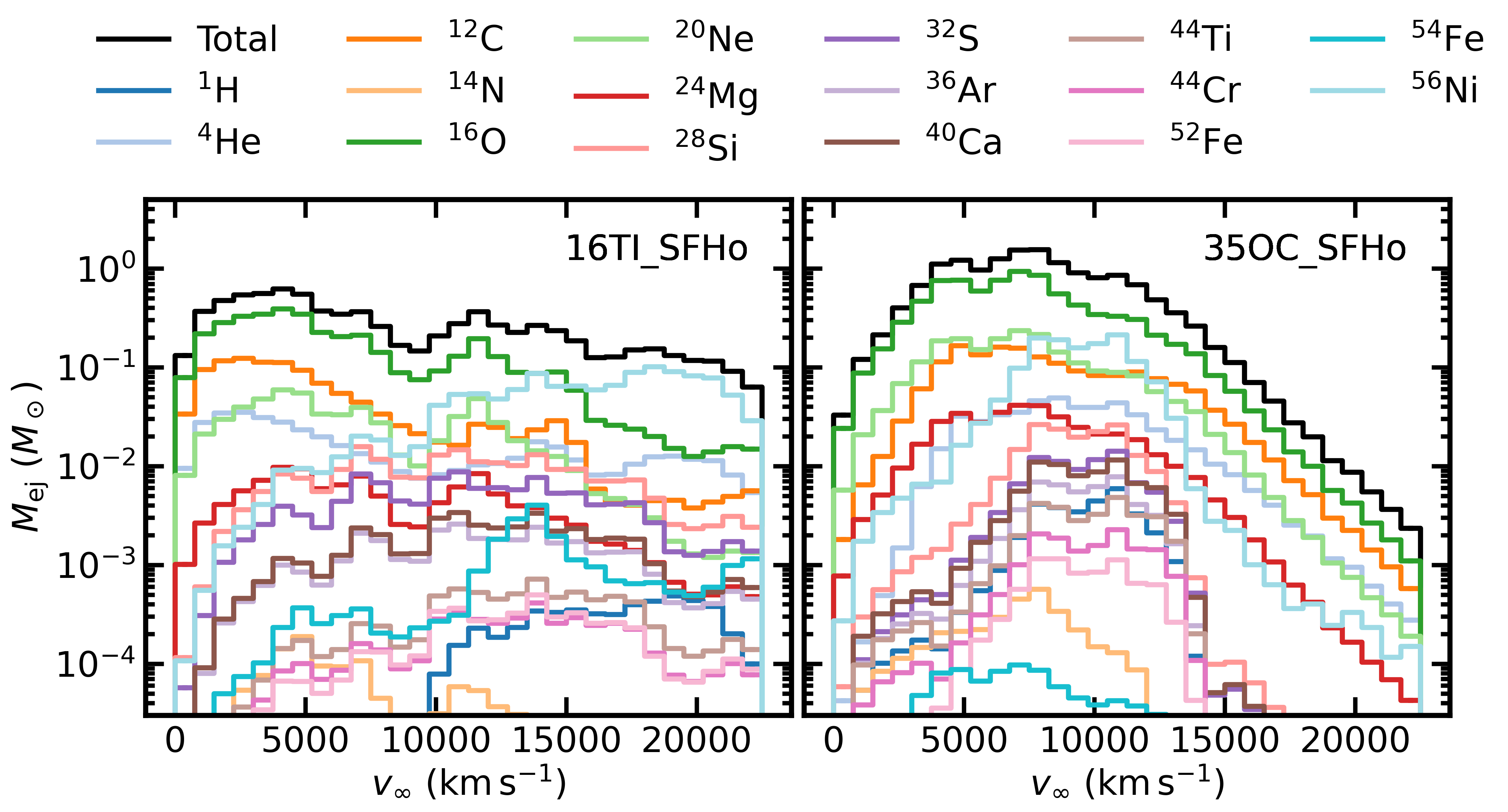}
    \caption{Unbound mass histograms of selected isotopes from the 19-isotope network, binned by asymptotic velocity (Equation~\ref{eq:v_infinity}) 
    for models \texttt{16TI\_SFHo} (left) and \texttt{35OC\_SFHo} (right). The $^1$H ejecta mass represents a combination of the p$^+$ and $^1$H ejecta masses.}
    \label{fig:vinf_hist}
\end{figure*}

\begin{figure*}
    \centering
    \includegraphics[width=\textwidth]{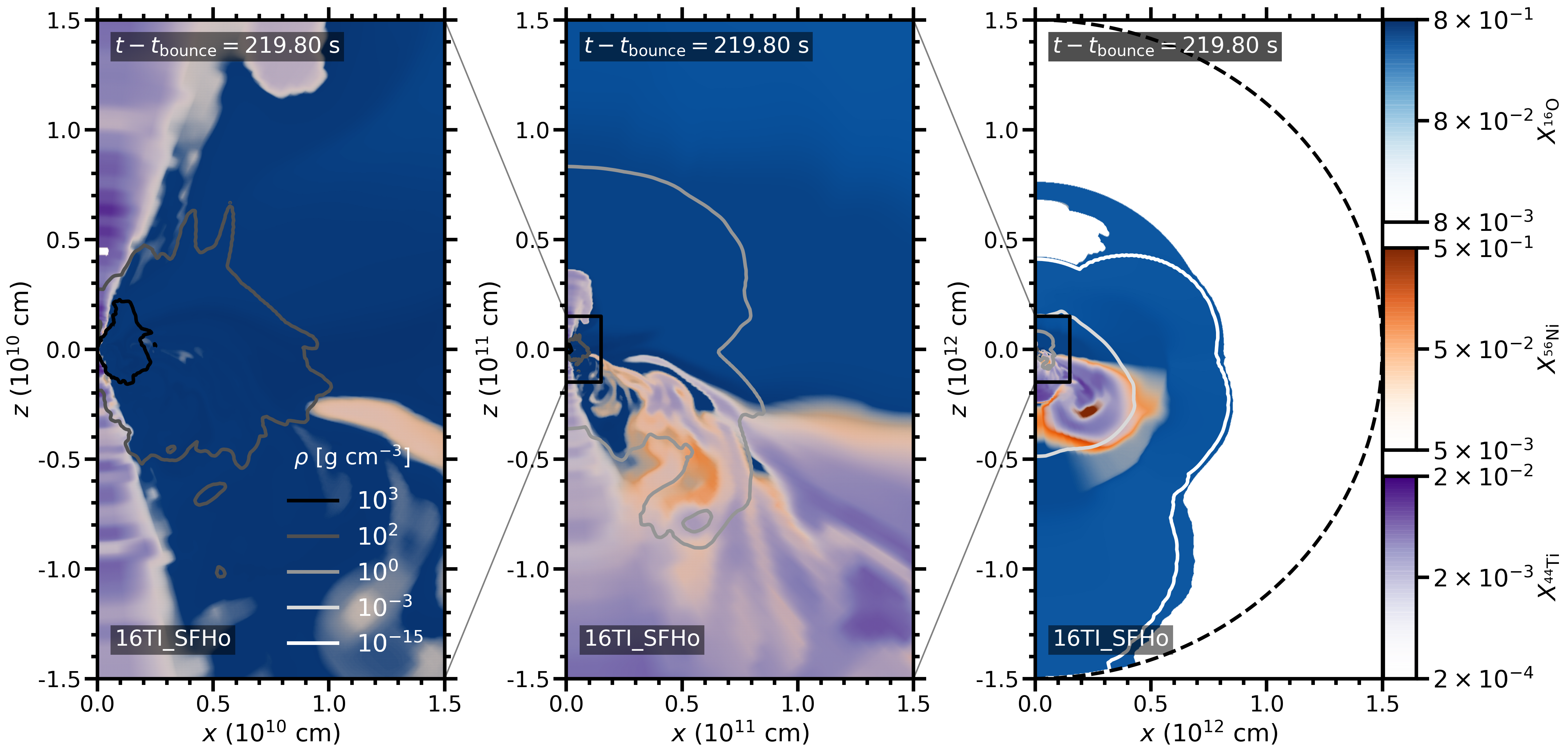}
    \caption{Mass fractions of $^{16}$O, $^{44}$Ti, and $^{56}$Ni at the last simulated time in model \texttt{16TI\_SFHo}, with density contours overlayed.
    Data is not shown when the atmospheric mass fraction $X_{\rm atm} > 0.01$ (as in, e.g., the white patch in the northern
    hemisphere of the rightmost panel). The outer domain edge is marked with a dashed line in the right panel, with the middle and left panels 
    representing subsequent zoom-ins towards the inner radial edge of the domain. 
    }
    \label{fig:comp_cmap}
\end{figure*}

As shown in Table~\ref{tab:supernova}, total mass ejection varies by $\sim2-20$\% between models with modified angular momentum and their 
fiducial models, with models with modified rotation profile always ejecting less unbound mass than the base models. This can be attributed 
in part to a loss of rotational support in the outermost layers of the progenitor star when modifying $j(M)$, which makes these layers 
more gravitationally bound and harder to eject. 

In contrast to ejected mass, the kinetic energies of the ejecta from models with modified rotation are always larger than their unmodified 
counterparts. We can attribute this to the higher extraction of accretion energy per unit mass by material that sinks deeper into the 
gravitational potential of the BH when keeping $r_{\rm circ}$ constant instead of increasing with enclosed mass. This higher accretion 
efficiency in models with constant $r_{\rm circ}$ can be seen by the higher viscous heating relative to their baseline models in 
Figure~\ref{fig:evolution}. 

As in Paper I, we define the asymptotic ejecta velocity using
\begin{equation}
\label{eq:v_infinity}
\frac{1}{2}v_\infty^2 = \max(Be,0),
\end{equation}
with $Be$ the Bernoulli parameter.
The mass-averaged asymptotic velocities $\langle v_\infty \rangle$ from models with modified angular momentum profiles are higher than
their base counterparts by $\sim 10-40$\%, leading to asymptotic kinetic energies higher by $\sim 30-60$\%. These increased 
asymptotic velocities of models with modified angular momentum still fall short of the velocities inferred from spectral line widths 
of Type Ic-BL SNe by a factor of $\sim 1.6-2$. Note that the disk outflow velocities may be further increased, relative to
what we find here, through energy deposition by a relativistic jet, or through magnetic driving by large-scale
magnetic fields. On the other hand, our $^{56}$Ni masses are sufficient to power a typical type Ic-BL SN light 
curve. A detailed comparison with observational data requires proper radiative transfer calculations on ejecta from our models, 
which will be left for future work.

\subsection{Analysis of the 19-Isotope Network Output}

Inclusion of the 19-isotope network in the \texttt{FLASH} models (Sec.~\ref{sec:computational_approach}), supplemented by the NSE solver 
for $T>T_{\rm NSE}$, provides mass fractions in every computational zone at each time step of the simulation, allowing for a detailed 
spatial and temporal nucleosynthesis analysis. Tables \ref{tab:isotope_mass_1} and \ref{tab:isotope_mass_2} show the mass ejected 
in various isotopes, along with their mass-weighted asymptotic velocities $\langle v_\infty \rangle$. All models produce
several solar masses of ${}^{16}$O ($3-8M_\odot$), the most abundant element in the ejecta, followed by ${}^{12}$C 
($1-2M_\odot$), ${}^{20}$Ne ($0.4-2M_\odot$), and ${}^{56}$Ni ($0.3-2M_\odot$). Other isotopes in the network have yields $<1M_\odot$. 

The ejected masses of $^{12}$C, $^{16}$O, and $^{20}$Ne correspond to $\sim 50$\% of their initial masses in the presupernova star, which proportionally 
tracks the ratio of total ejecta mass to initial presupernova mass, which is also $\sim 50\%$. Thus, we expect the majority of these elements to come from 
stellar layers ejected without much reprocessing. In contrast, the mass in $^{4}$He is $\sim 140$\% of the helium mass in the initial progenitor, 
thus collapsars are net producers of this element. 

Similarly, while initial $^{56}$Ni masses are negligible in the presupernova stars ($\sim 10^{-12}\,M_\odot$), our models produce $\sim 0.3-2.0\,\Msun$. 
Previous steady-state work on ${}^{56}$Ni production in collapsar disk winds finds a dependence on the entropy and expansion
time of the outflow \cite{surman_2011}. The $^{56}$Ni masses we see are consistent with the lower expansion velocity models of \cite{surman_2011}, 
given our accretion rates ($\sim 0.2-0.5\,\Msun\,$s$^{-1}$) and the mass weighted entropy per baryon of our outflows (Table~\ref{tab:sampling}).
For models with the same progenitor star, we find some correlation between the ejected
$^{56}$Ni mass (Table~\ref{tab:isotope_mass_2}) and the average asymptotic velocity of the outflow (Table~\ref{tab:sampling}), 
but not with the average entropy.

Figure \ref{fig:lagrangian_composition} shows the angle-averaged mass fraction of selected isotopes within unbound material, as a function of 
enclosed mass, for models \texttt{16TI\_SFHo} and \texttt{35OC\_SFHo}. The ejecta is well-mixed in Lagrangian mass coordinate, with most isotopes 
having a gradual stratification in radius. It is notable that $^{56}$Ni is present in most of the ejecta, with even an enhanced mass fraction 
in the outermost mass shells in model \texttt{16TI\_SFHo}. The angle-averaged velocity profile is consistent with Fig.~\ref{fig:model_histograms}, 
with most of the ejecta mass having velocity $\sim 0.03\,{\rm c} \simeq 10,000$\,km\,s$^{-1}$. Previous work investigating mixing in collapsar disk 
winds finds higher mixing efficiency with longer wind duration \cite{barnes_2023}, our models support this conclusion.

Figure \ref{fig:vinf_hist} shows histograms of the mass in individual isotopes as a function of asymptotic velocity $v_{\infty}$, for models 
\texttt{16TI\_SFHo} and model \texttt{35OC\_SFHo}. Again, we see that the ejecta is well mixed, with light (e.g., ${}^{12}$C) and heavy 
(e.g., ${}^{56}$Ni) elements spanning the entire range of expansion velocities. There is a slight preference for heavier isotopes being ejected 
with faster velocities and lighter isotopes with slower velocities (Tables \ref{tab:isotope_mass_1} and \ref{tab:isotope_mass_2}). The width of the 
asymptotic velocity distribution is broader for the \texttt{16TI\_SFHo} model, with more mass ejected 
at higher velocities, while \texttt{35OC\_SFHo} tends to have a narrower distribution with 
less material ejected at the highest speeds. The mass weighted averages $\langle v_\infty \rangle$ for each model are $\sim 2-3$ times smaller than 
those implied by the width of spectral features seen in type Ic-bl spectra \cite{modjaz_2016}, however we see that some fraction of the ejected 
isotopes reach asymptotic velocities in excess of $20,000\,$km$\,$s$^{-1}$. 

Figure \ref{fig:comp_cmap} shows the spatial distribution of $^{16}$O, $^{44}$Ti, and $^{56}$Ni in the 
ejecta from model \texttt{16TI\_SFHo} at the end of the simulation. Both $^{44}$Ti and $^{56}$Ni 
are largely co-spatial, produced in the expanding turbulent post-shock region, with $^{44}$Ti found somewhat 
deeper into the ejecta than $^{56}$Ni. $^{16}$O makes up the largest fraction of the ejecta, and is 
distributed mostly uniformly. The fact that $^{44}$Ti has a long radioactive decay timescale
and is spatially stratified offers favorable prospects to observationally probe a galactic collapsar remnant with
hard X-rays, as done with the CCSN remnant Cas A (e.g., \cite{grefenstette_2014}), in addition to more traditional
gamma-ray spectroscopy diagnostics (e.g., \cite{prantzos_2004,thielemann_2018}).

\begin{figure}[ht!]
    \centering
    \includegraphics[width=0.9\columnwidth]{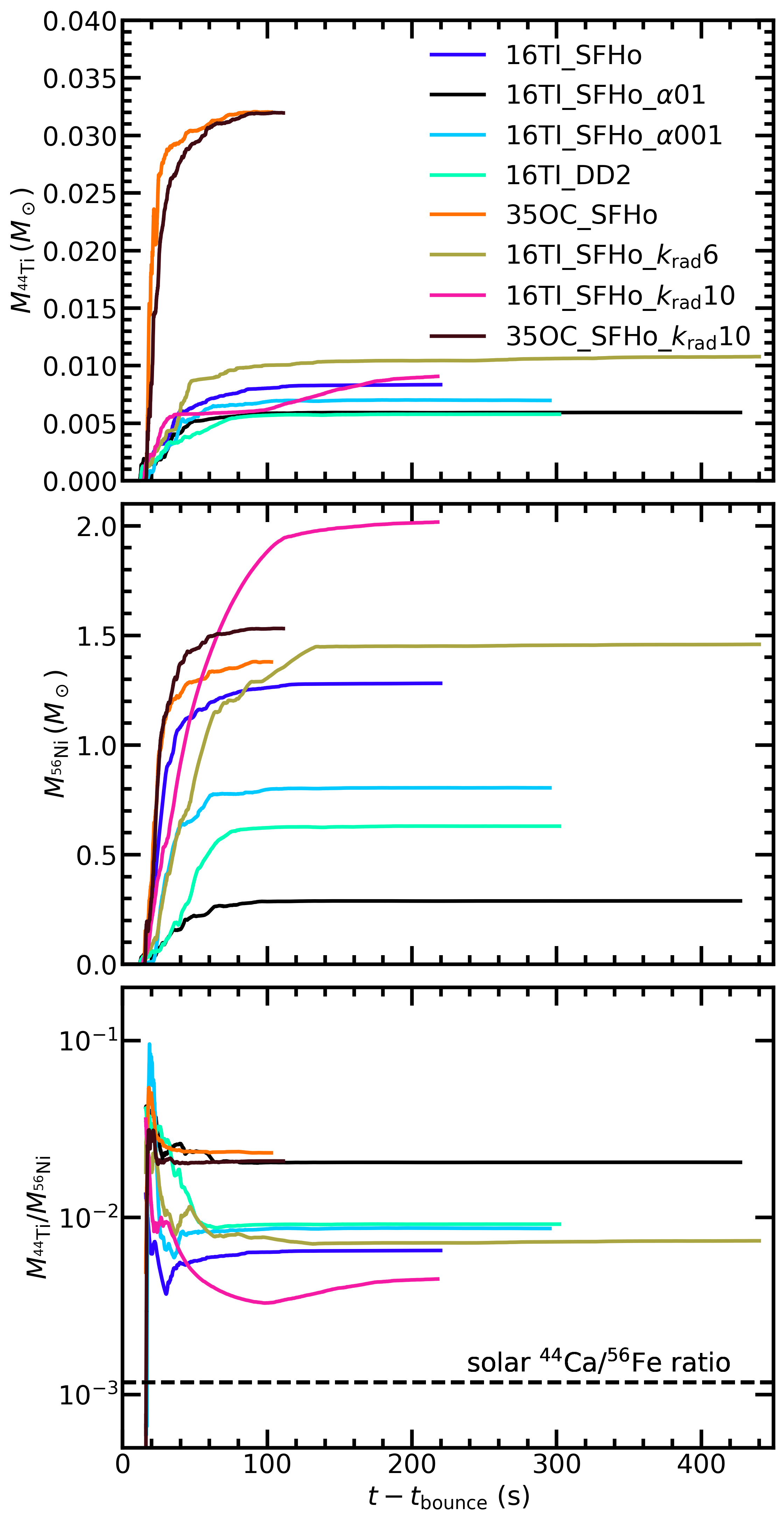}
    \caption{Evolution of the unbound mass in $^{44}$Ti (top) and $^{56}$Ni (middle), as well as their ratio $M_{^{44}\rm{Ti}}/M_{^{56}\rm{Ni}}$ (bottom), for all of our models.
             Also shown is the solar mass ratio of $^{44}$Ca$/^{56}$Fe (dashed line), corresponding to the endpoints of the $^{44}$Ti and $^{56}$Ni decay chains,  
             respectively, for reference. The bottom panel shows data for $t-t_{\rm bounce}>16\,$s, to improve visibility.}
    \label{fig:ti44_ni56}
\end{figure}

Figure \ref{fig:ti44_ni56} shows the evolution of the unbound $^{44}$Ti and $^{56}$Ni masses, as well as their ratio, in each model. The $^{44}$Ti masses 
are clustered by progenitor, with models based on \texttt{35OC} producing $\sim 3\E{-2}\,\Msun$ within $\sim 80\,$s 
post bounce, while models based on \texttt{16TI} are clustered around $5\E{-3}-10^{-2}\,\Msun$, with a 
non-uniform production timescale. The 
production of $^{56}$Ni is much more sensitive to model variations. 
In all of our models we see a super-solar $^{44}$Ti/$^{56}$Ni 
mass ratio, spanning the range $4\times 10^{-3} - 2.6\times 10^{-2}$, which is 4 to 26 times higher
than the solar mass ratio\footnote{We obtain the solar mass fraction of $^{44}$Ca/$^{56}$Fe 
using solar elemental abundances from \cite{scott_2015a,scott_2015b} and
isotopic fractions from \cite{meija_2016}.} of $^{44}$Ca/$^{56}$Fe ($\simeq 1.1\times 10^{-3}$). This is an expected
outcome of high explosion energies, which result in nuclear burning at lower densities than in normal CCSNe, enhancing
$\alpha$-rich freezout products \cite{nomoto_2017}.
In absolute terms, our models also overproduce $^{44}$Ti 
relative to typical values obtained in CCSNe models (e.g., \cite{sieverding_2023,wang_2024b}). 

\begin{figure*}
    \centering
    \includegraphics[width=\textwidth]{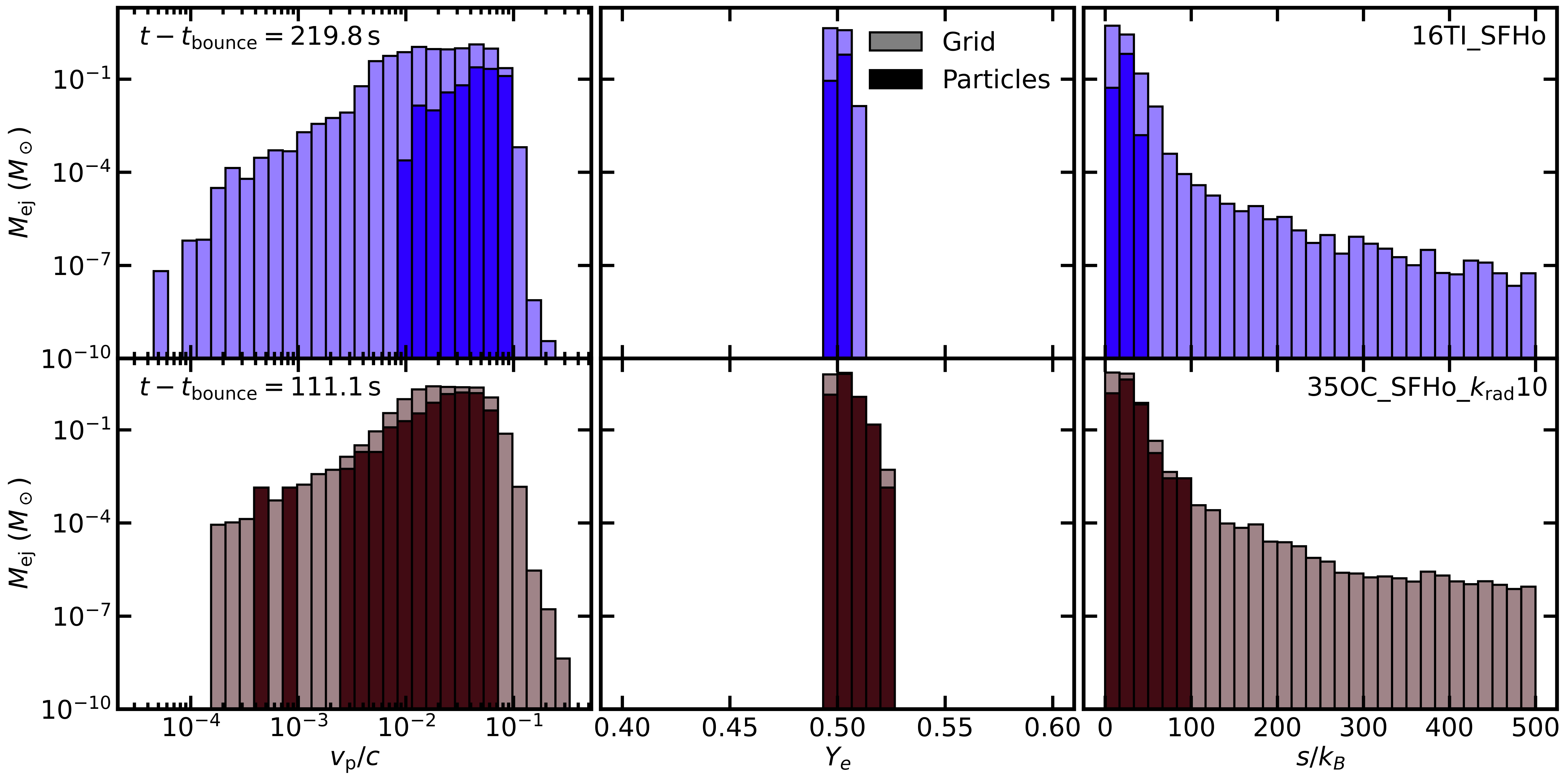}
    \caption{Unbound mass histograms binned by poloidal velocity (left), electron fraction (middle), and entropy per baryon (right), 
    for selected models. Light colors show results calculated by integrating unbound matter over the grid at the end of the simulation, 
    and dark colors show values obtained from unbound tracer particles which reach $T_{\rm max}>1\,$GK, remaining in the domain at the end of the simulation, illustrating 
    the level of sampling of the ejecta by the particles.}
    \label{fig:particle_histogram}
\end{figure*}

\begin{figure}
    \centering
    \includegraphics[width=\columnwidth]{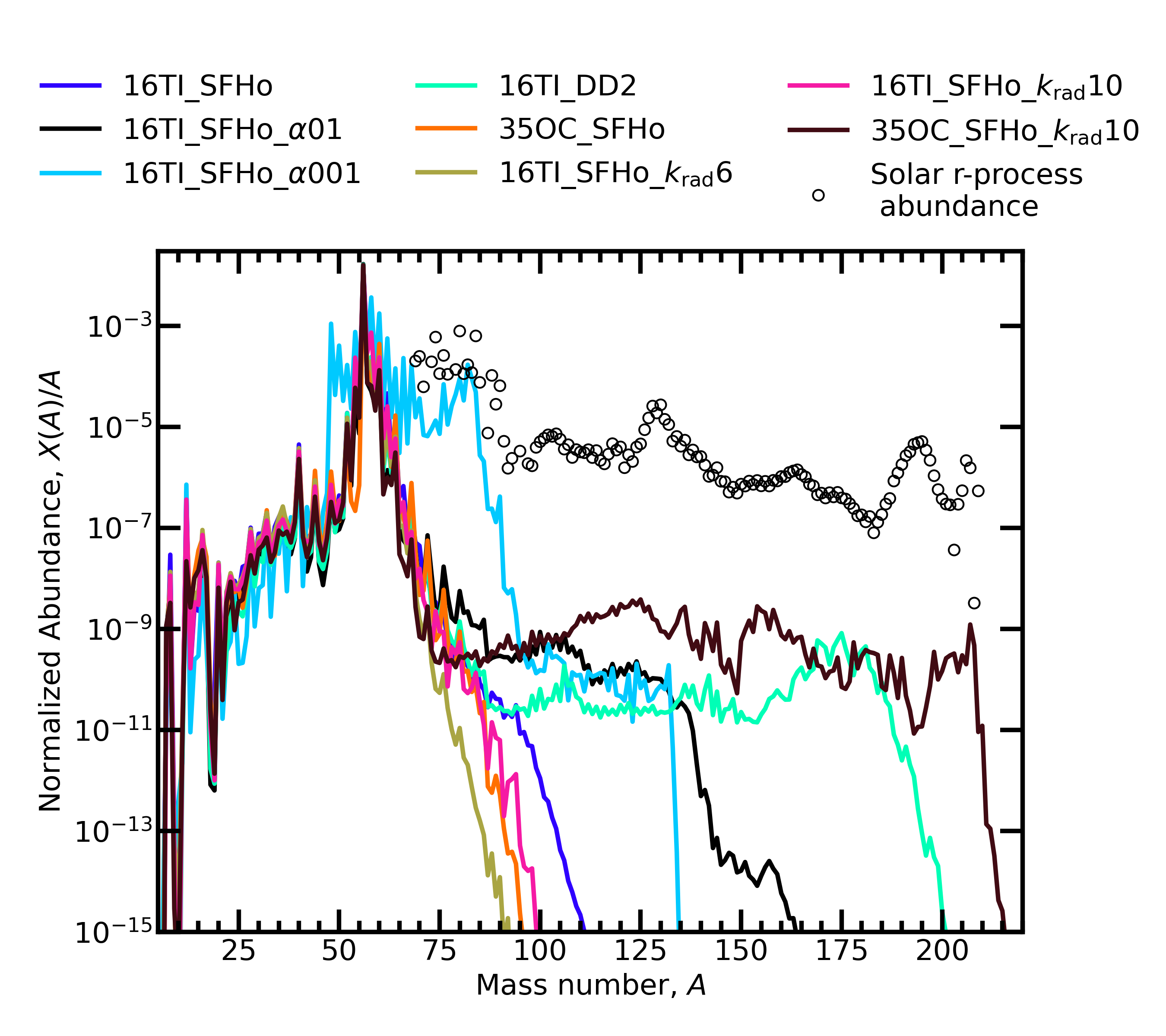}
    \caption{Isotopic abundances at $30\,$years from tracer particles evolved with \texttt{SkyNet}, for all models, as labelled.
    Abundances are obtained by dividing the isotopic mass fraction $X(A)$ by the mass number $A$, with the mass fractions adding up to unity.
    Open circles show the solar $r$-process abundance distribution from \cite{goriely_1999}, scaled to the first $r$-process peak ($A=82$) 
    from model \texttt{16TI\_SFHo\_$\alpha$001}, for reference.}
    \label{fig:skynet}
\end{figure}

\subsection{Nucleosynthesis in Post-Processing}

The tracer particles swept up by the shock wave that remain in the outflow in the simulations presented here and in Paper I, sample the
inner neutrino-reprocessed, turbulent region of the post shock region (Fig.~\ref{fig:particle_progression}, right panel). Figure~\ref{fig:particle_histogram} compares 
the unbound ejecta distribution obtained by volume-integrating across the grid, and by adding up the mass in tracer particles in selected models, 
illustrating the sampling of the total ejecta by particles. Sampling is incomplete, with particles accounting for a fraction $\sim 1-45$\% of the 
total unbound ejecta mass, depending on the model. Tracer particles tend to sample the peak of the ejecta distribution in velocity, electron 
fraction, and entropy, resulting in mass-weighted average asymptotic velocities similar or larger than the value calculated from 
the grid, in all but the \texttt{16TI\_DD2} model (Table \ref{tab:sampling}). The average electron fraction obtained from the particles and from the grid 
are very similar, owing to the narrow electron fraction distribution, and the average entropies sampled by the particles are higher in all models 
relative to the values from the grid.

\begin{sidewaystable}
\caption{Ejecta masses and mass-weighted average asymptotic velocities of selected isotopes from the 19-isotope nuclear network, for all models. 
Ejecta masses are given in units of $\Msun$, and asymptotic velocities in units of km$\,$s$^{-1}$. $^{1}$H ejecta mass and asymptotic velocity 
include a combination of p$^{+}$ and $^1$H from the nuclear network.}
\begin{ruledtabular}
\begin{tabular}{lcccccccccccccccc}
Model & \multicolumn{2}{c}{$^1$H} & \multicolumn{2}{c}{$^4$He} & \multicolumn{2}{c}{$^{12}$C} & \multicolumn{2}{c}{$^{14}$N} & \multicolumn{2}{c}{$^{16}$O} & \multicolumn{2}{c}{$^{20}$Ne} & 
\multicolumn{2}{c}{$^{24}$Mg} & \multicolumn{2}{c}{$^{28}$Si}  \\

& \multicolumn{1}{c}{$M_{\rm ej}$} & \multicolumn{1}{c}{$\langle v_\infty \rangle$} & \multicolumn{1}{c}{$M_{\rm ej}$} & \multicolumn{1}{c}{$\langle v_\infty \rangle$} & \multicolumn{1}{c}{$M_{\rm ej}$} & \multicolumn{1}{c}{$\langle v_\infty \rangle$} & \multicolumn{1}{c}{$M_{\rm ej}$} & \multicolumn{1}{c}{$\langle v_\infty \rangle$} & \multicolumn{1}{c}{$M_{\rm ej}$} & \multicolumn{1}{c}{$\langle v_\infty \rangle$}  & \multicolumn{1}{c}{$M_{\rm ej}$} & \multicolumn{1}{c}{$\langle v_\infty \rangle$}  & \multicolumn{1}{c}{$M_{\rm ej}$} & \multicolumn{1}{c}{$\langle v_\infty \rangle$}  & \multicolumn{1}{c}{$M_{\rm ej}$} & \multicolumn{1}{c}{$\langle v_\infty \rangle$}  \\ 

& \multicolumn{1}{c}{(M$_{\odot}$)} & \multicolumn{1}{c}{(km$\,$s$^{-1}$)} & \multicolumn{1}{c}{(M$_{\odot}$)} & \multicolumn{1}{c}{(km$\,$s$^{-1}$)} & \multicolumn{1}{c}{(M$_{\odot}$)} & \multicolumn{1}{c}{(km$\,$s$^{-1}$)} & \multicolumn{1}{c}{(M$_{\odot}$)} & \multicolumn{1}{c}{(km$\,$s$^{-1}$)} & \multicolumn{1}{c}{(M$_{\odot}$)} & \multicolumn{1}{c}{(km$\,$s$^{-1}$)}  & \multicolumn{1}{c}{(M$_{\odot}$)} & \multicolumn{1}{c}{(km$\,$s$^{-1}$)}  & \multicolumn{1}{c}{(M$_{\odot}$)} & \multicolumn{1}{c}{(km$\,$s$^{-1}$)}  & \multicolumn{1}{c}{(M$_{\odot}$)} & \multicolumn{1}{c}{(km$\,$s$^{-1}$)}  \\ \hline

\texttt{16TI\_SFHo}                    & $5.3e-3$ & $1.7e4$ & $4.6e-1$ & $9.4e3$ & $1.2$ & $6.2e3$ & $1.3e-3$ & $7.4e3$ & $4.0$ & $6.9e3$ & $6.3e-1$ & $7.5e3$ & $1.2e-1$ & $8.5e3$ & $2.2e-1$ & $1.1e4$ \\ 
\texttt{16TI\_SFHo\_$\alpha$01}        & $5.3e-4$ & $5.4e3$ & $3.9e-1$ & $5.8e3$ & $1.4$ & $5.3e3$ & $1.5e-3$ & $4.6e3$ & $5.5$ & $4.7e3$ & $1.1$    & $4.3e3$ & $1.7e-1$ & $4.3e3$ & $7.4e-2$ & $4.9e3$ \\ 
\texttt{16TI\_SFHo\_$\alpha$001}       & $1.5e-3$ & $1.1e4$ & $4.3e-1$ & $6.2e3$ & $1.3$ & $5.0e3$ & $1.1e-3$ & $5.9e3$ & $4.1$ & $5.2e3$ & $6.9e-1$ & $5.7e3$ & $1.1e-1$ & $6.0e3$ & $1.3e-1$ & $8.0e3$ \\ 
\texttt{16TI\_DD2}                     & $2.0e-3$ & $6.2e3$ & $4.3e-1$ & $6.7e3$ & $1.5$ & $6.1e3$ & $1.2e-3$ & $4.7e3$ & $5.3$ & $5.3e3$ & $9.6e-1$ & $4.7e3$ & $1.4e-1$ & $4.6e3$ & $1.3e-1$ & $6.6e3$ \\ 
\texttt{35OC\_SFHo}                    & $3.3e-2$ & $9.6e3$ & $5.0e-1$ & $9.3e3$ & $1.7$ & $8.4e3$ & $3.4e-3$ & $7.0e3$ & $8.4$ & $7.3e3$ & $2.1$    & $7.3e3$ & $4.0e-1$ & $7.8e3$ & $1.8e-1$ & $9.2e3$ \\ 
\texttt{16TI\_SFHo\_$k\_{\rm rad}$6}   & $1.5e-2$ & $1.8e4$ & $6.0e-1$ & $1.3e4$ & $1.1$ & $8.8e3$ & $4.8e-4$ & $1.4e4$ & $3.1$ & $1.0e4$ & $4.3e-1$ & $1.2e4$ & $8.9e-2$ & $1.3e4$ & $1.1e-1$ & $1.6e4$ \\ 
\texttt{16TI\_SFHo\_$k\_{\rm rad}$10}  & $7.5e-3$ & $1.5e4$ & $6.1e-1$ & $1.3e4$ & $1.1$ & $9.6e3$ & $4.7e-4$ & $1.1e4$ & $3.3$ & $9.7e3$ & $5.3e-1$ & $1.0e4$ & $1.0e-1$ & $1.1e4$ & $1.1e-1$ & $1.4e4$ \\ 
\texttt{35OC\_SFHo\_$k\_{\rm rad}$10}  & $4.5e-2$ & $1.3e4$ & $5.3e-1$ & $1.1e4$ & $1.6$ & $8.4e3$ & $3.2e-3$ & $8.6e3$ & $7.7$ & $7.9e3$ & $1.9$    & $7.8e3$ & $4.1e-1$ & $8.7e3$ & $2.5e-1$ & $~1.2e4$ 
\label{tab:isotope_mass_1}
\end{tabular}
\end{ruledtabular}
\end{sidewaystable}

\begin{sidewaystable}
\caption{Continuation of Table \ref{tab:isotope_mass_1}.}
\begin{ruledtabular}
\begin{tabular}{lcccccccccccccccc}
Model & \multicolumn{2}{c}{$^{32}$S} & \multicolumn{2}{c}{$^{36}$Ar} & \multicolumn{2}{c}{$^{40}$Ca} & \multicolumn{2}{c}{$^{44}$Ti} & \multicolumn{2}{c}{$^{48}$Cr} & \multicolumn{2}{c}{$^{52}$Fe} & \multicolumn{2}{c}{$^{54}$Fe} & \multicolumn{2}{c}{$^{56}$Ni} \\

& \multicolumn{1}{c}{$M_{\rm ej}$} & \multicolumn{1}{c}{$\langle v_\infty \rangle$} & \multicolumn{1}{c}{$M_{\rm ej}$} & \multicolumn{1}{c}{$\langle v_\infty \rangle$} & \multicolumn{1}{c}{$M_{\rm ej}$} & \multicolumn{1}{c}{$\langle v_\infty \rangle$} & \multicolumn{1}{c}{$M_{\rm ej}$} & \multicolumn{1}{c}{$\langle v_\infty \rangle$} & \multicolumn{1}{c}{$M_{\rm ej}$} & \multicolumn{1}{c}{$\langle v_\infty \rangle$}  & \multicolumn{1}{c}{$M_{\rm ej}$} & \multicolumn{1}{c}{$\langle v_\infty \rangle$}  & \multicolumn{1}{c}{$M_{\rm ej}$} & \multicolumn{1}{c}{$\langle v_\infty \rangle$}  & \multicolumn{1}{c}{$M_{\rm ej}$} & \multicolumn{1}{c}{$\langle v_\infty \rangle$}  \\

& \multicolumn{1}{c}{(M$_{\odot}$)} & \multicolumn{1}{c}{(km$\,$s$^{-1}$)} & \multicolumn{1}{c}{(M$_{\odot}$)} & \multicolumn{1}{c}{(km$\,$s$^{-1}$)} & \multicolumn{1}{c}{(M$_{\odot}$)} & \multicolumn{1}{c}{(km$\,$s$^{-1}$)} & \multicolumn{1}{c}{(M$_{\odot}$)} & \multicolumn{1}{c}{(km$\,$s$^{-1}$)} & \multicolumn{1}{c}{(M$_{\odot}$)} & \multicolumn{1}{c}{(km$\,$s$^{-1}$)}  & \multicolumn{1}{c}{(M$_{\odot}$)} & \multicolumn{1}{c}{(km$\,$s$^{-1}$)}  & \multicolumn{1}{c}{(M$_{\odot}$)} & \multicolumn{1}{c}{(km$\,$s$^{-1}$)}  & \multicolumn{1}{c}{(M$_{\odot}$)} & \multicolumn{1}{c}{(km$\,$s$^{-1}$)}  \\ \hline

\texttt{16TI\_SFHo}                    & $1.2e-1$ & $1.2e4$ & $3.5e-2$ & $1.2e4$ & $4.5e-2$ & $1.2e4$ & $8.4e-3$ & $1.3e4$ & $4.7e-3$ & $1.3e4$ & $4.9e-3$ & $1.4e4$ & $2.6e-2$ & $1.6e4$ & $1.3$ & $1.6e4$ \\ 
\texttt{16TI\_SFHo\_$\alpha$01}        & $3.2e-2$ & $4.8e3$ & $1.4e-2$ & $4.9e3$ & $2.1e-2$ & $5.0e3$ & $5.9e-3$ & $5.1e3$ & $3.2e-3$ & $5.1e3$ & $2.5e-3$ & $5.0e3$ & $2.7e-4$ & $4.4e3$ & $2.9e-1$ & $5.0e3$ \\ 
\texttt{16TI\_SFHo\_$\alpha$001}       & $7.8e-2$ & $8.3e3$ & $2.6e-2$ & $8.4e3$ & $3.4e-2$ & $8.2e3$ & $7.0e-3$ & $8.2e3$ & $3.7e-3$ & $8.3e3$ & $2.7e-3$ & $9.6e3$ & $2.0e-1$ & $8.1e3$ & $8.1e-1$ & $1.0e4$ \\ 
\texttt{16TI\_DD2}                     & $7.7e-2$ & $7.1e3$ & $2.4e-2$ & $6.8e3$ & $3.1e-2$ & $6.5e3$ & $5.8e-3$ & $5.3e3$ & $3.1e-3$ & $5.4e3$ & $3.8e-3$ & $6.9e3$ & $7.5e-4$ & $5.6e3$ & $6.3e-1$ & $7.0e3$ \\ 
\texttt{35OC\_SFHo}                    & $8.9e-2$ & $9.4e3$ & $5.1e-2$ & $9.4e3$ & $8.0e-2$ & $9.4e3$ & $3.2e-2$ & $9.6e3$ & $1.5e-2$ & $9.6e3$ & $8.0e-3$ & $9.5e3$ & $9.5e-4$ & $7.8e3$ & $1.4$ & $9.4e3$ \\ 
\texttt{16TI\_SFHo\_$k\_{\rm rad}$6}   & $5.1e-2$ & $1.7e4$ & $2.0e-2$ & $1.6e4$ & $3.4e-2$ & $1.6e4$ & $1.1e-2$ & $1.6e4$ & $6.1e-3$ & $1.6e4$ & $5.7e-3$ & $1.7e4$ & $4.5e-2$ & $1.9e4$ & $1.5$ & $1.8e4$ \\ 
\texttt{16TI\_SFHo\_$k\_{\rm rad}$10}  & $5.1e-2$ & $1.4e4$ & $2.0e-2$ & $1.4e4$ & $3.2e-2$ & $1.4e4$ & $9.1e-3$ & $1.4e4$ & $4.9e-3$ & $1.4e4$ & $5.5e-3$ & $1.4e4$ & $9.2e-3$ & $1.4e4$ & $2.0$ & $1.7e4$ \\ 
\texttt{35OC\_SFHo\_$k\_{\rm rad}$10}  & $1.1e-1$ & $1.2e4$ & $4.7e-2$ & $1.2e4$ & $8.2e-2$ & $1.2e4$ & $3.2e-2$ & $1.2e4$ & $1.8e-2$ & $1.2e4$ & $1.4e-2$ & $1.2e4$ & $8.7e-4$ & $8.0e3$ & $1.5$ & $1.3e4$ \\ 
\label{tab:isotope_mass_2}
\end{tabular}
\end{ruledtabular}
\end{sidewaystable}

\begin{table*}[]
\caption{Summary of tracer particle sampling of the ejecta. Columns from left to right show the number of unbound particles at the end of the simulation with $T_{\rm max}>1\,$GK
(all models are initialized with $10^4$ particles), total mass in unbound particles at the end of the simulation, total ejecta mass integrated from the grid, 
ratio of the mass in unbound particles to total ejecta mass (sampling percentage), mass-weighted average asymptotic velocity obtained from the particles and 
from the grid, mass-weighted average electron fraction from the particles and from the grid, and mass-weighted average entropy per baryon from the particles and grid, 
respectively.}
\begin{ruledtabular}
\begin{tabular}{lcccccccccc}
Model & $N_{\rm part,ej}$ & $M_{\rm ej,part}\,$ & $M_{\rm ej}\,$ & Sampling  & $\langle v_\infty \rangle_{\rm part}$ & $\langle v_\infty \rangle_{\rm grid}$ & $\langle Y_e \rangle_{\rm part}$ & $\langle Y_e \rangle_{\rm grid}$ & $\langle s/k_B \rangle_{\rm part}$ & $\langle s/k_B \rangle_{\rm grid}$ \\ 
& & ($\Msun$) & ($\Msun$) & (\%) & ($10^3$\,km\,s$^{-1}$) & ($10^3$\,km\,s$^{-1}$) & & & \\ \hline
\texttt{16TI\_SFHo}                  & $5846$ & $0.7$ & $8.2$  & $8.6$  & $1.6$ & $8.7$  & $0.50$ & $0.50$ & $23.8$ & $14.6$ \\
\texttt{16TI\_SFHo\_$\alpha$01}      & $6349$ & $0.8$ & $9.0$  & $8.7$  & $5.3$ & $4.8$  & $0.50$ & $0.50$ & $15.5$ & $10.7$ \\
\texttt{16TI\_SFHo\_$\alpha$001}     & $2096$ & $0.3$ & $7.9$  & $3.3$  & $7.6$ & $6.0$  & $0.49$ & $0.50$ & $17.5$ & $12.4$ \\
\texttt{16TI\_DD2}                   & $5160$ & $0.7$ & $9.2$  & $7.3$  & $4.3$ & $5.6$  & $0.50$ & $0.50$ & $15.6$ & $11.0$ \\
\texttt{35OC\_SFHo}                  & $716$  & $0.2$ & $15.1$ & $1.2$  & $9.5$ & $7.7$  & $0.51$ & $0.50$ & $25.1$ & $14.8$ \\
\texttt{16TI\_SFHo\_$k_{\rm rad}$6}  & $2881$ & $0.3$ & $7.0$  & $4.8$  & $1.2$ & $1.2$  & $0.50$ & $0.50$ & $25.4$ & $23.5$ \\
\texttt{16TI\_SFHo\_$k_{\rm rad}$10} & $4595$ & $0.5$ & $7.9$  & $6.9$  & $1.2$ & $1.2$  & $0.50$ & $0.50$ & $24.6$ & $19.9$ \\
\texttt{35OC\_SFHo\_$k_{\rm rad}$10} & $4636$ & $6.4$ & $14.3$ & $44.8$ & $8.6$ & $8.7$  & $0.50$ & $0.50$ & $22.9$ & $17.8$ 
\label{tab:sampling}
\end{tabular}
\end{ruledtabular}
\end{table*}

Figure \ref{fig:skynet} shows isotopic abundances from tracer particles evolved with \texttt{SkyNet}. As expected from the electron 
fraction distribution of the unbound ejecta being narrowly peaked around $Y_e = 0.5$ (Fig.~\ref{fig:model_histograms}), abundances 
are dominated by the iron group ($A\sim 50-60$), with steeply decreasing abundances for lighter and heavier elements, in a similar 
way as abundances from successful CCSNe (e.g., \cite{janka_2023}). The two base models \texttt{16TI\_SFHo} and \texttt{35OC\_SFHo}, 
as well as the 16TI models with modified rotation \texttt{16TI\_SFHo\_$k_{\rm rad}$6} and \texttt{16TI\_SFHo\_$k_{\rm rad}$10} do not 
produce any significant amount of elements beyond the iron group.

The low-viscosity model \texttt{16TI\_SFHo\_$\alpha$001} (Paper 1) stands out as the only one that produces significant amounts
of first $r$-process peak ejecta, with abundances decreasing sharply for heavier elements. This model produced the most neutron-rich ejecta
of our entire simulation set. The dependence of the $Y_e$ distribution on the strength of viscous angular momentum transport 
(also found by \cite{just_2022}) implies that the production of (at least) light $r$-process elements will also be sensitive to the 
character of angular momentum transport in the disk. This indicates that nucleosynthesis results from 3D GRMHD simulations that include the
relevant physics will likely differ in their exact abundance pattern relative to viscous hydrodynamic models like those presented here.

There is still some production of small amounts of heavy elements. Models \texttt{16TI\_DD2} and \texttt{16TI\_SFHo\_$\alpha$01}, produce 
elements beyond $A=130$, and model \texttt{35OC\_SFHo\_$k_{\rm rad}$10} yields a small abundance of elements up to and beyond $A=200$. 
None of the abundance patterns has an overall similarity to that of the $r$-process, although the undersampling of the ejecta 
by particles and the small amounts of elements heavier than the iron group do not allow us to make definitive statements about global 
abundance patterns from collapsar disk outflows. 

While the ratio of $r$-process peak elements to the iron peak (e.g., \cite{burbidge_1957}) 
is underproduced by a factor $10^1-10^3$ in models that produce the heaviest elements, 
the ratio of first $r$-process peak to iron is higher by a factor $100$ in the low-viscosity model \texttt{16TI\_SFHo\_$\alpha$001}. 
This suggests that small variations in the evolution of collapsars can lead to an intrinsic scatter in the abundance pattern of ejected material, 
particularly for $A \sim 80$.

\begin{figure}
    \centering
    \includegraphics[width=\columnwidth]{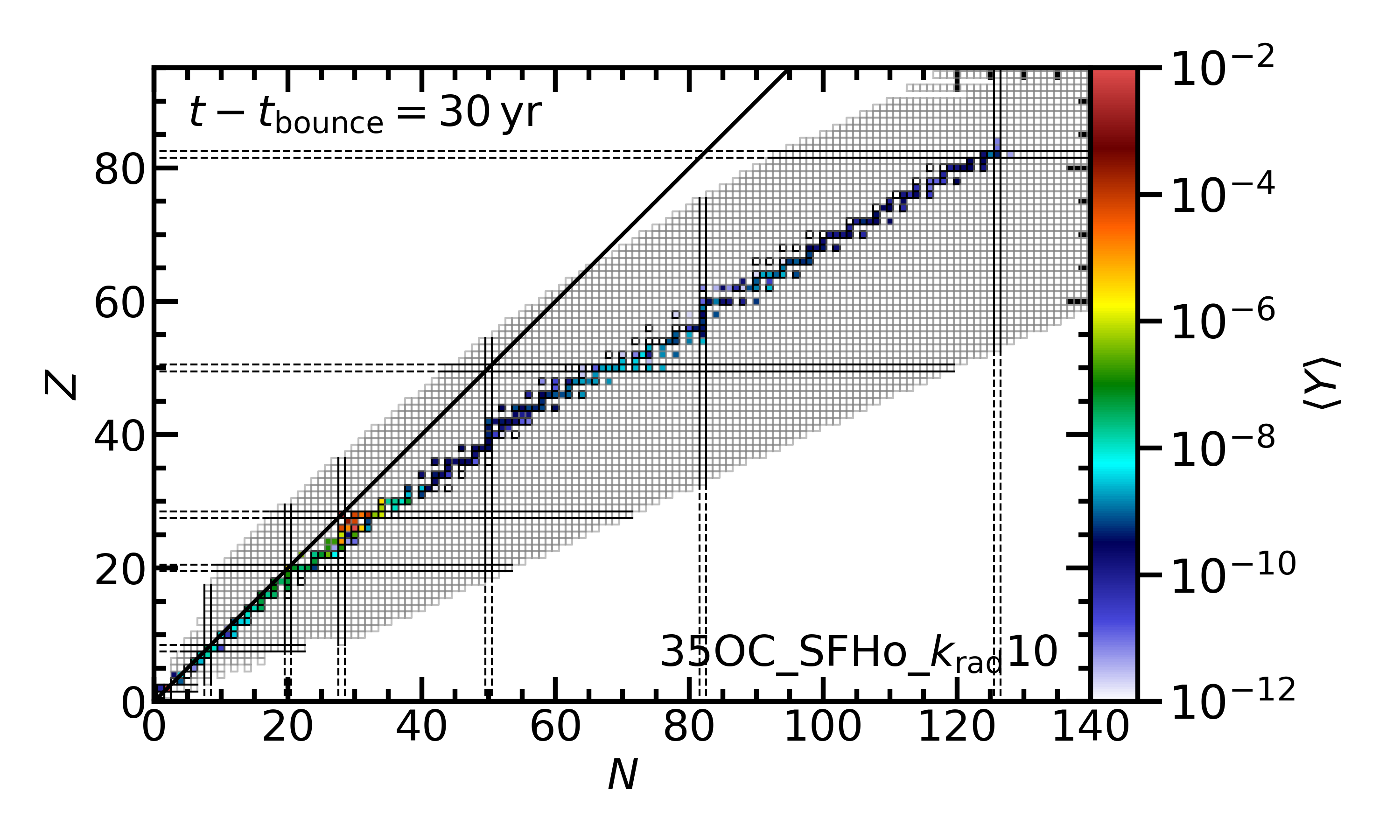}
    \caption{Average abundance $\langle Y \rangle$ of isotopes on the $(N,Z)$-plane, $30\,$years post-bounce,
    as evolved in \texttt{SkyNet} for model \texttt{35OC\_SFHo\_$k_{\rm rad}$10}. The final abundance of each simulation is 
    an average over the particle abundances, weighted by the number of baryons. Isotopes evolved in \texttt{SkyNet} are marked with 
    grey boxes, stable isotopes \citep{IAEA} are marked with black squares. Closed proton shells (corresponding to magic numbers 
    2, 8, 20, 28, 50, 82, and 126) are marked with horizontal black lines, and closed neutron shells are marked with vertical 
    black lines. The diagonal black line marks equal number of protons and neutrons in an isotope. }
    \label{fig:NZ_plane}
\end{figure}

\begin{figure*}
    \centering
    \includegraphics[width=\textwidth]{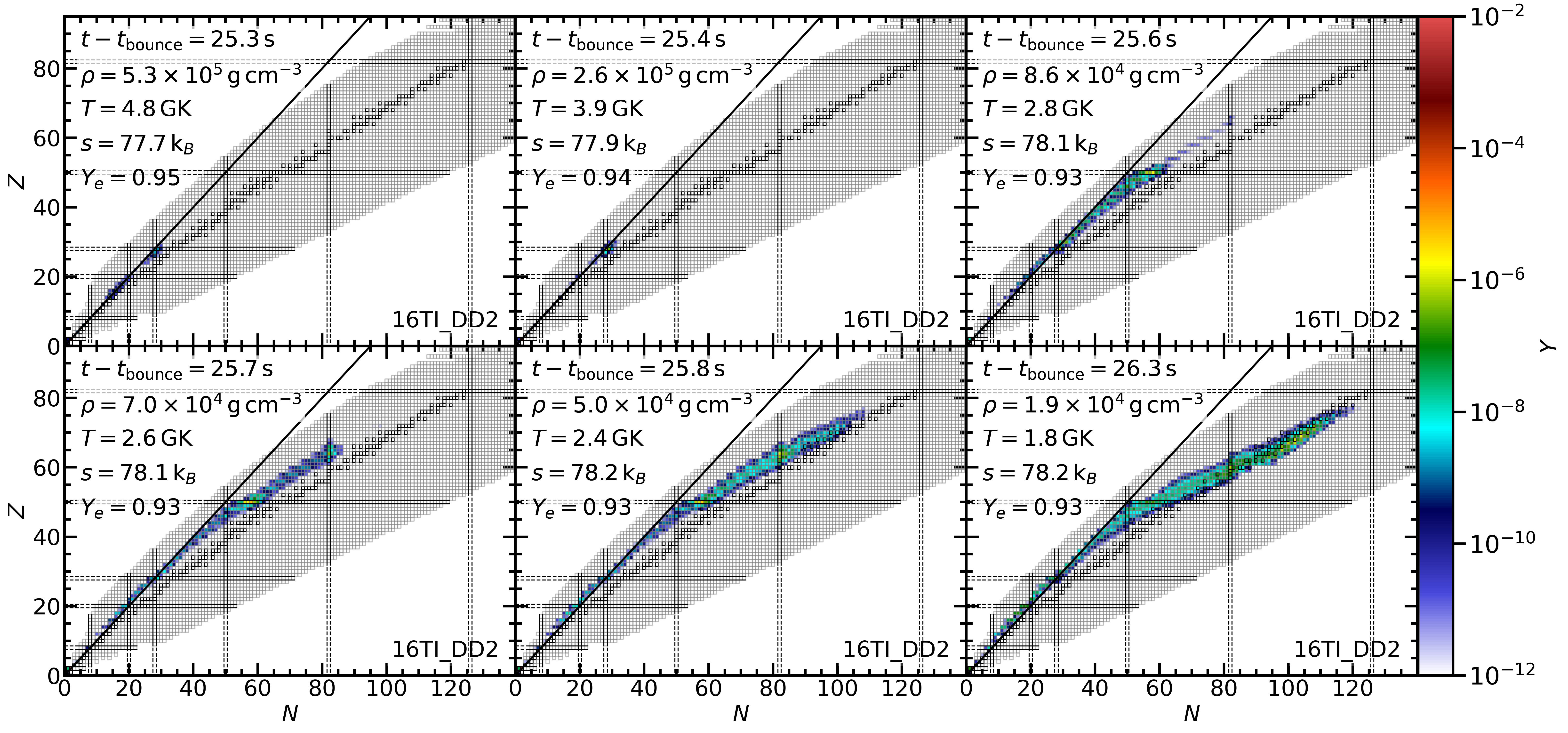}
    \caption{Time evolution of abundances on the $(N,Z)$-plane for the particle that produces the heaviest isotopes in model 
    \texttt{16TI\_DD2} (Fig.~\ref{fig:skynet}). The temperature, electron fraction (from \texttt{SkyNet}) and entropy per baryon of the tracer particle at each time are shown in each panel.}
    \label{fig:time_NZ_plane}
\end{figure*}

Our tracer particle abundances at $30$\,yr show production of stable isotopes on both the proton-rich and neutron-rich sides of the valley 
of stability (Figure~\ref{fig:NZ_plane}), suggesting the possibility of neutron- and proton captures occurring in some part of the outflow. 
Further investigation 
shows that heavy element production in models such as \texttt{16TI\_DD2} and \texttt{35OC\_SFHo\_$k_{\rm rad}$10} occurs via the $rp$-process 
\cite{wallace_1981}, building up elements via proton capture in a small number of 
tracer particles that sample fluid with an excess of free protons. Figure~\ref{fig:time_NZ_plane} shows 
the rapid buildup of heavy elements in one of the tracer particles from model 
\texttt{16TI\_DD2} that reaches $A\sim 200$, which evolves on the proton rich side of the valley of stability on a timescale of $\sim 1\,$s. 
This evolution in the $(N,Z)$-plane corresponds to a spike in the electron fraction (reaching $Y_e \sim 0.9$) into the proton-rich 
side, and a spike in the entropy (reaching $s/k_B \sim 80$) (Figure \ref{fig:time_NZ_plane}). 

Nucleosynthesis proceeds with 
alternating proton captures and $\beta^+$ decays up to $Z=50$ near the $Z=N$ line. Ref.~\cite{schatz_1997} showed that waiting points, corresponding
to the nuclear magic numbers along the proton drip line, create the need for several successive $\beta^+$-decays before further proton captures 
may be bridged by net proton capture, within a window of temperature $T\sim 1-2\,$GK, for densities $\rho \sim 10^6-10^7\,$g$\,$cm$^{-3}$. However, 
these thermodynamic ranges strongly depend on the nuclear $Q$-values for certain proton capture reactions, which are not known experimentally. Nonetheless,
this phenomenon may explain the ability of these few trajectories to exceed these waiting points. Thereafter, nucleosynthesis proceeds at a shallower angle across the $(N,Z)$-plane, towards- and
eventually across the valley of stability, likely due to $(\alpha,p)$ reactions on $rp$-process nuclei, similar to those described in \cite{wallace_1981}.

For completeness, we note that out of the particles producing heavy elements via the $rp$-process, only one does not
exceed our fiducial NSE transition temperature of $5$\,GK during its evolution. Nevertheless, that particle starts its \texttt{SkyNet} evolution in NSE from its maximum temperature of $4.7\,$GK, 
very close to the fiducial NSE temperature for the vast majority of post-processed particles.

\subsection{Optimal Initial Tracer Particle Placement} \label{sec:optimal_particle_placement}

The undersampling of the ejecta by tracer particles given our procedure for initial placement
motivates a brief discussion on how to improve this procedure in future collapsar simulations.
As discussed in Section \ref{sec:skynet}, particles are initialized in the domain pseudo-randomly between 
two radial bounds at the time of disk formation. Despite the formation of the shocked disk, accretion onto the BH continues 
predominantly through low density funnels near the poles during the first few seconds after disk formation. After the onset 
of shock expansion, significant accretion can continue through one hemisphere as the shock expands into the other. Of the 
initial $10^4$ tracer particles placed in each of our simulations, $37-93$\% are lost to accretion onto the BH. Which pole 
the shock wave expands into is not known a priori, and is set by the instability to axisymmetric perturbations as discussed 
in Paper I. Figure \ref{fig:particle_progression} illustrates this point for model \texttt{16TI\_SFHo}, showing particles which 
are accreted onto the BH (white dots) and those that make their way into the outflow (black dots). 

The radial range initially chosen is intended to maximize the presence of particles in the disk for the NDAF phase, during which 
the entirety of neutronization occurs (e.g., Paper I) and NSE is reached or closely approached, simplifying the handling of 
initial conditions for the post-processing nuclear reaction network. We choose the radial range for particle placement in our models 
by setting the free-fall timescale $t_{\rm ff}$ (as an estimate of the time it takes a particle to undergo infall and joining the disk) 
equal to the duration of the NDAF phase $\tau_{\rm NDAF}$, yielding a radius
\begin{eqnarray}
    R_{\rm ff} & = &\left[GM_{\rm BH}(t_{\rm df}) \tau_{\rm NDAF}^2\right]^{1/3}\\
               & \simeq & 10^9\,{\rm cm} \left(\frac{M_{\rm BH}(t_{\rm df})}{3M_\odot}\right)^{1/3}
                                         \left(\frac{\tau_{\rm NDAF}}{2\,{\rm s}}\right)^{2/3},\nonumber
\end{eqnarray}
where $M_{\rm BH}(t_{\rm df})$ is the lower limit to the BH mass when the disk is present. In most models, particles are initialized in the 
radial range $r=1.5\E{6}-3\E{9}\,$cm. 

Since outflows with the best chance of producing the conditions for the $r$-process originate during the NDAF phase, a better 
sampling of this outflow component requires more particles within $r \lesssim R_{\rm ff}$, given that the vast majority of 
these particles are lost to accretion, leaving only a very small percentage of the particles in the outflow. Better sampling 
of the entirety of the ejecta, particularly during the ADAF phase, requires increasing the outer radial limit for particle 
initialization. This, however, can run into memory limitations if too many particles are required for reasonable resolution in mass,
and also results in many particles never reaching NSE or even the explosive nucleosynthesis regime, introducing a dependence
on the initial abundance of the stellar progenitor and on how many isotopes are tracked in the hydrodynamic evolution (in our analysis, 
we discard particles that never exceed $1$\,GK).

\subsection{Comparison to previous work}

Nucleosynthesis in collapsars has been explored previously along two major branches: nucleosynthesis in jets, and in disk winds. 
While the inner accretion disk is capable of reaching the densities and temperatures needed for significant neutronization ($Y_e<0.25$) 
as needed for the production of heavy $r$-process elements, the question of whether that material makes its way into the outflow remains unsettled.  

Ref.~\cite{pruet_2003} examines neutronization of the accretion disk based on the steady state solutions of \cite{popham1999}, and argues, 
based on the expansion velocity of the jet, that $r$-process nucleosynthesis is possible. Varying the accretion rate and viscosity in the 
disk, they solve for the evolution of the electron fraction as governed by neutrino emission and absorption. In some models, disk material 
is significantly neutronized ($Y_e<0.25$), with the jet seen as a promising site for the production of $r$-process elements due to the high 
expansion speeds, allowing for the low electron fraction to be frozen out when ejected. Additionally, the disk wind is seen as a potentially 
larger source of $r$-process elements than the jet, if the entropy is higher than that seen 
in the center of the disk. Nevertheless, this work suggests that neutronization of the disk occurs primarily in at the inner radial edge in the 
mid-plane, making it very difficult for highly neutronized material to make its way into the outflow.

Nucleosynthesis in collapsar jets was studied by \cite{fujimoto_2007} using tracer particles on the simulations of \cite{fujimoto_2006}. The 
latter are axisymmetric, rotating MHD simulations with Newtonian gravity, with approximate neutrino effects, and produce magnetically-driven 
jets. Some particles reach electron fractions of $Y_e<0.25$, and produce isotopes up to and beyond the third $r$-process peak. Additionally, 
the production of light and heavy $p$-nuclei are found as well, suggesting some proton richness to the outflow.
The pseudo-Newtonian character of our simulation cannot produce a relativistic jet, and hence there is no direct comparison possible.

Ref.~\cite{ono_2012} performed nucleosynthetic calculations based on the axisymmetric Newtonian MHD simulations of \cite{ono_2009}, 
featuring collapsar jets. While the simulation only includes neutrino cooling, they too see sufficient neutronization to form elements 
up to the third $r$-process peak, but these third-peak elements are not significantly present in the final abundances. Similarly, \cite{nakamura_2013} 
performs axisymmetric relativistic MHD simulations of collapsar jets using a simplified neutrino scheme that is evolved on a coarser timescale 
than the hydrodynamics equations themselves. They also find that material in the jet reaches significant neutronization ($Y_e<0.25$), and the 
production of $r$-process elements beyond the third peak. 

More recently, nucleosynthesis in collapsar disk winds based on time-dependent 3D GRMHD simulations on a fixed Kerr metric and with a neutrino leakage scheme has been 
explored by \cite{siegel_2019}, starting from equilibrium tori. Based on the assumption that the disk wind overpowers the ram pressure 
from the infalling stellar mantle, the disk is embedded within a uniform low density medium. They find that the production of $r$-process 
elements is possible in the disk wind, and is highly accretion rate dependent. 

Ref.~\cite{siegel_2019} evolve their accretion disk in MHD, resolving the magneto-rotational instability directly, while we use the approximate 
alpha viscosity prescription of \cite{shakura1973}. BH accretion  disk winds driven by an alpha viscosity prescription have been compared (in the context of NS mergers) to MHD disk 
outflows by \cite{fernandez_2018}, with ISCO accretion rates, and viscous outflow rates reproducing the MHD outflow well in the ADAF phase  
($t\gtrsim 100\,$ms), with variations appearing on short timescales after the onset of the outflow due to transients associated to the chosen 
initial magnetic field geometry. Additionally, \cite{siegel_2019} evolves their disk in a uniform low density medium, while our disk is formed 
self-consistently within the star, with outflows making their way through the infalling stellar mantle. The main difference between our 
two approaches, in terms of the electron fraction distribution of the ejecta, stems from the mass accretion rate at the time of the ejection of 
matter from the disk. In our models, we see the degenerate conditions that lead to the neutronization of material predominantly in the midplane of 
the disk, also seen by \cite{siegel_2019}. However, the onset of mass ejection in our models does not occur until the transition to an ADAF phase, which 
corresponds to the end of neutronization of the disk. As a result, our outflowing material has electron fractions of $Y_e\sim 0.5$. The presence of a large-scale 
poloidal magnetic field, as well as the assumption of a uniform low density ambient medium in \cite{siegel_2019} allows for mass to be ejected earlier and with 
less resistance, leading to the ejection of material with $Y_e<0.25$. 

Ref.~\cite{janiuk_2019} performs 2D GRMHD simulations of BH accretion disk outflows with nucleosynthesis analysis. 
Like \cite{siegel_2019}, 
they embed an equilibrium torus in a low-density medium surrounding a central BH, but only accounting for neutrino cooling and evolving for $\sim 0.3\,$s. They find
that early outflows originating in the first $\sim 0.1\,$s may be composed of neutronized material varying in electron fraction from $\sim 0.2-0.35$, and through subsequent 
nucleosynthesis calculations, find the production of third peak $r$-process elements. Like with the
simulations of \cite{siegel_2019}, the main distinction with our models stems from neglecting the surrounding infalling stellar mantle and the presence of large-scale poloidal magnetic fields. 

Ref.~\cite{miller_2020} performs a similar 3D GRMHD simulation to that of \cite{siegel_2019}, but using Monte Carlo neutrino transport and evolving the 
torus for $\sim 150\,$ms. They obtain accretion rates  that match the $0.1\,\Msun\,$s$^{-1}$ of \cite{siegel_2019}, which allow
for significant neutrino cooling and thus neutronization. They find that there is no unbound material with electron 
fraction below $Y_e\sim 0.3$, however, and are thus unable to produce 3rd $r$-process peak elements. Additionally, they note that the assumption of neglecting 
the ram pressure from the infalling stellar mantle, as well as feedback effects from the jet, may have a significant effect on whether neutronized 
material is able to escape the star at all. Our results from Paper I support this conjecture. 

Ref.~\cite{zenati_2020} performs long-term ($\sim 100$\,s), axisymmetric, viscous hydrodynamic simulations of collapsar disks 
including the same 19-isotope nuclear network we use. Equilibrium tori, embedded in low-density atmospheres,
are constructed to match the post-circularization state of several progenitors, assuming an angular momentum profile that 
varies as a radial power-law in the core. The state of these disks is intended to match later times relative to disk 
formation and the NDAF phase that follows. This is reinforced by the peak densities in their disks being $\oforder{10^2}$
times smaller than those in the disks from our most comparable progenitor (\texttt{35OC}). Thus, they see disk outflows 
driven in the ADAF regime, with contributions from viscous and nuclear heating. Overall, our models eject more mass, 
and have higher explosion energies by roughly an order of magnitude, owing to the continual feeding of the accretion 
disk by the stellar mantle, which is not present in the models from \cite{zenati_2020}. The ${}^{56}$\,Ni mass
range that they obtain ($[0.6-7.0]\times 10^{-3}\,M_\odot$) is also smaller than what we find by a factor $\gtrsim 100$.


\section{Summary and Discussion \label{sec:summary}}

We have studied nucleosynthesis in the disk outflows from rapidly-rotating Wolf-Rayet stars that undergo
core-collapse and form a BH accretion disk. We evolve the stars from core-collapse to BH formation using a spherically-symmetric
general relativistic neutrino radiation hydrodynamics code with approximate rotation effects. We then map into an axisymmetric
viscous hydrodynamic code that includes Newtonian self-gravity and a pseudo-Newtonian potential for the BH, neutrino heating
and cooling via a lightbulb-type scheme, and the Helmholtz EOS. 

Simulations also include a 19-isotope nuclear reaction network, supplemented by an NSE solver for high temperatures, providing 
full temporal and spatial composition information for alpha chain and related elements. Additionally, we use passive tracer 
particles to sample neutrino-reprocessed matter, and post-process these trajectories with a large nuclear reaction network. 

Our models capture the self-consistent formation of a shocked, neutrino-cooled accretion disk that transitions into
an advective disk, within a collapsing star, following shock expansion until it breaks out from the stellar surface. 
The outflow from this disk is sufficiently energetic to explode the star. In Paper I, we reported on
a first set of 5 simulations that vary the progenitor star, nuclear EOS used prior to BH formation, and strength of viscous
angular momentum transport. In this follow-up paper, we carry out detailed nucleosynthesis analysis of the initial simulation
set, and present additional models that modify the rotation profile of progenitor stars, to maximize the exposure of 
circularized shells to significant neutrino reprocessing 
and thus to neutronization, as an optimistic upper limit to the 
dependence of neutron-rich matter generation on stellar rotation. 

Our main results are the following:
\newline

\noindent
1. -- The ejecta from all of our simulations is dominated by ${}^{16}$O, of which several $M_\odot$ are consistently produced. This
is followed by ${}^{12}$C, ${}^{20}$Ne, and ${}^{56}$Ni at $\sim 1\,M_\odot$ each. All other elements of the 19-isotope network are 
produced at a $<1M_\odot$ level (Tables~\ref{tab:isotope_mass_1}-\ref{tab:isotope_mass_2}). The ${}^{56}$Ni masses are consistent with 
the observed range from Ic-BL SNe, while average asymptotic velocities are lower by a factor $\sim 2$. The ejecta is well mixed in mass and 
in velocity (Figs.~\ref{fig:lagrangian_composition}-\ref{fig:vinf_hist}), although spatial stratification of heavy elements
is apparent (Fig.~\ref{fig:comp_cmap}). Production of ${}^{44}$Ti is super-solar (Fig.~\ref{fig:ti44_ni56}).
\newline

\noindent
2. -- Only one of our models, with the lowest viscosity, yields an outflow with sufficient neutrons to reach
the first $r$-process peak in significant amounts. All other models produce very small or negligible amounts of
elements beyond the iron group (Fig.~\ref{fig:skynet}). While the minimum electron fraction in the accretion disk drops to 
$Y_e < 0.25$ during the NDAF phase in all models, most of this material is near the central plane of the accretion disk, 
and is accreted onto the central BH, as reported in Paper I. The rapid expansion of the shock wave follows a transition to an 
ADAF phase with no additional neutronization (Figs.~\ref{fig:evolution}-\ref{fig:minYe}). Nevertheless, the fact that first 
$r$-process peak elements were produced due to a (small) change in viscosity suggests that an intrinsic scatter in abundances 
around $A\sim 80$ is expected due to variations between individual collapsar explosions in nature.
\newline

\noindent
3. -- A subset of our models produce small quantities of heavy elements (up to $A\sim 200$) via the $rp$-process and $(\alpha,p)$ reactions, 
in particles that sample proton-rich, high  entropy ($s/k_{\rm B}\sim 80$) ejecta (Figs.~\ref{fig:NZ_plane}-\ref{fig:time_NZ_plane}). 
While our tracer particles do not fully sample the entirety of the ejecta (Table~\ref{tab:sampling}), they trace matter that accretes 
into the disk during the NDAF phase and is thus subject to significant neutrino reprocessing. Our results suggest that 
collapsar disk winds may exhibit the conditions necessary for the $rp$-process to act. 
\newline

In addition to providing trajectories for nucleosynthesis analysis, our tracer particles show that 
while the most significant neutronization occurs in the midplane of the accretion disk near the BH, the majority of the disk 
outflow originates from the outer edge of the disk. Due to the ever-present central BH acting as a sink of matter 
in the center of the domain (Fig.~\ref{fig:particle_progression}), 
this material is easily accreted, and almost none of the neutronized material makes its way into the outflow. The large
fraction of tracer particles accreted into the BH in our simulations provides guidance on where in the star these initial
tracers must be located for optimal sampling (Sec.~\ref{sec:optimal_particle_placement}).  

The entropy per baryon of the ejected material covers a broad range, with mass ejected exceeding several hundred $k_B$. As our tracer 
particles only sample the low entropy end of the distribution, it is possible that some $r$-process elements are produced in this regime 
that were not sampled here. However, if the $r$-process indeed occurs in the high-entropy tail, the
amount of mass produced must be relatively small compared to the rest of the ejecta (Fig.~\ref{fig:model_histograms}).

The modification of the angular momentum profile in the new models presented here (Fig.~\ref{fig:radr_mgrav}) had at most a modest effect 
on the duration of the NDAF phase, delaying disk formation and the transition to the ADAF phase, while only extending the NDAF phase appreciably 
in one model (Fig.~\ref{fig:evolution}). This did not lead to a significant change in the minimum electron fraction in the outflow (Fig.~\ref{fig:minYe}), 
and yielded only trace amounts of elements beyond the iron group in another model 
(\texttt{35OC\_SFHo\_$k_{\rm rad}$10}; Fig.~\ref{fig:skynet}). We conclude that modification
of the rotation profile alone cannot lead to the production of significant amounts of $r$-process elements in the context
of viscous hydrodynamic evolution. 

The picture might change if MHD effects, beyond those modeled by viscous hydrodynamics, become important. Experience from BH accretion disks formed in NS mergers 
shows that significant ejection of matter during the NDAF phase is possible, but highly sensitive to the initial magnetic field geometry of the disk, 
requiring a large-scale poloidal component to generate magnetically-driven outflows in excess of those due to dissipation of magnetorotational turbulence \cite{christie_2019,FF22,hayashi_2023}.
Likewise, magnetorotational CCSNe simulations require strong, large-scale initial dipolar fields to reach the third $r$-process peak \cite{nishimura_2015,moesta_2018,reichert_2024,zha_2024}. 
Like in NS mergers, magnetic winds would not only carry away 
neutron-rich material, but would also be faster than outflows obtained in viscous hydrodynamics \cite{fernandez2019}, 
possibly ameliorating the velocity deficit of our ejecta relative to that inferred from Ic SNe spectra. 
A relativistic jet, likely magnetically-driven, can also contribute to speed up the disk outflow through energy deposition.
Ultimately, whether these rapid, magnetically-driven outflows take place in collapsars will depend on the magnetic field geometry and strength in the presupernova star. 


\appendix

\begin{acknowledgments}
We thank Steven Fahlman, Suhasini Rao, Brian Metzger, Rebecca Surman, Dan Kasen, and Ore Gottlieb 
for helpful discussions.
This research was supported by the Natural Sciences and Engineering Research
Council of Canada (NSERC) through Discovery Grant RGPIN-2022-03463. Support was
also provided by the Alberta Graduate Excellence Scholarship to CD.
The software used in this work was in part developed by the U.S Department of Energy (DOE)
NNSA-ASC OASCR Flash Center at the University of Chicago.
Data visualization was done in part using {\tt VisIt} \cite{VisIt}, which is supported
by DOE with funding from the Advanced Simulation and Computing Program
and the Scientific Discovery through Advanced Computing Program.
This research used storage resources of the
National Energy Research Scientific Computing Center
(NERSC), which is supported by the DOE Office of Science
under Contract No. DE-AC02-05CH11231 (repository m2058).
This research was enabled in part by computing and storage support
provided by Prairies DRI, BC DRI Group, Compute Ontario (computeontario.ca),
Calcul Qu\'ebec (www.calculquebec.ca) and the Digital Research Alliance of Canada (alliancecan.ca). 
Computations were performed on the Niagara supercomputer at the SciNet HPC Consortium \cite{SciNet,Niagara}. SciNet is 
funded by Innovation, Science and Economic Development Canada; the Digital Research Alliance of Canada; 
the Ontario Research Fund: Research Excellence; and the University of Toronto. 
\end{acknowledgments}


\bibliographystyle{apsrev4-2}
\bibliography{ms}

\begin{thebibliography}{87}%
\makeatletter
\providecommand \@ifxundefined [1]{%
 \@ifx{#1\undefined}
}%
\providecommand \@ifnum [1]{%
 \ifnum #1\expandafter \@firstoftwo
 \else \expandafter \@secondoftwo
 \fi
}%
\providecommand \@ifx [1]{%
 \ifx #1\expandafter \@firstoftwo
 \else \expandafter \@secondoftwo
 \fi
}%
\providecommand \natexlab [1]{#1}%
\providecommand \enquote  [1]{``#1''}%
\providecommand \bibnamefont  [1]{#1}%
\providecommand \bibfnamefont [1]{#1}%
\providecommand \citenamefont [1]{#1}%
\providecommand \href@noop [0]{\@secondoftwo}%
\providecommand \href [0]{\begingroup \@sanitize@url \@href}%
\providecommand \@href[1]{\@@startlink{#1}\@@href}%
\providecommand \@@href[1]{\endgroup#1\@@endlink}%
\providecommand \@sanitize@url [0]{\catcode `\\12\catcode `\$12\catcode `\&12\catcode `\#12\catcode `\^12\catcode `\_12\catcode `\%12\relax}%
\providecommand \@@startlink[1]{}%
\providecommand \@@endlink[0]{}%
\providecommand \url  [0]{\begingroup\@sanitize@url \@url }%
\providecommand \@url [1]{\endgroup\@href {#1}{\urlprefix }}%
\providecommand \urlprefix  [0]{URL }%
\providecommand \Eprint [0]{\href }%
\providecommand \doibase [0]{https://doi.org/}%
\providecommand \selectlanguage [0]{\@gobble}%
\providecommand \bibinfo  [0]{\@secondoftwo}%
\providecommand \bibfield  [0]{\@secondoftwo}%
\providecommand \translation [1]{[#1]}%
\providecommand \BibitemOpen [0]{}%
\providecommand \bibitemStop [0]{}%
\providecommand \bibitemNoStop [0]{.\EOS\space}%
\providecommand \EOS [0]{\spacefactor3000\relax}%
\providecommand \BibitemShut  [1]{\csname bibitem#1\endcsname}%
\let\auto@bib@innerbib\@empty
\bibitem [{\citenamefont {{Burbidge}}\ \emph {et~al.}(1957)\citenamefont {{Burbidge}}, \citenamefont {{Burbidge}}, \citenamefont {{Fowler}},\ and\ \citenamefont {{Hoyle}}}]{burbidge_1957}%
  \BibitemOpen
  \bibfield  {author} {\bibinfo {author} {\bibfnamefont {E.~M.}\ \bibnamefont {{Burbidge}}}, \bibinfo {author} {\bibfnamefont {G.~R.}\ \bibnamefont {{Burbidge}}}, \bibinfo {author} {\bibfnamefont {W.~A.}\ \bibnamefont {{Fowler}}},\ and\ \bibinfo {author} {\bibfnamefont {F.}~\bibnamefont {{Hoyle}}},\ }\href {https://doi.org/10.1103/RevModPhys.29.547} {\bibfield  {journal} {\bibinfo  {journal} {Reviews of Modern Physics}\ }\textbf {\bibinfo {volume} {29}},\ \bibinfo {pages} {547} (\bibinfo {year} {1957})}\BibitemShut {NoStop}%
\bibitem [{\citenamefont {{Johnson}}(2019)}]{johnson_2019}%
  \BibitemOpen
  \bibfield  {author} {\bibinfo {author} {\bibfnamefont {J.~A.}\ \bibnamefont {{Johnson}}},\ }\href {https://doi.org/10.1126/science.aau9540} {\bibfield  {journal} {\bibinfo  {journal} {Science}\ }\textbf {\bibinfo {volume} {363}},\ \bibinfo {pages} {474} (\bibinfo {year} {2019})}\BibitemShut {NoStop}%
\bibitem [{\citenamefont {Janka}\ and\ \citenamefont {Bauswein}(2023)}]{janka_2023}%
  \BibitemOpen
  \bibfield  {author} {\bibinfo {author} {\bibfnamefont {H.-T.}\ \bibnamefont {Janka}}\ and\ \bibinfo {author} {\bibfnamefont {A.}~\bibnamefont {Bauswein}},\ }\bibinfo {title} {Dynamics and equation of state dependencies of relevance for nucleosynthesis in supernovae and neutron star mergers},\ in\ \href {https://doi.org/10.1007/978-981-19-6345-2_93} {\emph {\bibinfo {booktitle} {Handbook of Nuclear Physics}}},\ \bibinfo {editor} {edited by\ \bibinfo {editor} {\bibfnamefont {I.}~\bibnamefont {Tanihata}}, \bibinfo {editor} {\bibfnamefont {H.}~\bibnamefont {Toki}},\ and\ \bibinfo {editor} {\bibfnamefont {T.}~\bibnamefont {Kajino}}}\ (\bibinfo  {publisher} {Springer Nature Singapore},\ \bibinfo {address} {Singapore},\ \bibinfo {year} {2023})\ pp.\ \bibinfo {pages} {4005--4102}\BibitemShut {NoStop}%
\bibitem [{\citenamefont {{Wanajo}}\ \emph {et~al.}(2018)\citenamefont {{Wanajo}}, \citenamefont {{M{\"u}ller}}, \citenamefont {{Janka}},\ and\ \citenamefont {{Heger}}}]{wanajo_2018}%
  \BibitemOpen
  \bibfield  {author} {\bibinfo {author} {\bibfnamefont {S.}~\bibnamefont {{Wanajo}}}, \bibinfo {author} {\bibfnamefont {B.}~\bibnamefont {{M{\"u}ller}}}, \bibinfo {author} {\bibfnamefont {H.-T.}\ \bibnamefont {{Janka}}},\ and\ \bibinfo {author} {\bibfnamefont {A.}~\bibnamefont {{Heger}}},\ }\href {https://doi.org/10.3847/1538-4357/aa9d97} {\bibfield  {journal} {\bibinfo  {journal} {\apj}\ }\textbf {\bibinfo {volume} {852}},\ \bibinfo {eid} {40} (\bibinfo {year} {2018})},\ \Eprint {https://arxiv.org/abs/1701.06786} {arXiv:1701.06786 [astro-ph.SR]} \BibitemShut {NoStop}%
\bibitem [{\citenamefont {{Witt}}\ \emph {et~al.}(2021)\citenamefont {{Witt}}, \citenamefont {{Psaltis}}, \citenamefont {{Yasin}}, \citenamefont {{Horn}}, \citenamefont {{Reichert}}, \citenamefont {{Kuroda}}, \citenamefont {{Obergaulinger}}, \citenamefont {{Couch}},\ and\ \citenamefont {{Arcones}}}]{witt_2021}%
  \BibitemOpen
  \bibfield  {author} {\bibinfo {author} {\bibfnamefont {M.}~\bibnamefont {{Witt}}}, \bibinfo {author} {\bibfnamefont {A.}~\bibnamefont {{Psaltis}}}, \bibinfo {author} {\bibfnamefont {H.}~\bibnamefont {{Yasin}}}, \bibinfo {author} {\bibfnamefont {C.}~\bibnamefont {{Horn}}}, \bibinfo {author} {\bibfnamefont {M.}~\bibnamefont {{Reichert}}}, \bibinfo {author} {\bibfnamefont {T.}~\bibnamefont {{Kuroda}}}, \bibinfo {author} {\bibfnamefont {M.}~\bibnamefont {{Obergaulinger}}}, \bibinfo {author} {\bibfnamefont {S.~M.}\ \bibnamefont {{Couch}}},\ and\ \bibinfo {author} {\bibfnamefont {A.}~\bibnamefont {{Arcones}}},\ }\href {https://doi.org/10.3847/1538-4357/ac1a6d} {\bibfield  {journal} {\bibinfo  {journal} {\apj}\ }\textbf {\bibinfo {volume} {921}},\ \bibinfo {eid} {19} (\bibinfo {year} {2021})},\ \Eprint {https://arxiv.org/abs/2107.00687} {arXiv:2107.00687 [astro-ph.HE]} \BibitemShut {NoStop}%
\bibitem [{\citenamefont {{Wang}}\ and\ \citenamefont {{Burrows}}(2023)}]{wang_2023}%
  \BibitemOpen
  \bibfield  {author} {\bibinfo {author} {\bibfnamefont {T.}~\bibnamefont {{Wang}}}\ and\ \bibinfo {author} {\bibfnamefont {A.}~\bibnamefont {{Burrows}}},\ }\href {https://doi.org/10.3847/1538-4357/ace7b2} {\bibfield  {journal} {\bibinfo  {journal} {\apj}\ }\textbf {\bibinfo {volume} {954}},\ \bibinfo {eid} {114} (\bibinfo {year} {2023})},\ \Eprint {https://arxiv.org/abs/2306.13712} {arXiv:2306.13712 [astro-ph.SR]} \BibitemShut {NoStop}%
\bibitem [{\citenamefont {{Mathews}}\ and\ \citenamefont {{Cowan}}(1990)}]{mathews_1990}%
  \BibitemOpen
  \bibfield  {author} {\bibinfo {author} {\bibfnamefont {G.~J.}\ \bibnamefont {{Mathews}}}\ and\ \bibinfo {author} {\bibfnamefont {J.~J.}\ \bibnamefont {{Cowan}}},\ }\href {https://doi.org/10.1038/345491a0} {\bibfield  {journal} {\bibinfo  {journal} {\nat}\ }\textbf {\bibinfo {volume} {345}},\ \bibinfo {pages} {491} (\bibinfo {year} {1990})}\BibitemShut {NoStop}%
\bibitem [{\citenamefont {{Woosley}}(1993)}]{woosley_1993}%
  \BibitemOpen
  \bibfield  {author} {\bibinfo {author} {\bibfnamefont {S.~E.}\ \bibnamefont {{Woosley}}},\ }\href {https://doi.org/10.1086/172359} {\bibfield  {journal} {\bibinfo  {journal} {ApJ}\ }\textbf {\bibinfo {volume} {405}},\ \bibinfo {pages} {273} (\bibinfo {year} {1993})}\BibitemShut {NoStop}%
\bibitem [{\citenamefont {{MacFadyen}}\ and\ \citenamefont {{Woosley}}(1999)}]{macfadyen_1999}%
  \BibitemOpen
  \bibfield  {author} {\bibinfo {author} {\bibfnamefont {A.~I.}\ \bibnamefont {{MacFadyen}}}\ and\ \bibinfo {author} {\bibfnamefont {S.~E.}\ \bibnamefont {{Woosley}}},\ }\href {https://doi.org/10.1086/307790} {\bibfield  {journal} {\bibinfo  {journal} {ApJ}\ }\textbf {\bibinfo {volume} {524}},\ \bibinfo {pages} {262} (\bibinfo {year} {1999})},\ \Eprint {https://arxiv.org/abs/astro-ph/9810274} {astro-ph/9810274} \BibitemShut {NoStop}%
\bibitem [{\citenamefont {{MacFadyen}}(2003)}]{macfadyen_2003}%
  \BibitemOpen
  \bibfield  {author} {\bibinfo {author} {\bibfnamefont {A.~I.}\ \bibnamefont {{MacFadyen}}},\ }in\ \href {https://doi.org/10.1007/10828549_14} {\emph {\bibinfo {booktitle} {From Twilight to Highlight: The Physics of Supernovae}}},\ \bibinfo {editor} {edited by\ \bibinfo {editor} {\bibfnamefont {W.}~\bibnamefont {{Hillebrandt}}}\ and\ \bibinfo {editor} {\bibfnamefont {B.}~\bibnamefont {{Leibundgut}}}}\ (\bibinfo {year} {2003})\ p.~\bibinfo {pages} {97},\ \Eprint {https://arxiv.org/abs/astro-ph/0301425} {arXiv:astro-ph/0301425 [astro-ph]} \BibitemShut {NoStop}%
\bibitem [{\citenamefont {{Kohri}}\ \emph {et~al.}(2005)\citenamefont {{Kohri}}, \citenamefont {{Narayan}},\ and\ \citenamefont {{Piran}}}]{kohri_2005}%
  \BibitemOpen
  \bibfield  {author} {\bibinfo {author} {\bibfnamefont {K.}~\bibnamefont {{Kohri}}}, \bibinfo {author} {\bibfnamefont {R.}~\bibnamefont {{Narayan}}},\ and\ \bibinfo {author} {\bibfnamefont {T.}~\bibnamefont {{Piran}}},\ }\href {https://doi.org/10.1086/431354} {\bibfield  {journal} {\bibinfo  {journal} {ApJ}\ }\textbf {\bibinfo {volume} {629}},\ \bibinfo {pages} {341} (\bibinfo {year} {2005})},\ \Eprint {https://arxiv.org/abs/astro-ph/0502470} {arXiv:astro-ph/0502470 [astro-ph]} \BibitemShut {NoStop}%
\bibitem [{\citenamefont {{Woosley}}\ and\ \citenamefont {{Howard}}(1978)}]{woosley_1978}%
  \BibitemOpen
  \bibfield  {author} {\bibinfo {author} {\bibfnamefont {S.~E.}\ \bibnamefont {{Woosley}}}\ and\ \bibinfo {author} {\bibfnamefont {W.~M.}\ \bibnamefont {{Howard}}},\ }\href {https://doi.org/10.1086/190501} {\bibfield  {journal} {\bibinfo  {journal} {\apjs}\ }\textbf {\bibinfo {volume} {36}},\ \bibinfo {pages} {285} (\bibinfo {year} {1978})}\BibitemShut {NoStop}%
\bibitem [{\citenamefont {{Howard}}\ \emph {et~al.}(1991)\citenamefont {{Howard}}, \citenamefont {{Meyer}},\ and\ \citenamefont {{Woosley}}}]{howard_1991}%
  \BibitemOpen
  \bibfield  {author} {\bibinfo {author} {\bibfnamefont {W.~M.}\ \bibnamefont {{Howard}}}, \bibinfo {author} {\bibfnamefont {B.~S.}\ \bibnamefont {{Meyer}}},\ and\ \bibinfo {author} {\bibfnamefont {S.~E.}\ \bibnamefont {{Woosley}}},\ }\href {https://doi.org/10.1086/186038} {\bibfield  {journal} {\bibinfo  {journal} {\apjl}\ }\textbf {\bibinfo {volume} {373}},\ \bibinfo {pages} {L5} (\bibinfo {year} {1991})}\BibitemShut {NoStop}%
\bibitem [{\citenamefont {{Wallace}}\ and\ \citenamefont {{Woosley}}(1981)}]{wallace_1981}%
  \BibitemOpen
  \bibfield  {author} {\bibinfo {author} {\bibfnamefont {R.~K.}\ \bibnamefont {{Wallace}}}\ and\ \bibinfo {author} {\bibfnamefont {S.~E.}\ \bibnamefont {{Woosley}}},\ }\href {https://doi.org/10.1086/190717} {\bibfield  {journal} {\bibinfo  {journal} {\apjs}\ }\textbf {\bibinfo {volume} {45}},\ \bibinfo {pages} {389} (\bibinfo {year} {1981})}\BibitemShut {NoStop}%
\bibitem [{\citenamefont {{Fr{\"o}hlich}}\ \emph {et~al.}(2006)\citenamefont {{Fr{\"o}hlich}}, \citenamefont {{Mart{\'\i}nez-Pinedo}}, \citenamefont {{Liebend{\"o}rfer}}, \citenamefont {{Thielemann}}, \citenamefont {{Bravo}}, \citenamefont {{Hix}}, \citenamefont {{Langanke}},\ and\ \citenamefont {{Zinner}}}]{frohlich_2006}%
  \BibitemOpen
  \bibfield  {author} {\bibinfo {author} {\bibfnamefont {C.}~\bibnamefont {{Fr{\"o}hlich}}}, \bibinfo {author} {\bibfnamefont {G.}~\bibnamefont {{Mart{\'\i}nez-Pinedo}}}, \bibinfo {author} {\bibfnamefont {M.}~\bibnamefont {{Liebend{\"o}rfer}}}, \bibinfo {author} {\bibfnamefont {F.~K.}\ \bibnamefont {{Thielemann}}}, \bibinfo {author} {\bibfnamefont {E.}~\bibnamefont {{Bravo}}}, \bibinfo {author} {\bibfnamefont {W.~R.}\ \bibnamefont {{Hix}}}, \bibinfo {author} {\bibfnamefont {K.}~\bibnamefont {{Langanke}}},\ and\ \bibinfo {author} {\bibfnamefont {N.~T.}\ \bibnamefont {{Zinner}}},\ }\href {https://doi.org/10.1103/PhysRevLett.96.142502} {\bibfield  {journal} {\bibinfo  {journal} {\prl}\ }\textbf {\bibinfo {volume} {96}},\ \bibinfo {eid} {142502} (\bibinfo {year} {2006})},\ \Eprint {https://arxiv.org/abs/astro-ph/0511376} {arXiv:astro-ph/0511376 [astro-ph]} \BibitemShut {NoStop}%
\bibitem [{\citenamefont {{Just}}\ \emph {et~al.}(2022)\citenamefont {{Just}}, \citenamefont {{Aloy}}, \citenamefont {{Obergaulinger}},\ and\ \citenamefont {{Nagataki}}}]{just_2022}%
  \BibitemOpen
  \bibfield  {author} {\bibinfo {author} {\bibfnamefont {O.}~\bibnamefont {{Just}}}, \bibinfo {author} {\bibfnamefont {M.~A.}\ \bibnamefont {{Aloy}}}, \bibinfo {author} {\bibfnamefont {M.}~\bibnamefont {{Obergaulinger}}},\ and\ \bibinfo {author} {\bibfnamefont {S.}~\bibnamefont {{Nagataki}}},\ }\href {https://doi.org/10.3847/2041-8213/ac83a1} {\bibfield  {journal} {\bibinfo  {journal} {\apj}\ }\textbf {\bibinfo {volume} {934}},\ \bibinfo {eid} {L30} (\bibinfo {year} {2022})},\ \Eprint {https://arxiv.org/abs/2205.14158} {arXiv:2205.14158 [astro-ph.HE]} \BibitemShut {NoStop}%
\bibitem [{\citenamefont {{Fujibayashi}}\ \emph {et~al.}(2023{\natexlab{a}})\citenamefont {{Fujibayashi}}, \citenamefont {{Sekiguchi}}, \citenamefont {{Shibata}},\ and\ \citenamefont {{Wanajo}}}]{fujibayashi_2022}%
  \BibitemOpen
  \bibfield  {author} {\bibinfo {author} {\bibfnamefont {S.}~\bibnamefont {{Fujibayashi}}}, \bibinfo {author} {\bibfnamefont {Y.}~\bibnamefont {{Sekiguchi}}}, \bibinfo {author} {\bibfnamefont {M.}~\bibnamefont {{Shibata}}},\ and\ \bibinfo {author} {\bibfnamefont {S.}~\bibnamefont {{Wanajo}}},\ }\href {https://doi.org/10.3847/1538-4357/acf5e5} {\bibfield  {journal} {\bibinfo  {journal} {\apj}\ }\textbf {\bibinfo {volume} {956}},\ \bibinfo {eid} {100} (\bibinfo {year} {2023}{\natexlab{a}})},\ \Eprint {https://arxiv.org/abs/2212.03958} {arXiv:2212.03958 [astro-ph.HE]} \BibitemShut {NoStop}%
\bibitem [{\citenamefont {{Pruet}}\ \emph {et~al.}(2003)\citenamefont {{Pruet}}, \citenamefont {{Woosley}},\ and\ \citenamefont {{Hoffman}}}]{pruet_2003}%
  \BibitemOpen
  \bibfield  {author} {\bibinfo {author} {\bibfnamefont {J.}~\bibnamefont {{Pruet}}}, \bibinfo {author} {\bibfnamefont {S.~E.}\ \bibnamefont {{Woosley}}},\ and\ \bibinfo {author} {\bibfnamefont {R.~D.}\ \bibnamefont {{Hoffman}}},\ }\href {https://doi.org/10.1086/367957} {\bibfield  {journal} {\bibinfo  {journal} {ApJ}\ }\textbf {\bibinfo {volume} {586}},\ \bibinfo {pages} {1254} (\bibinfo {year} {2003})},\ \Eprint {https://arxiv.org/abs/astro-ph/0209412} {arXiv:astro-ph/0209412 [astro-ph]} \BibitemShut {NoStop}%
\bibitem [{\citenamefont {{Fujimoto}}\ \emph {et~al.}(2004)\citenamefont {{Fujimoto}}, \citenamefont {{Hashimoto}}, \citenamefont {{Arai}},\ and\ \citenamefont {{Matsuba}}}]{fujimoto_2004}%
  \BibitemOpen
  \bibfield  {author} {\bibinfo {author} {\bibfnamefont {S.-i.}\ \bibnamefont {{Fujimoto}}}, \bibinfo {author} {\bibfnamefont {M.-a.}\ \bibnamefont {{Hashimoto}}}, \bibinfo {author} {\bibfnamefont {K.}~\bibnamefont {{Arai}}},\ and\ \bibinfo {author} {\bibfnamefont {R.}~\bibnamefont {{Matsuba}}},\ }\href {https://doi.org/10.1086/423778} {\bibfield  {journal} {\bibinfo  {journal} {\apj}\ }\textbf {\bibinfo {volume} {614}},\ \bibinfo {pages} {847} (\bibinfo {year} {2004})},\ \Eprint {https://arxiv.org/abs/astro-ph/0405510} {arXiv:astro-ph/0405510 [astro-ph]} \BibitemShut {NoStop}%
\bibitem [{\citenamefont {{Surman}}\ \emph {et~al.}(2006)\citenamefont {{Surman}}, \citenamefont {{McLaughlin}},\ and\ \citenamefont {{Hix}}}]{surman_2006}%
  \BibitemOpen
  \bibfield  {author} {\bibinfo {author} {\bibfnamefont {R.}~\bibnamefont {{Surman}}}, \bibinfo {author} {\bibfnamefont {G.~C.}\ \bibnamefont {{McLaughlin}}},\ and\ \bibinfo {author} {\bibfnamefont {W.~R.}\ \bibnamefont {{Hix}}},\ }\href {https://doi.org/10.1086/501116} {\bibfield  {journal} {\bibinfo  {journal} {\apj}\ }\textbf {\bibinfo {volume} {643}},\ \bibinfo {pages} {1057} (\bibinfo {year} {2006})},\ \Eprint {https://arxiv.org/abs/astro-ph/0509365} {arXiv:astro-ph/0509365 [astro-ph]} \BibitemShut {NoStop}%
\bibitem [{\citenamefont {{Fujimoto}}\ \emph {et~al.}(2007)\citenamefont {{Fujimoto}}, \citenamefont {{Hashimoto}}, \citenamefont {{Kotake}},\ and\ \citenamefont {{Yamada}}}]{fujimoto_2007}%
  \BibitemOpen
  \bibfield  {author} {\bibinfo {author} {\bibfnamefont {S.-i.}\ \bibnamefont {{Fujimoto}}}, \bibinfo {author} {\bibfnamefont {M.-a.}\ \bibnamefont {{Hashimoto}}}, \bibinfo {author} {\bibfnamefont {K.}~\bibnamefont {{Kotake}}},\ and\ \bibinfo {author} {\bibfnamefont {S.}~\bibnamefont {{Yamada}}},\ }\href {https://doi.org/10.1086/509908} {\bibfield  {journal} {\bibinfo  {journal} {\apj}\ }\textbf {\bibinfo {volume} {656}},\ \bibinfo {pages} {382} (\bibinfo {year} {2007})},\ \Eprint {https://arxiv.org/abs/astro-ph/0602460} {arXiv:astro-ph/0602460 [astro-ph]} \BibitemShut {NoStop}%
\bibitem [{\citenamefont {Ono}\ \emph {et~al.}(2012)\citenamefont {Ono}, \citenamefont {Hashimoto}, \citenamefont {Fujimoto}, \citenamefont {Kotake},\ and\ \citenamefont {Yamada}}]{ono_2012}%
  \BibitemOpen
  \bibfield  {author} {\bibinfo {author} {\bibfnamefont {M.}~\bibnamefont {Ono}}, \bibinfo {author} {\bibfnamefont {M.-a.}\ \bibnamefont {Hashimoto}}, \bibinfo {author} {\bibfnamefont {S.-i.}\ \bibnamefont {Fujimoto}}, \bibinfo {author} {\bibfnamefont {K.}~\bibnamefont {Kotake}},\ and\ \bibinfo {author} {\bibfnamefont {S.}~\bibnamefont {Yamada}},\ }\href {https://doi.org/10.1143/PTP.128.741} {\bibfield  {journal} {\bibinfo  {journal} {Progress of Theoretical Physics}\ }\textbf {\bibinfo {volume} {128}},\ \bibinfo {pages} {741} (\bibinfo {year} {2012})},\ \Eprint {https://arxiv.org/abs/https://academic.oup.com/ptp/article-pdf/128/4/741/5434393/128-4-741.pdf} {https://academic.oup.com/ptp/article-pdf/128/4/741/5434393/128-4-741.pdf} \BibitemShut {NoStop}%
\bibitem [{\citenamefont {NAKAMURA}\ \emph {et~al.}(2013)\citenamefont {NAKAMURA}, \citenamefont {KAJINO}, \citenamefont {MATHEWS}, \citenamefont {SATO},\ and\ \citenamefont {HARIKAE}}]{nakamura_2013}%
  \BibitemOpen
  \bibfield  {author} {\bibinfo {author} {\bibfnamefont {K.}~\bibnamefont {NAKAMURA}}, \bibinfo {author} {\bibfnamefont {T.}~\bibnamefont {KAJINO}}, \bibinfo {author} {\bibfnamefont {G.~J.}\ \bibnamefont {MATHEWS}}, \bibinfo {author} {\bibfnamefont {S.}~\bibnamefont {SATO}},\ and\ \bibinfo {author} {\bibfnamefont {S.}~\bibnamefont {HARIKAE}},\ }\href {https://doi.org/10.1142/S0218301313300221} {\bibfield  {journal} {\bibinfo  {journal} {International Journal of Modern Physics E}\ }\textbf {\bibinfo {volume} {22}},\ \bibinfo {pages} {1330022} (\bibinfo {year} {2013})},\ \Eprint {https://arxiv.org/abs/https://doi.org/10.1142/S0218301313300221} {https://doi.org/10.1142/S0218301313300221} \BibitemShut {NoStop}%
\bibitem [{\citenamefont {{Siegel}}\ \emph {et~al.}(2019)\citenamefont {{Siegel}}, \citenamefont {{Barnes}},\ and\ \citenamefont {{Metzger}}}]{siegel_2019}%
  \BibitemOpen
  \bibfield  {author} {\bibinfo {author} {\bibfnamefont {D.~M.}\ \bibnamefont {{Siegel}}}, \bibinfo {author} {\bibfnamefont {J.}~\bibnamefont {{Barnes}}},\ and\ \bibinfo {author} {\bibfnamefont {B.~D.}\ \bibnamefont {{Metzger}}},\ }\href {https://doi.org/10.1038/s41586-019-1136-0} {\bibfield  {journal} {\bibinfo  {journal} {Nature}\ }\textbf {\bibinfo {volume} {569}},\ \bibinfo {pages} {241} (\bibinfo {year} {2019})},\ \Eprint {https://arxiv.org/abs/1810.00098} {arXiv:1810.00098 [astro-ph.HE]} \BibitemShut {NoStop}%
\bibitem [{\citenamefont {{Miller}}\ \emph {et~al.}(2020)\citenamefont {{Miller}}, \citenamefont {{Sprouse}}, \citenamefont {{Fryer}}, \citenamefont {{Ryan}}, \citenamefont {{Dolence}}, \citenamefont {{Mumpower}},\ and\ \citenamefont {{Surman}}}]{miller_2020}%
  \BibitemOpen
  \bibfield  {author} {\bibinfo {author} {\bibfnamefont {J.~M.}\ \bibnamefont {{Miller}}}, \bibinfo {author} {\bibfnamefont {T.~M.}\ \bibnamefont {{Sprouse}}}, \bibinfo {author} {\bibfnamefont {C.~L.}\ \bibnamefont {{Fryer}}}, \bibinfo {author} {\bibfnamefont {B.~R.}\ \bibnamefont {{Ryan}}}, \bibinfo {author} {\bibfnamefont {J.~C.}\ \bibnamefont {{Dolence}}}, \bibinfo {author} {\bibfnamefont {M.~R.}\ \bibnamefont {{Mumpower}}},\ and\ \bibinfo {author} {\bibfnamefont {R.}~\bibnamefont {{Surman}}},\ }\href {https://doi.org/10.3847/1538-4357/abb4e3} {\bibfield  {journal} {\bibinfo  {journal} {\apj}\ }\textbf {\bibinfo {volume} {902}},\ \bibinfo {eid} {66} (\bibinfo {year} {2020})},\ \Eprint {https://arxiv.org/abs/1912.03378} {arXiv:1912.03378 [astro-ph.HE]} \BibitemShut {NoStop}%
\bibitem [{\citenamefont {{Barnes}}\ and\ \citenamefont {{Metzger}}(2022)}]{barnes_2022}%
  \BibitemOpen
  \bibfield  {author} {\bibinfo {author} {\bibfnamefont {J.}~\bibnamefont {{Barnes}}}\ and\ \bibinfo {author} {\bibfnamefont {B.~D.}\ \bibnamefont {{Metzger}}},\ }\href {https://doi.org/10.3847/2041-8213/ac9b41} {\bibfield  {journal} {\bibinfo  {journal} {\apjl}\ }\textbf {\bibinfo {volume} {939}},\ \bibinfo {eid} {L29} (\bibinfo {year} {2022})},\ \Eprint {https://arxiv.org/abs/2205.10421} {arXiv:2205.10421 [astro-ph.HE]} \BibitemShut {NoStop}%
\bibitem [{\citenamefont {{Rastinejad}}\ \emph {et~al.}(2023)\citenamefont {{Rastinejad}}, \citenamefont {{Fong}}, \citenamefont {{Levan}}, \citenamefont {{Tanvir}}, \citenamefont {{Kilpatrick}}, \citenamefont {{Fruchter}}, \citenamefont {{Anand}}, \citenamefont {{Bhirombhakdi}}, \citenamefont {{Covino}}, \citenamefont {{Fynbo}}, \citenamefont {{Halevi}}, \citenamefont {{Hartmann}}, \citenamefont {{Heintz}}, \citenamefont {{Izzo}}, \citenamefont {{Jakobsson}}, \citenamefont {{Lamb}}, \citenamefont {{Malesani}}, \citenamefont {{Melandri}}, \citenamefont {{Metzger}}, \citenamefont {{Milvang-Jensen}}, \citenamefont {{Pian}}, \citenamefont {{Pugliese}}, \citenamefont {{Rossi}}, \citenamefont {{Siegel}}, \citenamefont {{Singh}},\ and\ \citenamefont {{Stratta}}}]{rastinejad_2023}%
  \BibitemOpen
  \bibfield  {author} {\bibinfo {author} {\bibfnamefont {J.~C.}\ \bibnamefont {{Rastinejad}}}, \bibinfo {author} {\bibfnamefont {W.}~\bibnamefont {{Fong}}}, \bibinfo {author} {\bibfnamefont {A.~J.}\ \bibnamefont {{Levan}}}, \bibinfo {author} {\bibfnamefont {N.~R.}\ \bibnamefont {{Tanvir}}}, \bibinfo {author} {\bibfnamefont {C.~D.}\ \bibnamefont {{Kilpatrick}}}, \bibinfo {author} {\bibfnamefont {A.~S.}\ \bibnamefont {{Fruchter}}}, \bibinfo {author} {\bibfnamefont {S.}~\bibnamefont {{Anand}}}, \bibinfo {author} {\bibfnamefont {K.}~\bibnamefont {{Bhirombhakdi}}}, \bibinfo {author} {\bibfnamefont {S.}~\bibnamefont {{Covino}}}, \bibinfo {author} {\bibfnamefont {J.~P.~U.}\ \bibnamefont {{Fynbo}}}, \bibinfo {author} {\bibfnamefont {G.}~\bibnamefont {{Halevi}}}, \bibinfo {author} {\bibfnamefont {D.~H.}\ \bibnamefont {{Hartmann}}}, \bibinfo {author} {\bibfnamefont {K.~E.}\ \bibnamefont {{Heintz}}}, \bibinfo {author} {\bibfnamefont {L.}~\bibnamefont {{Izzo}}}, \bibinfo {author} {\bibfnamefont {P.}~\bibnamefont
  {{Jakobsson}}}, \bibinfo {author} {\bibfnamefont {G.~P.}\ \bibnamefont {{Lamb}}}, \bibinfo {author} {\bibfnamefont {D.~B.}\ \bibnamefont {{Malesani}}}, \bibinfo {author} {\bibfnamefont {A.}~\bibnamefont {{Melandri}}}, \bibinfo {author} {\bibfnamefont {B.~D.}\ \bibnamefont {{Metzger}}}, \bibinfo {author} {\bibfnamefont {B.}~\bibnamefont {{Milvang-Jensen}}}, \bibinfo {author} {\bibfnamefont {E.}~\bibnamefont {{Pian}}}, \bibinfo {author} {\bibfnamefont {G.}~\bibnamefont {{Pugliese}}}, \bibinfo {author} {\bibfnamefont {A.}~\bibnamefont {{Rossi}}}, \bibinfo {author} {\bibfnamefont {D.~M.}\ \bibnamefont {{Siegel}}}, \bibinfo {author} {\bibfnamefont {P.}~\bibnamefont {{Singh}}},\ and\ \bibinfo {author} {\bibfnamefont {G.}~\bibnamefont {{Stratta}}},\ }\href {https://doi.org/10.48550/arXiv.2312.04630} {\bibfield  {journal} {\bibinfo  {journal} {arXiv e-prints}\ ,\ \bibinfo {eid} {arXiv:2312.04630}} (\bibinfo {year} {2023})},\ \Eprint {https://arxiv.org/abs/2312.04630} {arXiv:2312.04630 [astro-ph.HE]} \BibitemShut
  {NoStop}%
\bibitem [{\citenamefont {{Fujibayashi}}\ \emph {et~al.}(2023{\natexlab{b}})\citenamefont {{Fujibayashi}}, \citenamefont {{Tsz-Lok Lam}}, \citenamefont {{Shibata}},\ and\ \citenamefont {{Sekiguchi}}}]{fujibayashi_2023}%
  \BibitemOpen
  \bibfield  {author} {\bibinfo {author} {\bibfnamefont {S.}~\bibnamefont {{Fujibayashi}}}, \bibinfo {author} {\bibfnamefont {A.}~\bibnamefont {{Tsz-Lok Lam}}}, \bibinfo {author} {\bibfnamefont {M.}~\bibnamefont {{Shibata}}},\ and\ \bibinfo {author} {\bibfnamefont {Y.}~\bibnamefont {{Sekiguchi}}},\ }\href {https://doi.org/10.48550/arXiv.2309.02161} {\bibfield  {journal} {\bibinfo  {journal} {arXiv e-prints}\ ,\ \bibinfo {eid} {arXiv:2309.02161}} (\bibinfo {year} {2023}{\natexlab{b}})},\ \Eprint {https://arxiv.org/abs/2309.02161} {arXiv:2309.02161 [astro-ph.HE]} \BibitemShut {NoStop}%
\bibitem [{\citenamefont {Dean}\ and\ \citenamefont {Fern\'andez}(2024)}]{paper_1}%
  \BibitemOpen
  \bibfield  {author} {\bibinfo {author} {\bibfnamefont {C.}~\bibnamefont {Dean}}\ and\ \bibinfo {author} {\bibfnamefont {R.}~\bibnamefont {Fern\'andez}},\ }\href {https://doi.org/10.1103/PhysRevD.109.083010} {\bibfield  {journal} {\bibinfo  {journal} {Phys. Rev. D}\ }\textbf {\bibinfo {volume} {109}},\ \bibinfo {pages} {083010} (\bibinfo {year} {2024})}\BibitemShut {NoStop}%
\bibitem [{\citenamefont {{Woosley}}\ and\ \citenamefont {{Heger}}(2006)}]{woosley_2006b}%
  \BibitemOpen
  \bibfield  {author} {\bibinfo {author} {\bibfnamefont {S.~E.}\ \bibnamefont {{Woosley}}}\ and\ \bibinfo {author} {\bibfnamefont {A.}~\bibnamefont {{Heger}}},\ }\href {https://doi.org/10.1086/498500} {\bibfield  {journal} {\bibinfo  {journal} {\apj}\ }\textbf {\bibinfo {volume} {637}},\ \bibinfo {pages} {914} (\bibinfo {year} {2006})},\ \Eprint {https://arxiv.org/abs/astro-ph/0508175} {arXiv:astro-ph/0508175 [astro-ph]} \BibitemShut {NoStop}%
\bibitem [{\citenamefont {{O'Connor}}\ and\ \citenamefont {{Ott}}(2010)}]{oconnor_2010}%
  \BibitemOpen
  \bibfield  {author} {\bibinfo {author} {\bibfnamefont {E.}~\bibnamefont {{O'Connor}}}\ and\ \bibinfo {author} {\bibfnamefont {C.~D.}\ \bibnamefont {{Ott}}},\ }\href {https://doi.org/10.1088/0264-9381/27/11/114103} {\bibfield  {journal} {\bibinfo  {journal} {Classical and Quantum Gravity}\ }\textbf {\bibinfo {volume} {27}},\ \bibinfo {eid} {114103} (\bibinfo {year} {2010})},\ \Eprint {https://arxiv.org/abs/0912.2393} {arXiv:0912.2393 [astro-ph.HE]} \BibitemShut {NoStop}%
\bibitem [{\citenamefont {{Fryxell}}\ \emph {et~al.}(2000)\citenamefont {{Fryxell}}, \citenamefont {{Olson}}, \citenamefont {{Ricker}}, \citenamefont {{Timmes}}, \citenamefont {{Zingale}}, \citenamefont {{Lamb}}, \citenamefont {{MacNeice}}, \citenamefont {{Rosner}}, \citenamefont {{Truran}},\ and\ \citenamefont {{Tufo}}}]{fryxell00}%
  \BibitemOpen
  \bibfield  {author} {\bibinfo {author} {\bibfnamefont {B.}~\bibnamefont {{Fryxell}}}, \bibinfo {author} {\bibfnamefont {K.}~\bibnamefont {{Olson}}}, \bibinfo {author} {\bibfnamefont {P.}~\bibnamefont {{Ricker}}}, \bibinfo {author} {\bibfnamefont {F.~X.}\ \bibnamefont {{Timmes}}}, \bibinfo {author} {\bibfnamefont {M.}~\bibnamefont {{Zingale}}}, \bibinfo {author} {\bibfnamefont {D.~Q.}\ \bibnamefont {{Lamb}}}, \bibinfo {author} {\bibfnamefont {P.}~\bibnamefont {{MacNeice}}}, \bibinfo {author} {\bibfnamefont {R.}~\bibnamefont {{Rosner}}}, \bibinfo {author} {\bibfnamefont {J.~W.}\ \bibnamefont {{Truran}}},\ and\ \bibinfo {author} {\bibfnamefont {H.}~\bibnamefont {{Tufo}}},\ }\href {https://doi.org/10.1086/317361} {\bibfield  {journal} {\bibinfo  {journal} {ApJS}\ }\textbf {\bibinfo {volume} {131}},\ \bibinfo {pages} {273} (\bibinfo {year} {2000})}\BibitemShut {NoStop}%
\bibitem [{\citenamefont {Dubey}\ \emph {et~al.}(2009)\citenamefont {Dubey}, \citenamefont {Antypas}, \citenamefont {Ganapathy}, \citenamefont {Reid}, \citenamefont {Riley}, \citenamefont {Sheeler}, \citenamefont {Siegel},\ and\ \citenamefont {Weide}}]{dubey2009}%
  \BibitemOpen
  \bibfield  {author} {\bibinfo {author} {\bibfnamefont {A.}~\bibnamefont {Dubey}}, \bibinfo {author} {\bibfnamefont {K.}~\bibnamefont {Antypas}}, \bibinfo {author} {\bibfnamefont {M.~K.}\ \bibnamefont {Ganapathy}}, \bibinfo {author} {\bibfnamefont {L.~B.}\ \bibnamefont {Reid}}, \bibinfo {author} {\bibfnamefont {K.}~\bibnamefont {Riley}}, \bibinfo {author} {\bibfnamefont {D.}~\bibnamefont {Sheeler}}, \bibinfo {author} {\bibfnamefont {A.}~\bibnamefont {Siegel}},\ and\ \bibinfo {author} {\bibfnamefont {K.}~\bibnamefont {Weide}},\ }\href {https://doi.org/DOI: 10.1016/j.parco.2009.08.001} {\bibfield  {journal} {\bibinfo  {journal} {J. Par. Comp.}\ }\textbf {\bibinfo {volume} {35}},\ \bibinfo {pages} {512 } (\bibinfo {year} {2009})}\BibitemShut {NoStop}%
\bibitem [{\citenamefont {{Artemova}}\ \emph {et~al.}(1996)\citenamefont {{Artemova}}, \citenamefont {{Bjoernsson}},\ and\ \citenamefont {{Novikov}}}]{artemova1996}%
  \BibitemOpen
  \bibfield  {author} {\bibinfo {author} {\bibfnamefont {I.~V.}\ \bibnamefont {{Artemova}}}, \bibinfo {author} {\bibfnamefont {G.}~\bibnamefont {{Bjoernsson}}},\ and\ \bibinfo {author} {\bibfnamefont {I.~D.}\ \bibnamefont {{Novikov}}},\ }\href {https://doi.org/10.1086/177084} {\bibfield  {journal} {\bibinfo  {journal} {ApJ}\ }\textbf {\bibinfo {volume} {461}},\ \bibinfo {pages} {565} (\bibinfo {year} {1996})}\BibitemShut {NoStop}%
\bibitem [{\citenamefont {{Timmes}}\ and\ \citenamefont {{Swesty}}(2000)}]{timmes2000}%
  \BibitemOpen
  \bibfield  {author} {\bibinfo {author} {\bibfnamefont {F.~X.}\ \bibnamefont {{Timmes}}}\ and\ \bibinfo {author} {\bibfnamefont {F.~D.}\ \bibnamefont {{Swesty}}},\ }\href {https://doi.org/10.1086/313304} {\bibfield  {journal} {\bibinfo  {journal} {ApJS}\ }\textbf {\bibinfo {volume} {126}},\ \bibinfo {pages} {501} (\bibinfo {year} {2000})}\BibitemShut {NoStop}%
\bibitem [{\citenamefont {{Weaver}}\ \emph {et~al.}(1978)\citenamefont {{Weaver}}, \citenamefont {{Zimmerman}},\ and\ \citenamefont {{Woosley}}}]{weaver1978}%
  \BibitemOpen
  \bibfield  {author} {\bibinfo {author} {\bibfnamefont {T.~A.}\ \bibnamefont {{Weaver}}}, \bibinfo {author} {\bibfnamefont {G.~B.}\ \bibnamefont {{Zimmerman}}},\ and\ \bibinfo {author} {\bibfnamefont {S.~E.}\ \bibnamefont {{Woosley}}},\ }\href {https://doi.org/10.1086/156569} {\bibfield  {journal} {\bibinfo  {journal} {ApJ}\ }\textbf {\bibinfo {volume} {225}},\ \bibinfo {pages} {1021} (\bibinfo {year} {1978})}\BibitemShut {NoStop}%
\bibitem [{\citenamefont {{Seitenzahl}}\ \emph {et~al.}(2008)\citenamefont {{Seitenzahl}}, \citenamefont {{Timmes}}, \citenamefont {{Marin-Lafl{\`e}che}}, \citenamefont {{Brown}}, \citenamefont {{Magkotsios}},\ and\ \citenamefont {{Truran}}}]{seitenzahl_2008}%
  \BibitemOpen
  \bibfield  {author} {\bibinfo {author} {\bibfnamefont {I.~R.}\ \bibnamefont {{Seitenzahl}}}, \bibinfo {author} {\bibfnamefont {F.~X.}\ \bibnamefont {{Timmes}}}, \bibinfo {author} {\bibfnamefont {A.}~\bibnamefont {{Marin-Lafl{\`e}che}}}, \bibinfo {author} {\bibfnamefont {E.}~\bibnamefont {{Brown}}}, \bibinfo {author} {\bibfnamefont {G.}~\bibnamefont {{Magkotsios}}},\ and\ \bibinfo {author} {\bibfnamefont {J.}~\bibnamefont {{Truran}}},\ }\href {https://doi.org/10.1086/592501} {\bibfield  {journal} {\bibinfo  {journal} {\apjl}\ }\textbf {\bibinfo {volume} {685}},\ \bibinfo {pages} {L129} (\bibinfo {year} {2008})},\ \Eprint {https://arxiv.org/abs/0808.2033} {arXiv:0808.2033 [astro-ph]} \BibitemShut {NoStop}%
\bibitem [{\citenamefont {Lippuner}\ and\ \citenamefont {Roberts}(2017)}]{lippuner_2017_skynet}%
  \BibitemOpen
  \bibfield  {author} {\bibinfo {author} {\bibfnamefont {J.}~\bibnamefont {Lippuner}}\ and\ \bibinfo {author} {\bibfnamefont {L.~F.}\ \bibnamefont {Roberts}},\ }\href {https://doi.org/10.3847/1538-4365/aa94cb} {\bibfield  {journal} {\bibinfo  {journal} {The Astrophysical Journal Supplement Series}\ }\textbf {\bibinfo {volume} {233}},\ \bibinfo {pages} {18} (\bibinfo {year} {2017})}\BibitemShut {NoStop}%
\bibitem [{\citenamefont {{Lippuner}}\ \emph {et~al.}(2017)\citenamefont {{Lippuner}}, \citenamefont {{Fern{\'a}ndez}}, \citenamefont {{Roberts}}, \citenamefont {{Foucart}}, \citenamefont {{Kasen}}, \citenamefont {{Metzger}},\ and\ \citenamefont {{Ott}}}]{lippuner_2017}%
  \BibitemOpen
  \bibfield  {author} {\bibinfo {author} {\bibfnamefont {J.}~\bibnamefont {{Lippuner}}}, \bibinfo {author} {\bibfnamefont {R.}~\bibnamefont {{Fern{\'a}ndez}}}, \bibinfo {author} {\bibfnamefont {L.~F.}\ \bibnamefont {{Roberts}}}, \bibinfo {author} {\bibfnamefont {F.}~\bibnamefont {{Foucart}}}, \bibinfo {author} {\bibfnamefont {D.}~\bibnamefont {{Kasen}}}, \bibinfo {author} {\bibfnamefont {B.~D.}\ \bibnamefont {{Metzger}}},\ and\ \bibinfo {author} {\bibfnamefont {C.~D.}\ \bibnamefont {{Ott}}},\ }\href {https://doi.org/10.1093/mnras/stx1987} {\bibfield  {journal} {\bibinfo  {journal} {MNRAS}\ }\textbf {\bibinfo {volume} {472}},\ \bibinfo {pages} {904} (\bibinfo {year} {2017})},\ \Eprint {https://arxiv.org/abs/1703.06216} {arXiv:1703.06216 [astro-ph.HE]} \BibitemShut {NoStop}%
\bibitem [{\citenamefont {{Fern{\'a}ndez}}\ \emph {et~al.}(2020)\citenamefont {{Fern{\'a}ndez}}, \citenamefont {{Foucart}},\ and\ \citenamefont {{Lippuner}}}]{Fernandez2020BHNS}%
  \BibitemOpen
  \bibfield  {author} {\bibinfo {author} {\bibfnamefont {R.}~\bibnamefont {{Fern{\'a}ndez}}}, \bibinfo {author} {\bibfnamefont {F.}~\bibnamefont {{Foucart}}},\ and\ \bibinfo {author} {\bibfnamefont {J.}~\bibnamefont {{Lippuner}}},\ }\href {https://doi.org/10.1093/mnras/staa2209} {\bibfield  {journal} {\bibinfo  {journal} {\mnras}\ }\textbf {\bibinfo {volume} {497}},\ \bibinfo {pages} {3221} (\bibinfo {year} {2020})},\ \Eprint {https://arxiv.org/abs/2005.14208} {arXiv:2005.14208 [astro-ph.HE]} \BibitemShut {NoStop}%
\bibitem [{\citenamefont {{Fern{\'a}ndez}}\ \emph {et~al.}(2022)\citenamefont {{Fern{\'a}ndez}}, \citenamefont {{Richers}}, \citenamefont {{Mulyk}},\ and\ \citenamefont {{Fahlman}}}]{fernandez_2022}%
  \BibitemOpen
  \bibfield  {author} {\bibinfo {author} {\bibfnamefont {R.}~\bibnamefont {{Fern{\'a}ndez}}}, \bibinfo {author} {\bibfnamefont {S.}~\bibnamefont {{Richers}}}, \bibinfo {author} {\bibfnamefont {N.}~\bibnamefont {{Mulyk}}},\ and\ \bibinfo {author} {\bibfnamefont {S.}~\bibnamefont {{Fahlman}}},\ }\href {https://doi.org/10.1103/PhysRevD.106.103003} {\bibfield  {journal} {\bibinfo  {journal} {\prd}\ }\textbf {\bibinfo {volume} {106}},\ \bibinfo {eid} {103003} (\bibinfo {year} {2022})},\ \Eprint {https://arxiv.org/abs/2207.10680} {arXiv:2207.10680 [astro-ph.HE]} \BibitemShut {NoStop}%
\bibitem [{\citenamefont {{Cyburt}}\ \emph {et~al.}(2010)\citenamefont {{Cyburt}}, \citenamefont {{Amthor}}, \citenamefont {{Ferguson}}, \citenamefont {{Meisel}}, \citenamefont {{Smith}}, \citenamefont {{Warren}}, \citenamefont {{Heger}}, \citenamefont {{Hoffman}}, \citenamefont {{Rauscher}}, \citenamefont {{Sakharuk}}, \citenamefont {{Schatz}}, \citenamefont {{Thielemann}},\ and\ \citenamefont {{Wiescher}}}]{cyburt_2010}%
  \BibitemOpen
  \bibfield  {author} {\bibinfo {author} {\bibfnamefont {R.~H.}\ \bibnamefont {{Cyburt}}}, \bibinfo {author} {\bibfnamefont {A.~M.}\ \bibnamefont {{Amthor}}}, \bibinfo {author} {\bibfnamefont {R.}~\bibnamefont {{Ferguson}}}, \bibinfo {author} {\bibfnamefont {Z.}~\bibnamefont {{Meisel}}}, \bibinfo {author} {\bibfnamefont {K.}~\bibnamefont {{Smith}}}, \bibinfo {author} {\bibfnamefont {S.}~\bibnamefont {{Warren}}}, \bibinfo {author} {\bibfnamefont {A.~e.}\ \bibnamefont {{Heger}}}, \bibinfo {author} {\bibfnamefont {R.~D.}\ \bibnamefont {{Hoffman}}}, \bibinfo {author} {\bibfnamefont {T.}~\bibnamefont {{Rauscher}}}, \bibinfo {author} {\bibfnamefont {A.~e.}\ \bibnamefont {{Sakharuk}}}, \bibinfo {author} {\bibfnamefont {H.}~\bibnamefont {{Schatz}}}, \bibinfo {author} {\bibfnamefont {F.~K.}\ \bibnamefont {{Thielemann}}},\ and\ \bibinfo {author} {\bibfnamefont {M.}~\bibnamefont {{Wiescher}}},\ }\href {https://doi.org/10.1088/0067-0049/189/1/240} {\bibfield  {journal} {\bibinfo  {journal} {\apjs}\ }\textbf {\bibinfo
  {volume} {189}},\ \bibinfo {pages} {240} (\bibinfo {year} {2010})}\BibitemShut {NoStop}%
\bibitem [{\citenamefont {{Frankel}}\ and\ \citenamefont {{Metropolis}}(1947)}]{frankel_1947}%
  \BibitemOpen
  \bibfield  {author} {\bibinfo {author} {\bibfnamefont {S.}~\bibnamefont {{Frankel}}}\ and\ \bibinfo {author} {\bibfnamefont {N.}~\bibnamefont {{Metropolis}}},\ }\href {https://doi.org/10.1103/PhysRev.72.914} {\bibfield  {journal} {\bibinfo  {journal} {Physical Review}\ }\textbf {\bibinfo {volume} {72}},\ \bibinfo {pages} {914} (\bibinfo {year} {1947})}\BibitemShut {NoStop}%
\bibitem [{\citenamefont {{Mamdouh}}\ \emph {et~al.}(2001)\citenamefont {{Mamdouh}}, \citenamefont {{Pearson}}, \citenamefont {{Rayet}},\ and\ \citenamefont {{Tondeur}}}]{mamdouh_2001}%
  \BibitemOpen
  \bibfield  {author} {\bibinfo {author} {\bibfnamefont {A.}~\bibnamefont {{Mamdouh}}}, \bibinfo {author} {\bibfnamefont {J.~M.}\ \bibnamefont {{Pearson}}}, \bibinfo {author} {\bibfnamefont {M.}~\bibnamefont {{Rayet}}},\ and\ \bibinfo {author} {\bibfnamefont {F.}~\bibnamefont {{Tondeur}}},\ }\href {https://doi.org/10.1016/S0375-9474(00)00358-4} {\bibfield  {journal} {\bibinfo  {journal} {Nuclear Physics A}\ }\textbf {\bibinfo {volume} {679}},\ \bibinfo {pages} {337} (\bibinfo {year} {2001})},\ \Eprint {https://arxiv.org/abs/nucl-th/0010093} {arXiv:nucl-th/0010093 [nucl-th]} \BibitemShut {NoStop}%
\bibitem [{\citenamefont {{Wahl}}(2002)}]{wahl_2002}%
  \BibitemOpen
  \bibfield  {author} {\bibinfo {author} {\bibfnamefont {A.~C.}\ \bibnamefont {{Wahl}}},\ }\href@noop {} {\emph {\bibinfo {title} {{Technical Report LA-13928. Systematics of fission-product yields}}}}\ (\bibinfo  {publisher} {Los Alamos National Laboratory},\ \bibinfo {address} {Los Alamos, NM},\ \bibinfo {year} {2002})\BibitemShut {NoStop}%
\bibitem [{\citenamefont {{Panov}}\ \emph {et~al.}(2010)\citenamefont {{Panov}}, \citenamefont {{Korneev}}, \citenamefont {{Rauscher}}, \citenamefont {{Mart{\'\i}nez-Pinedo}}, \citenamefont {{Keli{\'c}-Heil}}, \citenamefont {{Zinner}},\ and\ \citenamefont {{Thielemann}}}]{panov_2010}%
  \BibitemOpen
  \bibfield  {author} {\bibinfo {author} {\bibfnamefont {I.~V.}\ \bibnamefont {{Panov}}}, \bibinfo {author} {\bibfnamefont {I.~Y.}\ \bibnamefont {{Korneev}}}, \bibinfo {author} {\bibfnamefont {T.}~\bibnamefont {{Rauscher}}}, \bibinfo {author} {\bibfnamefont {G.}~\bibnamefont {{Mart{\'\i}nez-Pinedo}}}, \bibinfo {author} {\bibfnamefont {A.}~\bibnamefont {{Keli{\'c}-Heil}}}, \bibinfo {author} {\bibfnamefont {N.~T.}\ \bibnamefont {{Zinner}}},\ and\ \bibinfo {author} {\bibfnamefont {F.~K.}\ \bibnamefont {{Thielemann}}},\ }\href {https://doi.org/10.1051/0004-6361/200911967} {\bibfield  {journal} {\bibinfo  {journal} {Astronomy \& Astrophysics}\ }\textbf {\bibinfo {volume} {513}},\ \bibinfo {eid} {A61} (\bibinfo {year} {2010})},\ \Eprint {https://arxiv.org/abs/0911.2181} {arXiv:0911.2181 [astro-ph.SR]} \BibitemShut {NoStop}%
\bibitem [{\citenamefont {{Fuller}}\ \emph {et~al.}(1982)\citenamefont {{Fuller}}, \citenamefont {{Fowler}},\ and\ \citenamefont {{Newman}}}]{fuller_1982}%
  \BibitemOpen
  \bibfield  {author} {\bibinfo {author} {\bibfnamefont {G.~M.}\ \bibnamefont {{Fuller}}}, \bibinfo {author} {\bibfnamefont {W.~A.}\ \bibnamefont {{Fowler}}},\ and\ \bibinfo {author} {\bibfnamefont {M.~J.}\ \bibnamefont {{Newman}}},\ }\href {https://doi.org/10.1086/190779} {\bibfield  {journal} {\bibinfo  {journal} {\apjs}\ }\textbf {\bibinfo {volume} {48}},\ \bibinfo {pages} {279} (\bibinfo {year} {1982})}\BibitemShut {NoStop}%
\bibitem [{\citenamefont {{Oda}}\ \emph {et~al.}(1994)\citenamefont {{Oda}}, \citenamefont {{Hino}}, \citenamefont {{Muto}}, \citenamefont {{Takahara}},\ and\ \citenamefont {{Sato}}}]{oda_1994}%
  \BibitemOpen
  \bibfield  {author} {\bibinfo {author} {\bibfnamefont {T.}~\bibnamefont {{Oda}}}, \bibinfo {author} {\bibfnamefont {M.}~\bibnamefont {{Hino}}}, \bibinfo {author} {\bibfnamefont {K.}~\bibnamefont {{Muto}}}, \bibinfo {author} {\bibfnamefont {M.}~\bibnamefont {{Takahara}}},\ and\ \bibinfo {author} {\bibfnamefont {K.}~\bibnamefont {{Sato}}},\ }\href {https://doi.org/10.1006/adnd.1994.1007} {\bibfield  {journal} {\bibinfo  {journal} {Atomic Data and Nuclear Data Tables}\ }\textbf {\bibinfo {volume} {56}},\ \bibinfo {pages} {231} (\bibinfo {year} {1994})}\BibitemShut {NoStop}%
\bibitem [{\citenamefont {{Langanke}}\ and\ \citenamefont {{Mart{\'\i}nez-Pinedo}}(2000)}]{langanke_2000}%
  \BibitemOpen
  \bibfield  {author} {\bibinfo {author} {\bibfnamefont {K.}~\bibnamefont {{Langanke}}}\ and\ \bibinfo {author} {\bibfnamefont {G.}~\bibnamefont {{Mart{\'\i}nez-Pinedo}}},\ }\href {https://doi.org/10.1016/S0375-9474(00)00131-7} {\bibfield  {journal} {\bibinfo  {journal} {Nuclear Physics A}\ }\textbf {\bibinfo {volume} {673}},\ \bibinfo {pages} {481} (\bibinfo {year} {2000})},\ \Eprint {https://arxiv.org/abs/nucl-th/0001018} {arXiv:nucl-th/0001018 [nucl-th]} \BibitemShut {NoStop}%
\bibitem [{\citenamefont {{M{\"o}ller}}\ \emph {et~al.}(2016)\citenamefont {{M{\"o}ller}}, \citenamefont {{Sierk}}, \citenamefont {{Ichikawa}},\ and\ \citenamefont {{Sagawa}}}]{moeller_2016}%
  \BibitemOpen
  \bibfield  {author} {\bibinfo {author} {\bibfnamefont {P.}~\bibnamefont {{M{\"o}ller}}}, \bibinfo {author} {\bibfnamefont {A.~J.}\ \bibnamefont {{Sierk}}}, \bibinfo {author} {\bibfnamefont {T.}~\bibnamefont {{Ichikawa}}},\ and\ \bibinfo {author} {\bibfnamefont {H.}~\bibnamefont {{Sagawa}}},\ }\href {https://doi.org/10.1016/j.adt.2015.10.002} {\bibfield  {journal} {\bibinfo  {journal} {Atomic Data and Nuclear Data Tables}\ }\textbf {\bibinfo {volume} {109}},\ \bibinfo {pages} {1} (\bibinfo {year} {2016})},\ \Eprint {https://arxiv.org/abs/1508.06294} {arXiv:1508.06294 [nucl-th]} \BibitemShut {NoStop}%
\bibitem [{\citenamefont {{Lippuner}}\ and\ \citenamefont {{Roberts}}(2015)}]{lippuner_2015}%
  \BibitemOpen
  \bibfield  {author} {\bibinfo {author} {\bibfnamefont {J.}~\bibnamefont {{Lippuner}}}\ and\ \bibinfo {author} {\bibfnamefont {L.~F.}\ \bibnamefont {{Roberts}}},\ }\href {https://doi.org/10.1088/0004-637X/815/2/82} {\bibfield  {journal} {\bibinfo  {journal} {\apj}\ }\textbf {\bibinfo {volume} {815}},\ \bibinfo {eid} {82} (\bibinfo {year} {2015})},\ \Eprint {https://arxiv.org/abs/1508.03133} {arXiv:1508.03133 [astro-ph.HE]} \BibitemShut {NoStop}%
\bibitem [{\citenamefont {{Narayan}}\ and\ \citenamefont {{Yi}}(1994)}]{narayan_1994}%
  \BibitemOpen
  \bibfield  {author} {\bibinfo {author} {\bibfnamefont {R.}~\bibnamefont {{Narayan}}}\ and\ \bibinfo {author} {\bibfnamefont {I.}~\bibnamefont {{Yi}}},\ }\href {https://doi.org/10.1086/187381} {\bibfield  {journal} {\bibinfo  {journal} {\apjl}\ }\textbf {\bibinfo {volume} {428}},\ \bibinfo {pages} {L13} (\bibinfo {year} {1994})},\ \Eprint {https://arxiv.org/abs/astro-ph/9403052} {arXiv:astro-ph/9403052 [astro-ph]} \BibitemShut {NoStop}%
\bibitem [{\citenamefont {{Chen}}\ and\ \citenamefont {{Beloborodov}}(2007)}]{chen_2007}%
  \BibitemOpen
  \bibfield  {author} {\bibinfo {author} {\bibfnamefont {W.-X.}\ \bibnamefont {{Chen}}}\ and\ \bibinfo {author} {\bibfnamefont {A.~M.}\ \bibnamefont {{Beloborodov}}},\ }\href {https://doi.org/10.1086/508923} {\bibfield  {journal} {\bibinfo  {journal} {\apj}\ }\textbf {\bibinfo {volume} {657}},\ \bibinfo {pages} {383} (\bibinfo {year} {2007})},\ \Eprint {https://arxiv.org/abs/astro-ph/0607145} {arXiv:astro-ph/0607145 [astro-ph]} \BibitemShut {NoStop}%
\bibitem [{\citenamefont {{De}}\ and\ \citenamefont {{Siegel}}(2021)}]{de_2021}%
  \BibitemOpen
  \bibfield  {author} {\bibinfo {author} {\bibfnamefont {S.}~\bibnamefont {{De}}}\ and\ \bibinfo {author} {\bibfnamefont {D.~M.}\ \bibnamefont {{Siegel}}},\ }\href {https://doi.org/10.3847/1538-4357/ac110b} {\bibfield  {journal} {\bibinfo  {journal} {\apj}\ }\textbf {\bibinfo {volume} {921}},\ \bibinfo {eid} {94} (\bibinfo {year} {2021})},\ \Eprint {https://arxiv.org/abs/2011.07176} {arXiv:2011.07176 [astro-ph.HE]} \BibitemShut {NoStop}%
\bibitem [{\citenamefont {{Surman}}\ \emph {et~al.}(2011)\citenamefont {{Surman}}, \citenamefont {{McLaughlin}},\ and\ \citenamefont {{Sabbatino}}}]{surman_2011}%
  \BibitemOpen
  \bibfield  {author} {\bibinfo {author} {\bibfnamefont {R.}~\bibnamefont {{Surman}}}, \bibinfo {author} {\bibfnamefont {G.~C.}\ \bibnamefont {{McLaughlin}}},\ and\ \bibinfo {author} {\bibfnamefont {N.}~\bibnamefont {{Sabbatino}}},\ }\href {https://doi.org/10.1088/0004-637X/743/2/155} {\bibfield  {journal} {\bibinfo  {journal} {\apj}\ }\textbf {\bibinfo {volume} {743}},\ \bibinfo {eid} {155} (\bibinfo {year} {2011})},\ \Eprint {https://arxiv.org/abs/1112.2673} {arXiv:1112.2673 [astro-ph.HE]} \BibitemShut {NoStop}%
\bibitem [{\citenamefont {{Barnes}}\ and\ \citenamefont {{Duffell}}(2023)}]{barnes_2023}%
  \BibitemOpen
  \bibfield  {author} {\bibinfo {author} {\bibfnamefont {J.}~\bibnamefont {{Barnes}}}\ and\ \bibinfo {author} {\bibfnamefont {P.~C.}\ \bibnamefont {{Duffell}}},\ }\href {https://doi.org/10.3847/1538-4357/acdb67} {\bibfield  {journal} {\bibinfo  {journal} {\apj}\ }\textbf {\bibinfo {volume} {952}},\ \bibinfo {eid} {96} (\bibinfo {year} {2023})},\ \Eprint {https://arxiv.org/abs/2305.00056} {arXiv:2305.00056 [astro-ph.HE]} \BibitemShut {NoStop}%
\bibitem [{\citenamefont {{Modjaz}}\ \emph {et~al.}(2016)\citenamefont {{Modjaz}}, \citenamefont {{Liu}}, \citenamefont {{Bianco}},\ and\ \citenamefont {{Graur}}}]{modjaz_2016}%
  \BibitemOpen
  \bibfield  {author} {\bibinfo {author} {\bibfnamefont {M.}~\bibnamefont {{Modjaz}}}, \bibinfo {author} {\bibfnamefont {Y.~Q.}\ \bibnamefont {{Liu}}}, \bibinfo {author} {\bibfnamefont {F.~B.}\ \bibnamefont {{Bianco}}},\ and\ \bibinfo {author} {\bibfnamefont {O.}~\bibnamefont {{Graur}}},\ }\href {https://doi.org/10.3847/0004-637X/832/2/108} {\bibfield  {journal} {\bibinfo  {journal} {\apj}\ }\textbf {\bibinfo {volume} {832}},\ \bibinfo {eid} {108} (\bibinfo {year} {2016})},\ \Eprint {https://arxiv.org/abs/1509.07124} {arXiv:1509.07124 [astro-ph.HE]} \BibitemShut {NoStop}%
\bibitem [{\citenamefont {{Grefenstette}}\ \emph {et~al.}(2014)\citenamefont {{Grefenstette}}, \citenamefont {{Harrison}}, \citenamefont {{Boggs}}, \citenamefont {{Reynolds}}, \citenamefont {{Fryer}}, \citenamefont {{Madsen}}, \citenamefont {{Wik}}, \citenamefont {{Zoglauer}}, \citenamefont {{Ellinger}}, \citenamefont {{Alexander}}, \citenamefont {{An}}, \citenamefont {{Barret}}, \citenamefont {{Christensen}}, \citenamefont {{Craig}}, \citenamefont {{Forster}}, \citenamefont {{Giommi}}, \citenamefont {{Hailey}}, \citenamefont {{Hornstrup}}, \citenamefont {{Kaspi}}, \citenamefont {{Kitaguchi}}, \citenamefont {{Koglin}}, \citenamefont {{Mao}}, \citenamefont {{Miyasaka}}, \citenamefont {{Mori}}, \citenamefont {{Perri}}, \citenamefont {{Pivovaroff}}, \citenamefont {{Puccetti}}, \citenamefont {{Rana}}, \citenamefont {{Stern}}, \citenamefont {{Westergaard}},\ and\ \citenamefont {{Zhang}}}]{grefenstette_2014}%
  \BibitemOpen
  \bibfield  {author} {\bibinfo {author} {\bibfnamefont {B.~W.}\ \bibnamefont {{Grefenstette}}}, \bibinfo {author} {\bibfnamefont {F.~A.}\ \bibnamefont {{Harrison}}}, \bibinfo {author} {\bibfnamefont {S.~E.}\ \bibnamefont {{Boggs}}}, \bibinfo {author} {\bibfnamefont {S.~P.}\ \bibnamefont {{Reynolds}}}, \bibinfo {author} {\bibfnamefont {C.~L.}\ \bibnamefont {{Fryer}}}, \bibinfo {author} {\bibfnamefont {K.~K.}\ \bibnamefont {{Madsen}}}, \bibinfo {author} {\bibfnamefont {D.~R.}\ \bibnamefont {{Wik}}}, \bibinfo {author} {\bibfnamefont {A.}~\bibnamefont {{Zoglauer}}}, \bibinfo {author} {\bibfnamefont {C.~I.}\ \bibnamefont {{Ellinger}}}, \bibinfo {author} {\bibfnamefont {D.~M.}\ \bibnamefont {{Alexander}}}, \bibinfo {author} {\bibfnamefont {H.}~\bibnamefont {{An}}}, \bibinfo {author} {\bibfnamefont {D.}~\bibnamefont {{Barret}}}, \bibinfo {author} {\bibfnamefont {F.~E.}\ \bibnamefont {{Christensen}}}, \bibinfo {author} {\bibfnamefont {W.~W.}\ \bibnamefont {{Craig}}}, \bibinfo {author} {\bibfnamefont
  {K.}~\bibnamefont {{Forster}}}, \bibinfo {author} {\bibfnamefont {P.}~\bibnamefont {{Giommi}}}, \bibinfo {author} {\bibfnamefont {C.~J.}\ \bibnamefont {{Hailey}}}, \bibinfo {author} {\bibfnamefont {A.}~\bibnamefont {{Hornstrup}}}, \bibinfo {author} {\bibfnamefont {V.~M.}\ \bibnamefont {{Kaspi}}}, \bibinfo {author} {\bibfnamefont {T.}~\bibnamefont {{Kitaguchi}}}, \bibinfo {author} {\bibfnamefont {J.~E.}\ \bibnamefont {{Koglin}}}, \bibinfo {author} {\bibfnamefont {P.~H.}\ \bibnamefont {{Mao}}}, \bibinfo {author} {\bibfnamefont {H.}~\bibnamefont {{Miyasaka}}}, \bibinfo {author} {\bibfnamefont {K.}~\bibnamefont {{Mori}}}, \bibinfo {author} {\bibfnamefont {M.}~\bibnamefont {{Perri}}}, \bibinfo {author} {\bibfnamefont {M.~J.}\ \bibnamefont {{Pivovaroff}}}, \bibinfo {author} {\bibfnamefont {S.}~\bibnamefont {{Puccetti}}}, \bibinfo {author} {\bibfnamefont {V.}~\bibnamefont {{Rana}}}, \bibinfo {author} {\bibfnamefont {D.}~\bibnamefont {{Stern}}}, \bibinfo {author} {\bibfnamefont {N.~J.}\ \bibnamefont
  {{Westergaard}}},\ and\ \bibinfo {author} {\bibfnamefont {W.~W.}\ \bibnamefont {{Zhang}}},\ }\href {https://doi.org/10.1038/nature12997} {\bibfield  {journal} {\bibinfo  {journal} {\nat}\ }\textbf {\bibinfo {volume} {506}},\ \bibinfo {pages} {339} (\bibinfo {year} {2014})},\ \Eprint {https://arxiv.org/abs/1403.4978} {arXiv:1403.4978 [astro-ph.HE]} \BibitemShut {NoStop}%
\bibitem [{\citenamefont {{Prantzos}}(2004)}]{prantzos_2004}%
  \BibitemOpen
  \bibfield  {author} {\bibinfo {author} {\bibfnamefont {N.}~\bibnamefont {{Prantzos}}},\ }in\ \href {https://doi.org/10.48550/arXiv.astro-ph/0404501} {\emph {\bibinfo {booktitle} {5th INTEGRAL Workshop on the INTEGRAL Universe}}},\ \bibinfo {series} {ESA Special Publication}, Vol.\ \bibinfo {volume} {552},\ \bibinfo {editor} {edited by\ \bibinfo {editor} {\bibfnamefont {V.}~\bibnamefont {{Schoenfelder}}}, \bibinfo {editor} {\bibfnamefont {G.}~\bibnamefont {{Lichti}}},\ and\ \bibinfo {editor} {\bibfnamefont {C.}~\bibnamefont {{Winkler}}}}\ (\bibinfo {year} {2004})\ p.~\bibinfo {pages} {15},\ \Eprint {https://arxiv.org/abs/astro-ph/0404501} {arXiv:astro-ph/0404501 [astro-ph]} \BibitemShut {NoStop}%
\bibitem [{\citenamefont {{Thielemann}}\ \emph {et~al.}(2018)\citenamefont {{Thielemann}}, \citenamefont {{Isern}}, \citenamefont {{Perego}},\ and\ \citenamefont {{von Ballmoos}}}]{thielemann_2018}%
  \BibitemOpen
  \bibfield  {author} {\bibinfo {author} {\bibfnamefont {F.-K.}\ \bibnamefont {{Thielemann}}}, \bibinfo {author} {\bibfnamefont {J.}~\bibnamefont {{Isern}}}, \bibinfo {author} {\bibfnamefont {A.}~\bibnamefont {{Perego}}},\ and\ \bibinfo {author} {\bibfnamefont {P.}~\bibnamefont {{von Ballmoos}}},\ }\href {https://doi.org/10.1007/s11214-018-0494-5} {\bibfield  {journal} {\bibinfo  {journal} {\ssr}\ }\textbf {\bibinfo {volume} {214}},\ \bibinfo {eid} {62} (\bibinfo {year} {2018})}\BibitemShut {NoStop}%
\bibitem [{\citenamefont {{Scott}}\ \emph {et~al.}(2015{\natexlab{a}})\citenamefont {{Scott}}, \citenamefont {{Grevesse}}, \citenamefont {{Asplund}}, \citenamefont {{Sauval}}, \citenamefont {{Lind}}, \citenamefont {{Takeda}}, \citenamefont {{Collet}}, \citenamefont {{Trampedach}},\ and\ \citenamefont {{Hayek}}}]{scott_2015a}%
  \BibitemOpen
  \bibfield  {author} {\bibinfo {author} {\bibfnamefont {P.}~\bibnamefont {{Scott}}}, \bibinfo {author} {\bibfnamefont {N.}~\bibnamefont {{Grevesse}}}, \bibinfo {author} {\bibfnamefont {M.}~\bibnamefont {{Asplund}}}, \bibinfo {author} {\bibfnamefont {A.~J.}\ \bibnamefont {{Sauval}}}, \bibinfo {author} {\bibfnamefont {K.}~\bibnamefont {{Lind}}}, \bibinfo {author} {\bibfnamefont {Y.}~\bibnamefont {{Takeda}}}, \bibinfo {author} {\bibfnamefont {R.}~\bibnamefont {{Collet}}}, \bibinfo {author} {\bibfnamefont {R.}~\bibnamefont {{Trampedach}}},\ and\ \bibinfo {author} {\bibfnamefont {W.}~\bibnamefont {{Hayek}}},\ }\href {https://doi.org/10.1051/0004-6361/201424109} {\bibfield  {journal} {\bibinfo  {journal} {\aap}\ }\textbf {\bibinfo {volume} {573}},\ \bibinfo {eid} {A25} (\bibinfo {year} {2015}{\natexlab{a}})},\ \Eprint {https://arxiv.org/abs/1405.0279} {arXiv:1405.0279 [astro-ph.SR]} \BibitemShut {NoStop}%
\bibitem [{\citenamefont {{Scott}}\ \emph {et~al.}(2015{\natexlab{b}})\citenamefont {{Scott}}, \citenamefont {{Asplund}}, \citenamefont {{Grevesse}}, \citenamefont {{Bergemann}},\ and\ \citenamefont {{Sauval}}}]{scott_2015b}%
  \BibitemOpen
  \bibfield  {author} {\bibinfo {author} {\bibfnamefont {P.}~\bibnamefont {{Scott}}}, \bibinfo {author} {\bibfnamefont {M.}~\bibnamefont {{Asplund}}}, \bibinfo {author} {\bibfnamefont {N.}~\bibnamefont {{Grevesse}}}, \bibinfo {author} {\bibfnamefont {M.}~\bibnamefont {{Bergemann}}},\ and\ \bibinfo {author} {\bibfnamefont {A.~J.}\ \bibnamefont {{Sauval}}},\ }\href {https://doi.org/10.1051/0004-6361/201424110} {\bibfield  {journal} {\bibinfo  {journal} {\aap}\ }\textbf {\bibinfo {volume} {573}},\ \bibinfo {eid} {A26} (\bibinfo {year} {2015}{\natexlab{b}})},\ \Eprint {https://arxiv.org/abs/1405.0287} {arXiv:1405.0287 [astro-ph.SR]} \BibitemShut {NoStop}%
\bibitem [{\citenamefont {Meija}\ \emph {et~al.}(2016)\citenamefont {Meija}, \citenamefont {Coplen}, \citenamefont {Berglund}, \citenamefont {Brand}, \citenamefont {Bièvre}, \citenamefont {Gröning}, \citenamefont {Holden}, \citenamefont {Irrgeher}, \citenamefont {Loss}, \citenamefont {Walczyk},\ and\ \citenamefont {Prohaska}}]{meija_2016}%
  \BibitemOpen
  \bibfield  {author} {\bibinfo {author} {\bibfnamefont {J.}~\bibnamefont {Meija}}, \bibinfo {author} {\bibfnamefont {T.~B.}\ \bibnamefont {Coplen}}, \bibinfo {author} {\bibfnamefont {M.}~\bibnamefont {Berglund}}, \bibinfo {author} {\bibfnamefont {W.~A.}\ \bibnamefont {Brand}}, \bibinfo {author} {\bibfnamefont {P.~D.}\ \bibnamefont {Bièvre}}, \bibinfo {author} {\bibfnamefont {M.}~\bibnamefont {Gröning}}, \bibinfo {author} {\bibfnamefont {N.~E.}\ \bibnamefont {Holden}}, \bibinfo {author} {\bibfnamefont {J.}~\bibnamefont {Irrgeher}}, \bibinfo {author} {\bibfnamefont {R.~D.}\ \bibnamefont {Loss}}, \bibinfo {author} {\bibfnamefont {T.}~\bibnamefont {Walczyk}},\ and\ \bibinfo {author} {\bibfnamefont {T.}~\bibnamefont {Prohaska}},\ }\href {https://doi.org/doi:10.1515/pac-2015-0503} {\bibfield  {journal} {\bibinfo  {journal} {Pure and Applied Chemistry}\ }\textbf {\bibinfo {volume} {88}},\ \bibinfo {pages} {293} (\bibinfo {year} {2016})}\BibitemShut {NoStop}%
\bibitem [{\citenamefont {{Nomoto}}(2017)}]{nomoto_2017}%
  \BibitemOpen
  \bibfield  {author} {\bibinfo {author} {\bibfnamefont {K.}~\bibnamefont {{Nomoto}}},\ }in\ \href {https://doi.org/10.1007/978-3-319-21846-5_86} {\emph {\bibinfo {booktitle} {Handbook of Supernovae}}},\ \bibinfo {editor} {edited by\ \bibinfo {editor} {\bibfnamefont {A.~W.}\ \bibnamefont {{Alsabti}}}\ and\ \bibinfo {editor} {\bibfnamefont {P.}~\bibnamefont {{Murdin}}}}\ (\bibinfo  {publisher} {Springer, Cham},\ \bibinfo {year} {2017})\ p.\ \bibinfo {pages} {1931}\BibitemShut {NoStop}%
\bibitem [{\citenamefont {{Sieverding}}\ \emph {et~al.}(2023)\citenamefont {{Sieverding}}, \citenamefont {{Kresse}},\ and\ \citenamefont {{Janka}}}]{sieverding_2023}%
  \BibitemOpen
  \bibfield  {author} {\bibinfo {author} {\bibfnamefont {A.}~\bibnamefont {{Sieverding}}}, \bibinfo {author} {\bibfnamefont {D.}~\bibnamefont {{Kresse}}},\ and\ \bibinfo {author} {\bibfnamefont {H.-T.}\ \bibnamefont {{Janka}}},\ }\href {https://doi.org/10.3847/2041-8213/ad045b} {\bibfield  {journal} {\bibinfo  {journal} {\apjl}\ }\textbf {\bibinfo {volume} {957}},\ \bibinfo {eid} {L25} (\bibinfo {year} {2023})},\ \Eprint {https://arxiv.org/abs/2308.09659} {arXiv:2308.09659 [astro-ph.HE]} \BibitemShut {NoStop}%
\bibitem [{\citenamefont {{Wang}}\ and\ \citenamefont {{Burrows}}(2024)}]{wang_2024b}%
  \BibitemOpen
  \bibfield  {author} {\bibinfo {author} {\bibfnamefont {T.}~\bibnamefont {{Wang}}}\ and\ \bibinfo {author} {\bibfnamefont {A.}~\bibnamefont {{Burrows}}},\ }\href {https://doi.org/10.48550/arXiv.2406.13746} {\bibfield  {journal} {\bibinfo  {journal} {arXiv e-prints}\ ,\ \bibinfo {eid} {arXiv:2406.13746}} (\bibinfo {year} {2024})},\ \Eprint {https://arxiv.org/abs/2406.13746} {arXiv:2406.13746 [astro-ph.HE]} \BibitemShut {NoStop}%
\bibitem [{\citenamefont {{Goriely}}(1999)}]{goriely_1999}%
  \BibitemOpen
  \bibfield  {author} {\bibinfo {author} {\bibfnamefont {S.}~\bibnamefont {{Goriely}}},\ }\href@noop {} {\bibfield  {journal} {\bibinfo  {journal} {\aap}\ }\textbf {\bibinfo {volume} {342}},\ \bibinfo {pages} {881} (\bibinfo {year} {1999})}\BibitemShut {NoStop}%
\bibitem [{\citenamefont {Schatz}\ \emph {et~al.}(1997)\citenamefont {Schatz}, \citenamefont {Aprahamian}, \citenamefont {Brown}, \citenamefont {Görres}, \citenamefont {Herndl}, \citenamefont {Kratz}, \citenamefont {Möller}, \citenamefont {Pfeiffer}, \citenamefont {Rauscher}, \citenamefont {Rembges}, \citenamefont {Thielemann}, \citenamefont {Wiescher},\ and\ \citenamefont {{van Wormer}}}]{schatz_1997}%
  \BibitemOpen
  \bibfield  {author} {\bibinfo {author} {\bibfnamefont {H.}~\bibnamefont {Schatz}}, \bibinfo {author} {\bibfnamefont {A.}~\bibnamefont {Aprahamian}}, \bibinfo {author} {\bibfnamefont {B.}~\bibnamefont {Brown}}, \bibinfo {author} {\bibfnamefont {J.}~\bibnamefont {Görres}}, \bibinfo {author} {\bibfnamefont {H.}~\bibnamefont {Herndl}}, \bibinfo {author} {\bibfnamefont {K.-L.}\ \bibnamefont {Kratz}}, \bibinfo {author} {\bibfnamefont {P.}~\bibnamefont {Möller}}, \bibinfo {author} {\bibfnamefont {B.}~\bibnamefont {Pfeiffer}}, \bibinfo {author} {\bibfnamefont {T.}~\bibnamefont {Rauscher}}, \bibinfo {author} {\bibfnamefont {J.}~\bibnamefont {Rembges}}, \bibinfo {author} {\bibfnamefont {F.-K.}\ \bibnamefont {Thielemann}}, \bibinfo {author} {\bibfnamefont {M.}~\bibnamefont {Wiescher}},\ and\ \bibinfo {author} {\bibfnamefont {L.}~\bibnamefont {{van Wormer}}},\ }\href {https://doi.org/https://doi.org/10.1016/S0375-9474(97)00283-2} {\bibfield  {journal} {\bibinfo  {journal} {Nuclear Physics A}\ }\textbf {\bibinfo
  {volume} {621}},\ \bibinfo {pages} {417} (\bibinfo {year} {1997})},\ \bibinfo {note} {nuclei in the Cosmos}\BibitemShut {NoStop}%
\bibitem [{IAE(2023)}]{IAEA}%
  \BibitemOpen
  \href@noop {} {\bibinfo {title} {Evaluated nuclear structure data file (ensdf)}},\ \bibinfo {howpublished} {\url{https://www.nndc.bnl.gov/ensdfarchivals/}} (\bibinfo {year} {2023}),\ \bibinfo {note} {accessed: 2024-04-30}\BibitemShut {NoStop}%
\bibitem [{\citenamefont {{Popham}}\ \emph {et~al.}(1999)\citenamefont {{Popham}}, \citenamefont {{Woosley}},\ and\ \citenamefont {{Fryer}}}]{popham1999}%
  \BibitemOpen
  \bibfield  {author} {\bibinfo {author} {\bibfnamefont {R.}~\bibnamefont {{Popham}}}, \bibinfo {author} {\bibfnamefont {S.~E.}\ \bibnamefont {{Woosley}}},\ and\ \bibinfo {author} {\bibfnamefont {C.}~\bibnamefont {{Fryer}}},\ }\href@noop {} {\bibfield  {journal} {\bibinfo  {journal} {ApJ}\ }\textbf {\bibinfo {volume} {518}},\ \bibinfo {pages} {356} (\bibinfo {year} {1999})}\BibitemShut {NoStop}%
\bibitem [{\citenamefont {{Fujimoto}}\ \emph {et~al.}(2006)\citenamefont {{Fujimoto}}, \citenamefont {{Kotake}}, \citenamefont {{Yamada}}, \citenamefont {{Hashimoto}},\ and\ \citenamefont {{Sato}}}]{fujimoto_2006}%
  \BibitemOpen
  \bibfield  {author} {\bibinfo {author} {\bibfnamefont {S.-i.}\ \bibnamefont {{Fujimoto}}}, \bibinfo {author} {\bibfnamefont {K.}~\bibnamefont {{Kotake}}}, \bibinfo {author} {\bibfnamefont {S.}~\bibnamefont {{Yamada}}}, \bibinfo {author} {\bibfnamefont {M.-a.}\ \bibnamefont {{Hashimoto}}},\ and\ \bibinfo {author} {\bibfnamefont {K.}~\bibnamefont {{Sato}}},\ }\href {https://doi.org/10.1086/503624} {\bibfield  {journal} {\bibinfo  {journal} {ApJ}\ }\textbf {\bibinfo {volume} {644}},\ \bibinfo {pages} {1040} (\bibinfo {year} {2006})},\ \Eprint {https://arxiv.org/abs/astro-ph/0602457} {arXiv:astro-ph/0602457 [astro-ph]} \BibitemShut {NoStop}%
\bibitem [{\citenamefont {{Ono}}\ \emph {et~al.}(2009)\citenamefont {{Ono}}, \citenamefont {{Hashimoto}}, \citenamefont {{Fujimoto}}, \citenamefont {{Kotake}},\ and\ \citenamefont {{Yamada}}}]{ono_2009}%
  \BibitemOpen
  \bibfield  {author} {\bibinfo {author} {\bibfnamefont {M.}~\bibnamefont {{Ono}}}, \bibinfo {author} {\bibfnamefont {M.}~\bibnamefont {{Hashimoto}}}, \bibinfo {author} {\bibfnamefont {S.}~\bibnamefont {{Fujimoto}}}, \bibinfo {author} {\bibfnamefont {K.}~\bibnamefont {{Kotake}}},\ and\ \bibinfo {author} {\bibfnamefont {S.}~\bibnamefont {{Yamada}}},\ }\href {https://doi.org/10.1143/PTP.122.755} {\bibfield  {journal} {\bibinfo  {journal} {Progress of Theoretical Physics}\ }\textbf {\bibinfo {volume} {122}},\ \bibinfo {pages} {755} (\bibinfo {year} {2009})}\BibitemShut {NoStop}%
\bibitem [{\citenamefont {{Shakura}}\ and\ \citenamefont {{Sunyaev}}(1973)}]{shakura1973}%
  \BibitemOpen
  \bibfield  {author} {\bibinfo {author} {\bibfnamefont {N.~I.}\ \bibnamefont {{Shakura}}}\ and\ \bibinfo {author} {\bibfnamefont {R.~A.}\ \bibnamefont {{Sunyaev}}},\ }\href@noop {} {\bibfield  {journal} {\bibinfo  {journal} {A\&A}\ }\textbf {\bibinfo {volume} {24}},\ \bibinfo {pages} {337} (\bibinfo {year} {1973})}\BibitemShut {NoStop}%
\bibitem [{\citenamefont {Fern{\'{a}}ndez}\ \emph {et~al.}(2018)\citenamefont {Fern{\'{a}}ndez}, \citenamefont {Quataert}, \citenamefont {Kashiyama},\ and\ \citenamefont {Coughlin}}]{fernandez_2018}%
  \BibitemOpen
  \bibfield  {author} {\bibinfo {author} {\bibfnamefont {R.}~\bibnamefont {Fern{\'{a}}ndez}}, \bibinfo {author} {\bibfnamefont {E.}~\bibnamefont {Quataert}}, \bibinfo {author} {\bibfnamefont {K.}~\bibnamefont {Kashiyama}},\ and\ \bibinfo {author} {\bibfnamefont {E.~R.}\ \bibnamefont {Coughlin}},\ }\href {https://doi.org/10.1093/mnras/sty306} {\bibfield  {journal} {\bibinfo  {journal} {Monthly Notices of the Royal Astronomical Society}\ }\textbf {\bibinfo {volume} {476}},\ \bibinfo {pages} {2366} (\bibinfo {year} {2018})}\BibitemShut {NoStop}%
\bibitem [{\citenamefont {{Janiuk}}(2019)}]{janiuk_2019}%
  \BibitemOpen
  \bibfield  {author} {\bibinfo {author} {\bibfnamefont {A.}~\bibnamefont {{Janiuk}}},\ }\href {https://doi.org/10.3847/1538-4357/ab3349} {\bibfield  {journal} {\bibinfo  {journal} {\apj}\ }\textbf {\bibinfo {volume} {882}},\ \bibinfo {eid} {163} (\bibinfo {year} {2019})},\ \Eprint {https://arxiv.org/abs/1907.00809} {arXiv:1907.00809 [astro-ph.HE]} \BibitemShut {NoStop}%
\bibitem [{\citenamefont {{Zenati}}\ \emph {et~al.}(2020)\citenamefont {{Zenati}}, \citenamefont {{Siegel}}, \citenamefont {{Metzger}},\ and\ \citenamefont {{Perets}}}]{zenati_2020}%
  \BibitemOpen
  \bibfield  {author} {\bibinfo {author} {\bibfnamefont {Y.}~\bibnamefont {{Zenati}}}, \bibinfo {author} {\bibfnamefont {D.~M.}\ \bibnamefont {{Siegel}}}, \bibinfo {author} {\bibfnamefont {B.~D.}\ \bibnamefont {{Metzger}}},\ and\ \bibinfo {author} {\bibfnamefont {H.~B.}\ \bibnamefont {{Perets}}},\ }\href {https://doi.org/10.1093/mnras/staa3002} {\bibfield  {journal} {\bibinfo  {journal} {\mnras}\ }\textbf {\bibinfo {volume} {499}},\ \bibinfo {pages} {4097} (\bibinfo {year} {2020})},\ \Eprint {https://arxiv.org/abs/2008.04309} {arXiv:2008.04309 [astro-ph.SR]} \BibitemShut {NoStop}%
\bibitem [{\citenamefont {{Christie}}\ \emph {et~al.}(2019)\citenamefont {{Christie}}, \citenamefont {{Lalakos}}, \citenamefont {{Tchekhovskoy}}, \citenamefont {{Fern{\'a}ndez}}, \citenamefont {{Foucart}}, \citenamefont {{Quataert}},\ and\ \citenamefont {{Kasen}}}]{christie_2019}%
  \BibitemOpen
  \bibfield  {author} {\bibinfo {author} {\bibfnamefont {I.~M.}\ \bibnamefont {{Christie}}}, \bibinfo {author} {\bibfnamefont {A.}~\bibnamefont {{Lalakos}}}, \bibinfo {author} {\bibfnamefont {A.}~\bibnamefont {{Tchekhovskoy}}}, \bibinfo {author} {\bibfnamefont {R.}~\bibnamefont {{Fern{\'a}ndez}}}, \bibinfo {author} {\bibfnamefont {F.}~\bibnamefont {{Foucart}}}, \bibinfo {author} {\bibfnamefont {E.}~\bibnamefont {{Quataert}}},\ and\ \bibinfo {author} {\bibfnamefont {D.}~\bibnamefont {{Kasen}}},\ }\href {https://doi.org/10.1093/mnras/stz2552} {\bibfield  {journal} {\bibinfo  {journal} {\mnras}\ }\textbf {\bibinfo {volume} {490}},\ \bibinfo {pages} {4811} (\bibinfo {year} {2019})},\ \Eprint {https://arxiv.org/abs/1907.02079} {arXiv:1907.02079 [astro-ph.HE]} \BibitemShut {NoStop}%
\bibitem [{\citenamefont {{Fahlman}}\ and\ \citenamefont {{Fern{\'a}ndez}}(2022)}]{FF22}%
  \BibitemOpen
  \bibfield  {author} {\bibinfo {author} {\bibfnamefont {S.}~\bibnamefont {{Fahlman}}}\ and\ \bibinfo {author} {\bibfnamefont {R.}~\bibnamefont {{Fern{\'a}ndez}}},\ }\href {https://doi.org/10.1093/mnras/stac948} {\bibfield  {journal} {\bibinfo  {journal} {\mnras}\ }\textbf {\bibinfo {volume} {513}},\ \bibinfo {pages} {2689} (\bibinfo {year} {2022})},\ \Eprint {https://arxiv.org/abs/2204.03005} {arXiv:2204.03005 [astro-ph.HE]} \BibitemShut {NoStop}%
\bibitem [{\citenamefont {{Hayashi}}\ \emph {et~al.}(2023)\citenamefont {{Hayashi}}, \citenamefont {{Kiuchi}}, \citenamefont {{Kyutoku}}, \citenamefont {{Sekiguchi}},\ and\ \citenamefont {{Shibata}}}]{hayashi_2023}%
  \BibitemOpen
  \bibfield  {author} {\bibinfo {author} {\bibfnamefont {K.}~\bibnamefont {{Hayashi}}}, \bibinfo {author} {\bibfnamefont {K.}~\bibnamefont {{Kiuchi}}}, \bibinfo {author} {\bibfnamefont {K.}~\bibnamefont {{Kyutoku}}}, \bibinfo {author} {\bibfnamefont {Y.}~\bibnamefont {{Sekiguchi}}},\ and\ \bibinfo {author} {\bibfnamefont {M.}~\bibnamefont {{Shibata}}},\ }\href {https://doi.org/10.1103/PhysRevD.107.123001} {\bibfield  {journal} {\bibinfo  {journal} {\prd}\ }\textbf {\bibinfo {volume} {107}},\ \bibinfo {eid} {123001} (\bibinfo {year} {2023})},\ \Eprint {https://arxiv.org/abs/2211.07158} {arXiv:2211.07158 [astro-ph.HE]} \BibitemShut {NoStop}%
\bibitem [{\citenamefont {{Nishimura}}\ \emph {et~al.}(2015)\citenamefont {{Nishimura}}, \citenamefont {{Takiwaki}},\ and\ \citenamefont {{Thielemann}}}]{nishimura_2015}%
  \BibitemOpen
  \bibfield  {author} {\bibinfo {author} {\bibfnamefont {N.}~\bibnamefont {{Nishimura}}}, \bibinfo {author} {\bibfnamefont {T.}~\bibnamefont {{Takiwaki}}},\ and\ \bibinfo {author} {\bibfnamefont {F.-K.}\ \bibnamefont {{Thielemann}}},\ }\href {https://doi.org/10.1088/0004-637X/810/2/109} {\bibfield  {journal} {\bibinfo  {journal} {\apj}\ }\textbf {\bibinfo {volume} {810}},\ \bibinfo {eid} {109} (\bibinfo {year} {2015})},\ \Eprint {https://arxiv.org/abs/1501.06567} {arXiv:1501.06567 [astro-ph.SR]} \BibitemShut {NoStop}%
\bibitem [{\citenamefont {{M{\"o}sta}}\ \emph {et~al.}(2018)\citenamefont {{M{\"o}sta}}, \citenamefont {{Roberts}}, \citenamefont {{Halevi}}, \citenamefont {{Ott}}, \citenamefont {{Lippuner}}, \citenamefont {{Haas}},\ and\ \citenamefont {{Schnetter}}}]{moesta_2018}%
  \BibitemOpen
  \bibfield  {author} {\bibinfo {author} {\bibfnamefont {P.}~\bibnamefont {{M{\"o}sta}}}, \bibinfo {author} {\bibfnamefont {L.~F.}\ \bibnamefont {{Roberts}}}, \bibinfo {author} {\bibfnamefont {G.}~\bibnamefont {{Halevi}}}, \bibinfo {author} {\bibfnamefont {C.~D.}\ \bibnamefont {{Ott}}}, \bibinfo {author} {\bibfnamefont {J.}~\bibnamefont {{Lippuner}}}, \bibinfo {author} {\bibfnamefont {R.}~\bibnamefont {{Haas}}},\ and\ \bibinfo {author} {\bibfnamefont {E.}~\bibnamefont {{Schnetter}}},\ }\href {https://doi.org/10.3847/1538-4357/aad6ec} {\bibfield  {journal} {\bibinfo  {journal} {\apj}\ }\textbf {\bibinfo {volume} {864}},\ \bibinfo {eid} {171} (\bibinfo {year} {2018})},\ \Eprint {https://arxiv.org/abs/1712.09370} {arXiv:1712.09370 [astro-ph.HE]} \BibitemShut {NoStop}%
\bibitem [{\citenamefont {{Reichert}}\ \emph {et~al.}(2024)\citenamefont {{Reichert}}, \citenamefont {{Bugli}}, \citenamefont {{Guilet}}, \citenamefont {{Obergaulinger}}, \citenamefont {{Aloy}},\ and\ \citenamefont {{Arcones}}}]{reichert_2024}%
  \BibitemOpen
  \bibfield  {author} {\bibinfo {author} {\bibfnamefont {M.}~\bibnamefont {{Reichert}}}, \bibinfo {author} {\bibfnamefont {M.}~\bibnamefont {{Bugli}}}, \bibinfo {author} {\bibfnamefont {J.}~\bibnamefont {{Guilet}}}, \bibinfo {author} {\bibfnamefont {M.}~\bibnamefont {{Obergaulinger}}}, \bibinfo {author} {\bibfnamefont {M.~{\'A}.}\ \bibnamefont {{Aloy}}},\ and\ \bibinfo {author} {\bibfnamefont {A.}~\bibnamefont {{Arcones}}},\ }\href {https://doi.org/10.1093/mnras/stae561} {\bibfield  {journal} {\bibinfo  {journal} {\mnras}\ }\textbf {\bibinfo {volume} {529}},\ \bibinfo {pages} {3197} (\bibinfo {year} {2024})},\ \Eprint {https://arxiv.org/abs/2401.14402} {arXiv:2401.14402 [astro-ph.HE]} \BibitemShut {NoStop}%
\bibitem [{\citenamefont {{Zha}}\ \emph {et~al.}(2024)\citenamefont {{Zha}}, \citenamefont {{M{\"u}ller}},\ and\ \citenamefont {{Powell}}}]{zha_2024}%
  \BibitemOpen
  \bibfield  {author} {\bibinfo {author} {\bibfnamefont {S.}~\bibnamefont {{Zha}}}, \bibinfo {author} {\bibfnamefont {B.}~\bibnamefont {{M{\"u}ller}}},\ and\ \bibinfo {author} {\bibfnamefont {J.}~\bibnamefont {{Powell}}},\ }\href {https://doi.org/10.3847/1538-4357/ad4ae7} {\bibfield  {journal} {\bibinfo  {journal} {\apj}\ }\textbf {\bibinfo {volume} {969}},\ \bibinfo {eid} {141} (\bibinfo {year} {2024})},\ \Eprint {https://arxiv.org/abs/2403.02072} {arXiv:2403.02072 [astro-ph.HE]} \BibitemShut {NoStop}%
\bibitem [{\citenamefont {{Fern{\'a}ndez}}\ \emph {et~al.}(2019)\citenamefont {{Fern{\'a}ndez}}, \citenamefont {{Tchekhovskoy}}, \citenamefont {{Quataert}}, \citenamefont {{Foucart}},\ and\ \citenamefont {{Kasen}}}]{fernandez2019}%
  \BibitemOpen
  \bibfield  {author} {\bibinfo {author} {\bibfnamefont {R.}~\bibnamefont {{Fern{\'a}ndez}}}, \bibinfo {author} {\bibfnamefont {A.}~\bibnamefont {{Tchekhovskoy}}}, \bibinfo {author} {\bibfnamefont {E.}~\bibnamefont {{Quataert}}}, \bibinfo {author} {\bibfnamefont {F.}~\bibnamefont {{Foucart}}},\ and\ \bibinfo {author} {\bibfnamefont {D.}~\bibnamefont {{Kasen}}},\ }\href {https://doi.org/10.1093/mnras/sty2932} {\bibfield  {journal} {\bibinfo  {journal} {MNRAS}\ }\textbf {\bibinfo {volume} {482}},\ \bibinfo {pages} {3373} (\bibinfo {year} {2019})},\ \Eprint {https://arxiv.org/abs/1808.00461} {arXiv:1808.00461 [astro-ph.HE]} \BibitemShut {NoStop}%
\bibitem [{\citenamefont {Childs}\ \emph {et~al.}(2012)\citenamefont {Childs} \emph {et~al.}}]{VisIt}%
  \BibitemOpen
  \bibfield  {author} {\bibinfo {author} {\bibfnamefont {H.}~\bibnamefont {Childs}} \emph {et~al.},\ }in\ \href {https://escholarship.org/uc/item/69r5m58v} {\emph {\bibinfo {booktitle} {{High Performance Visualization--Enabling Extreme-Scale Scientific Insight}}}}\ (\bibinfo  {publisher} {eScholarship, University of California},\ \bibinfo {year} {2012})\ pp.\ \bibinfo {pages} {357--372}\BibitemShut {NoStop}%
\bibitem [{\citenamefont {{Loken}}\ \emph {et~al.}(2010)\citenamefont {{Loken}}, \citenamefont {{Gruner}}, \citenamefont {{Groer}}, \citenamefont {{Peltier}}, \citenamefont {{Bunn}}, \citenamefont {{Craig}}, \citenamefont {{Henriques}}, \citenamefont {{Dempsey}}, \citenamefont {{Yu}}, \citenamefont {{Chen}}, \citenamefont {{Dursi}}, \citenamefont {{Chong}}, \citenamefont {{Northrup}}, \citenamefont {{Pinto}}, \citenamefont {{Knecht}},\ and\ \citenamefont {{Van Zon}}}]{SciNet}%
  \BibitemOpen
  \bibfield  {author} {\bibinfo {author} {\bibfnamefont {C.}~\bibnamefont {{Loken}}}, \bibinfo {author} {\bibfnamefont {D.}~\bibnamefont {{Gruner}}}, \bibinfo {author} {\bibfnamefont {L.}~\bibnamefont {{Groer}}}, \bibinfo {author} {\bibfnamefont {R.}~\bibnamefont {{Peltier}}}, \bibinfo {author} {\bibfnamefont {N.}~\bibnamefont {{Bunn}}}, \bibinfo {author} {\bibfnamefont {M.}~\bibnamefont {{Craig}}}, \bibinfo {author} {\bibfnamefont {T.}~\bibnamefont {{Henriques}}}, \bibinfo {author} {\bibfnamefont {J.}~\bibnamefont {{Dempsey}}}, \bibinfo {author} {\bibfnamefont {C.-H.}\ \bibnamefont {{Yu}}}, \bibinfo {author} {\bibfnamefont {J.}~\bibnamefont {{Chen}}}, \bibinfo {author} {\bibfnamefont {L.~J.}\ \bibnamefont {{Dursi}}}, \bibinfo {author} {\bibfnamefont {J.}~\bibnamefont {{Chong}}}, \bibinfo {author} {\bibfnamefont {S.}~\bibnamefont {{Northrup}}}, \bibinfo {author} {\bibfnamefont {J.}~\bibnamefont {{Pinto}}}, \bibinfo {author} {\bibfnamefont {N.}~\bibnamefont {{Knecht}}},\ and\ \bibinfo {author} {\bibfnamefont
  {R.}~\bibnamefont {{Van Zon}}},\ }in\ \href {https://doi.org/10.1088/1742-6596/256/1/012026} {\emph {\bibinfo {booktitle} {Journal of Physics Conference Series}}},\ \bibinfo {series} {Journal of Physics Conference Series}, Vol.\ \bibinfo {volume} {256}\ (\bibinfo {year} {2010})\ p.\ \bibinfo {pages} {012026}\BibitemShut {NoStop}%
\bibitem [{\citenamefont {Ponce}\ \emph {et~al.}(2019)\citenamefont {Ponce}, \citenamefont {van Zon}, \citenamefont {Northrup}, \citenamefont {Gruner}, \citenamefont {Chen}, \citenamefont {Ertinaz}, \citenamefont {Fedoseev}, \citenamefont {Groer}, \citenamefont {Mao}, \citenamefont {Mundim}, \citenamefont {Nolta}, \citenamefont {Pinto}, \citenamefont {Saldarriaga}, \citenamefont {Slavnic}, \citenamefont {Spence}, \citenamefont {Yu},\ and\ \citenamefont {Peltier}}]{Niagara}%
  \BibitemOpen
  \bibfield  {author} {\bibinfo {author} {\bibfnamefont {M.}~\bibnamefont {Ponce}}, \bibinfo {author} {\bibfnamefont {R.}~\bibnamefont {van Zon}}, \bibinfo {author} {\bibfnamefont {S.}~\bibnamefont {Northrup}}, \bibinfo {author} {\bibfnamefont {D.}~\bibnamefont {Gruner}}, \bibinfo {author} {\bibfnamefont {J.}~\bibnamefont {Chen}}, \bibinfo {author} {\bibfnamefont {F.}~\bibnamefont {Ertinaz}}, \bibinfo {author} {\bibfnamefont {A.}~\bibnamefont {Fedoseev}}, \bibinfo {author} {\bibfnamefont {L.}~\bibnamefont {Groer}}, \bibinfo {author} {\bibfnamefont {F.}~\bibnamefont {Mao}}, \bibinfo {author} {\bibfnamefont {B.~C.}\ \bibnamefont {Mundim}}, \bibinfo {author} {\bibfnamefont {M.}~\bibnamefont {Nolta}}, \bibinfo {author} {\bibfnamefont {J.}~\bibnamefont {Pinto}}, \bibinfo {author} {\bibfnamefont {M.}~\bibnamefont {Saldarriaga}}, \bibinfo {author} {\bibfnamefont {V.}~\bibnamefont {Slavnic}}, \bibinfo {author} {\bibfnamefont {E.}~\bibnamefont {Spence}}, \bibinfo {author} {\bibfnamefont {C.-H.}\ \bibnamefont {Yu}},\
  and\ \bibinfo {author} {\bibfnamefont {W.~R.}\ \bibnamefont {Peltier}},\ }in\ \href {https://doi.org/10.1145/3332186.3332195} {\emph {\bibinfo {booktitle} {Proceedings of the Practice and Experience in Advanced Research Computing on Rise of the Machines (Learning)}}},\ \bibinfo {series and number} {PEARC '19}\ (\bibinfo  {publisher} {Association for Computing Machinery},\ \bibinfo {address} {New York, NY, USA},\ \bibinfo {year} {2019})\BibitemShut {NoStop}%
\end{thebibliography}%

\end{document}